\RequirePackage{fix-cm}

\documentclass[11pt]{article}

\usepackage{arxiv}
\makeatletter
\input{size11.clo}
\makeatother
\AtBeginDocument{
  \newgeometry{
    textheight=9in,
    textwidth=5.75in,
    top=1in,
    headheight=16.5pt,
    headsep=25pt,
    footskip=30pt
  }
}

\usepackage{microtype}
\usepackage{xurl}
\usepackage{amsmath}
\usepackage{amssymb}
\usepackage{mathtools}
\usepackage{booktabs}
\usepackage{graphicx}
\graphicspath{{./figures/}}
\usepackage[numbers,sort&compress]{natbib}
\usepackage{authblk}

\usepackage{placeins}
\usepackage{algorithm}
\usepackage{algorithmic}

\makeatletter
\long\def\@makecaption#1#2{%
  \vskip\abovecaptionskip
  \begingroup
    \fontsize{9.6}{11}\selectfont
    \sbox\@tempboxa{#1: #2}%
    \ifdim\wd\@tempboxa >\hsize
      #1: #2\par
    \else
      \global\@minipagefalse
      \hb@xt@\hsize{\hfil\box\@tempboxa\hfil}%
    \fi
  \endgroup
  \vskip\belowcaptionskip}
\makeatother

\usepackage{hyperref}
\usepackage[noabbrev]{cleveref}

\crefname{section}{Section}{Sections}
\Crefname{section}{Section}{Sections}
\crefname{subappendix}{Appendix}{Appendices}
\Crefname{subappendix}{Appendix}{Appendices}

\title{A renormalization-group hierarchy of stochastic effective dynamics learned through path integrals}

\renewcommand{\shorttitle}{A renormalization-group hierarchy of effective dynamics}

\author[1,*]{Yuki Yasuda}
\author[2]{Tobias Bischoff}

\affil[1]{\footnotesize Research Institute for Earth and Information Sciences, Japan Agency for Marine-Earth Science and Technology, Yokohama, Japan}

\affil[2]{\footnotesize Aeolus Labs, San Francisco, United States}

\affil[*]{\footnotesize 3173-25 Showa-machi, Kanazawa-ku, Yokohama, Kanagawa 2360001, Japan. \url{yuki.yasuda@jamstec.go.jp}}

\date{}

\begin{document}

\maketitle

\begin{abstract}
This study combines a stochastic renormalization group (RG) in space with a path-integral description of the time evolution, giving a spatial hierarchy of coarse-grained dynamics together with the distribution over spatiotemporal paths.
The RG is defined as a diffusion process that applies scale-dependent Laplacian damping together with additive Gaussian noise.
The time evolution at each spatial scale is formulated as an Onsager--Machlup action, with a drift (i.e., the predictor) and white noise whose amplitude is fixed by the RG.
The predictor of these dynamics is optimized by minimizing the Kullback--Leibler divergence between the path distributions from the RG and from the path-integral description.
The optimal predictor contains the score function that connects the spatial scales, so the predictor and the score are two aspects of the same multiscale path formulation.
The formulation unifies simulation using the predictor, unconditional generation using the score, and super-resolution using both.
The predictor has no closed form, so it is computed by a neural network in two realizations that differ only in how the score is computed.
The first realization obtains the score by automatic differentiation of the path distribution, remaining faithful to that formulation.
The second obtains the score as an additional network output trained by denoising, at a lower computational cost.
These two realizations are validated through numerical experiments on two representative multiscale systems, the Kolmogorov flow and the two-timescale Lorenz-96 model.
\end{abstract}

\keywords{renormalization group \and path integrals \and coarse-graining \and stochastic dynamics \and diffusion models \and deep learning}

\section{Introduction}
\label{sec:introduction}

Many natural and engineered systems exhibit multiscale structures, where components at different scales interact \citep{Levin92,BarYam+04,Weinan+11,Newman+12,Alexakis+Biferale18}.
Such a system is generally described at a finite resolution, but a simple truncation at that resolution removes the influence of the small scales on the large-scale components.
For example, in the atmosphere, convection and clouds affect the large-scale flows, and numerical weather models parameterize these processes in terms of the resolved variables \citep{Stevens+Bony13,Bony+15,Berner+17}.
This kind of effective modeling is thus necessary to incorporate the small-scale effects into the large-scale dynamics.

In multiscale systems, however, clear scale separation is not always possible.
Physically, adjacent scales interact and sometimes form a hierarchy \citep{Ravasz+Barabasi03,Buzsaki+Draguhn04,Berloff+21}.
Practically, the scale of interest may be determined by the purpose of the modeling rather than by the nature of the system.
Although any scale may be chosen, the effective models at different scales are not independent of one another.
The model at a coarser scale should follow from the model that includes the finer scales.
An effective theory should be parameterized by the scale of interest.

In statistical physics, the renormalization group (RG) provides such a parameterized theory.
By progressively eliminating small-scale degrees of freedom while retaining their influence, the RG yields a family of effective models indexed by scale \citep{Wilson71}.
When scales are strongly coupled, this elimination cannot rely on perturbation methods.
Nonperturbative formulations are necessary \citep[e.g.,][]{Berges+02}, but they can be computationally expensive.
Recently, nonperturbative RG has been formulated as a continuous stochastic process in field space, so that the observables of the effective theory can be computed by sampling rather than by evaluating the effective action \citep{Carosso20,Monthus25}.
Furthermore, such stochastic RG flows can be computed numerically with diffusion models in deep learning \citep{Cotler+Rezchikov23a}.
The same technique has also been used to generate data such as images \citep{Sheshmani+25,Masuki+Ashida25}.
These works treat static data and do not examine how the state at a given spatial scale evolves in physical time.

The time evolution at each spatial scale may be described as a probability distribution over trajectories, because the stochastic RG gives a distribution of fields at that scale \citep{Carosso20,Cotler+Rezchikov23a,Monthus25}.
Interestingly, a similar view appears in effective theories not directly tied to RG, where the eliminated small scales act stochastically on the retained scales \citep{Zwanzig61,Mori65}.
However, little attention has been paid to a framework that provides both the hierarchy of spatial scales and the distribution over temporal trajectories at each scale, in a form to which deep learning can be applied.

\begin{figure}[t]
    \begin{center}
        \includegraphics[width=\textwidth]{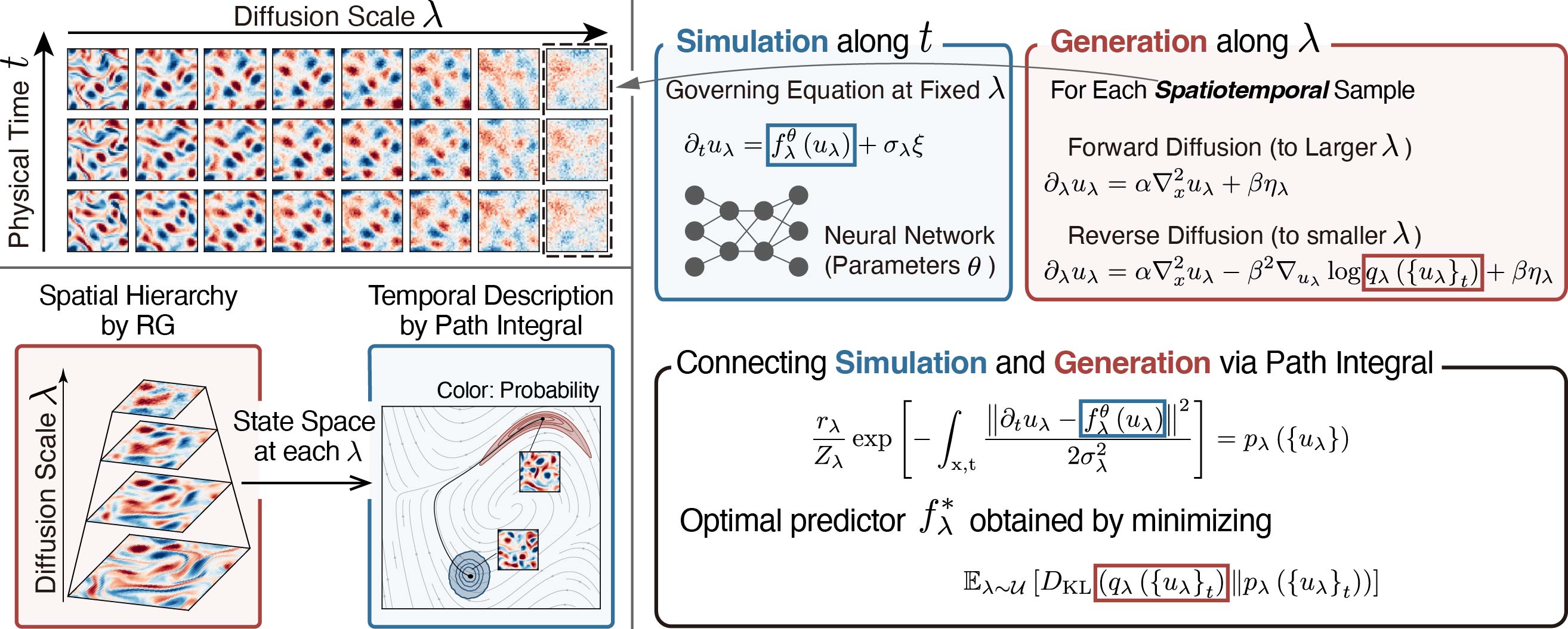} 
        \caption{
        Schematic of the proposed framework.
        Top left: the diffusion scale $\lambda$ (horizontal) indexes the spatial scale of coarse-graining, with small scales eliminated as $\lambda$ increases, while physical time $t$ (vertical) is the axis along which the dynamics evolve.
        Bottom left: the renormalization group (RG) gives a hierarchy of spatial scales, and the time evolution at each $\lambda$ is described by a path integral over trajectories in state space.
        Blue: \emph{simulation}, the forward evolution in $t$ at a fixed $\lambda$, whose drift is represented by a neural network.
        Red: \emph{generation}, the reverse-$\lambda$ integration of a whole spatiotemporal sample, which reverses the coarse-graining, from large $\lambda$ (coarse) back to small $\lambda$ (fine).
        Bottom right: the path integral links simulation and generation, and the optimal drift $f_\lambda^*$ is obtained by minimizing the Kullback--Leibler divergence between the path densities.
        }
        \label{fig:schematic}
    \end{center}
\end{figure}

This study proposes a framework that constructs the hierarchy of spatial scales with the RG and describes the time evolution at each scale as a probability distribution over trajectories (Fig.~\ref{fig:schematic}).
To specify this distribution, we introduce a path integral with the Onsager--Machlup action \citep{Onsager+Machlup53a}.
In this integral, the term that cannot be evaluated analytically is approximated from data by deep learning.
Since the framework spans both spatial scale and physical time, a single trained network unifies simulation at each scale, generation of spatiotemporal samples, and super-resolution.

The remainder of this paper is organized as follows.
\Cref{sec:framework} proposes the theoretical framework, and \Cref{sec:realizations} gives two computational realizations.
\Cref{sec:experiments} reports numerical experiments on the Kolmogorov flow and the Lorenz-96 model.
\Cref{sec:related-work} positions this work relative to previous studies, \Cref{sec:discussion} discusses the implications of our framework and future directions, and \Cref{sec:conclusions} presents the conclusions.
Theoretical details are collected in the \nameref{sec:appendix}, and implementation details and additional results in the supplementary material.

\section{Proposed framework}
\label{sec:framework}

\subsection{RG coarse-graining of spatiotemporal paths}
\label{subsec:rg-coarse-graining}

The stochastic formulations of RG define coarse-graining of static fields as a diffusion process along a scale parameter $\lambda$ \citep{Carosso20,Cotler+Rezchikov23a,Monthus25}.
We extend this construction to spatiotemporal paths by applying the same forward process to the field at each physical time $t$ independently, to preserve causality.

Let $u_0(t)$ denote the field at the finest scale, $\lambda = 0$, at time $t$.
The spatial argument is suppressed for notational simplicity, so that $u_\lambda(t)$ stands for $u_\lambda(x, t)$, where $x$ is the spatial coordinate in $d$ dimensions.
The field is also taken to be a scalar for simplicity.
For a vector field with several channels, each operator acts channel-wise and squared quantities become squared Euclidean norms over the channels.
Both $x$ and $t$ are discretized on the data grid, and the continuous notation used throughout is shorthand for the discrete sums or finite differences, as detailed at the beginning of the \nameref{sec:appendix}.

Our starting point is the probability density $q_0$ of the paths $\{u_0\}_t$, where a path is the collection of $u_0(t)$ over the time steps of some window.
In practice, $q_0$ is given only through its samples (i.e., the data), and its analytical form may be unknown, hence the use of deep learning to approximate $q_0$ (\Cref{sec:realizations}).
Coarse-graining $q_0$ along $\lambda$ yields a scale-indexed family of densities $\{q_\lambda\}$, which constitutes the hierarchy of spatial scales.

To construct this hierarchy, we apply the following forward process at each $t$:
\begin{equation}
\label{eq:forward-rg-sde}
    \partial_\lambda u_\lambda(t) = \alpha \nabla_x^2 u_\lambda(t) + \beta \eta_\lambda(t),
\end{equation}
where $\alpha, \beta > 0$ are constants, $\nabla_x^2$ is the spatial Laplacian, and $\eta_\lambda(t)$ is white Gaussian noise in $x$, $t$, and $\lambda$.
The Laplacian damps small scales, and the noise makes $u_\lambda(t)$ a random variable even for a given $u_0(t)$.
The noise represents the statistics of the eliminated scales \citep{Carosso20,Cotler+Rezchikov23a}.
In contrast, $\beta = 0$ gives deterministic smoothing and does not represent these statistics.
The roles of the two terms in Eq.~(\ref{eq:forward-rg-sde}) are discussed in \Cref{subsec:intermediate-states} and examined numerically in supplementary material Sections~S3.3 and S3.4.

Equation~(\ref{eq:forward-rg-sde}) has the analytic solution
\begin{equation}
\label{eq:ou-solution-spatiotemporal}
    u_\lambda(t) = \mathcal{C}_\lambda u_0(t) + \sqrt{\Sigma_\lambda}\,\epsilon(t),
\end{equation}
where $\mathcal{C}_\lambda := e^{\alpha\lambda\nabla_x^2}$ is the coarse-graining operator, $\Sigma_\lambda$ is the spatial covariance induced by the noise, and $\epsilon(t)$ is a standard Gaussian variable independent at each $x$ and $t$.
In Fourier space, $\mathcal{C}_\lambda$ multiplies the mode at wavenumber $k$ by $e^{-\alpha\lambda|k|^2}$, where $|k|$ is the magnitude of the wavenumber vector.
Modes with $|k| \gg k_{\mathrm{cut}} := 1/\sqrt{\alpha\lambda}$ (small scales) are eliminated, so $\lambda$ indexes the spatial scale of coarse-graining, with a cutoff wavelength proportional to $\sqrt{\alpha\lambda}$.
Note that coarse-graining is defined on a \emph{fixed} spatial grid, as in the stochastic RG and the RG-based diffusion models \citep{Carosso20,Cotler+Rezchikov23a,Monthus25}.
The coarse-graining formalism is detailed in \Cref{subsec:coarse-graining-operator-and-covariance-matrix}.

The forward process acts independently at each $t$, and thus the conditional density of a coarse-grained path $\{u_\lambda\}_t$ given the fine-resolution path $\{u_0\}_t$ factorizes over time:
\begin{equation}
\label{eq:conditional-density}
    q_\lambda(\{u_\lambda\}_t \mid \{u_0\}_t) = \prod_{t} \mathcal{N}\!\left(u_\lambda(t);\, \mathcal{C}_\lambda u_0(t),\, \Sigma_\lambda\right).
\end{equation}
We write the fine-resolution path density as $q_0(\{u_0\}_t) = Z_0^{-1}\exp[-S_0(\{u_0\}_t)]$, where $Z_0$ normalizes the density.
For deterministic dynamics, this expression is formal, since $q_0$ is then a delta function concentrated on the trajectories.
The coarse-grained path density is given by
\begin{equation}
\label{eq:joint-probability-by-RG}
    q_\lambda(\{u_\lambda\}_t) = \int \mathcal{D}u_0 \, q_\lambda(\{u_\lambda\}_t \mid \{u_0\}_t)\, q_0(\{u_0\}_t),
\end{equation}
where $\mathcal{D}u_0$ denotes the path measure, the product of $\mathrm{d}u_0$ over all spatiotemporal grid points.
Although the conditional density in Eq.~(\ref{eq:conditional-density}) factorizes over time, the marginal $q_\lambda(\{u_\lambda\}_t)$ cannot, because it inherits the temporal correlations of $q_0(\{u_0\}_t)$.
The Gaussian kernel $q_\lambda(\{u_\lambda\}_t \mid \{u_0\}_t)$ in Eq.~(\ref{eq:joint-probability-by-RG}) is normalized, so the convolution preserves the normalization \citep{Carosso20}.
The coarse-grained density is written as $q_\lambda(\{u_\lambda\}_t) = Z_0^{-1}\exp[-S_\lambda(\{u_\lambda\}_t)]$, with the same $Z_0$ for all $\lambda$.
We call $S_\lambda$ the effective action over paths at scale $\lambda$.

The convolution in Eq.~(\ref{eq:joint-probability-by-RG}) has a semigroup structure in $\lambda$: coarse-graining to $\lambda$ and then to $\lambda'$ equals coarse-graining to $\lambda + \lambda'$ (\Cref{subsec:coarse-graining-operator-and-covariance-matrix}).
In this sense it is an RG transformation \citep{Carosso20,Monthus25}.
We use the term RG for this successive elimination of small scales.
Unlike conventional RG, no rescaling of fields or wavenumbers is applied \citep{Carosso20,Monthus25}, though rescaling can be added when fixed points and critical exponents are of interest \citep{Carosso20,Cotler+Rezchikov23a}.
Without rescaling, the distribution becomes increasingly Gaussian as $\lambda$ grows \citep{Carosso20}.
We take the distribution at $\lambda = 1$ to be approximately Gaussian, which gives the initial condition of the reverse process introduced below.

The spatial hierarchy is also traversed from coarse to fine, to restore the eliminated components.
This inverse of coarse-graining is given by the reverse equation with respect to $\lambda$ \citep{Anderson82,Song+21,Cotler+Rezchikov23a},
\begin{equation}
\label{eq:reverse-rg-sde}
    \partial_\lambda u_\lambda(t) = \alpha \nabla_x^2 u_\lambda(t) - \beta^2 \nabla_{u_\lambda(t)} \ln q_\lambda(\{u_\lambda\}_t) + \beta \eta_\lambda(t),
\end{equation}
which is applied simultaneously at every $t$ of the path $\{u_\lambda\}_t$ and integrated from large to small $\lambda$.
Here, $\nabla_{u_\lambda(t)}$ denotes the gradient with respect to the field values at time $t$.
When Eq.~(\ref{eq:reverse-rg-sde}) is applied at another time $t'$, the gradient is taken at that time, $\nabla_{u_\lambda(t')} \ln q_\lambda(\{u_\lambda\}_t)$.
The distribution of the reverse process at each $\lambda$ coincides with that of the forward process, $q_\lambda(\{u_\lambda\}_t)$ \citep{Anderson82,Song+21,Cotler+Rezchikov23a}.
Started from the Gaussian at $\lambda = 1$, the reverse equation yields paths distributed approximately as $q_0(\{u_0\}_t)$.
Started from an intermediate $\lambda > 0$, it restores the components eliminated up to that scale (\Cref{subsec:effective-theories}).
The gradient $\nabla_{u_\lambda(t)} \ln q_\lambda(\{u_\lambda\}_t)$ in Eq.~(\ref{eq:reverse-rg-sde}), however, involves the unknown $q_\lambda(\{u_\lambda\}_t)$.
The following subsections give a tractable approximation of $q_\lambda(\{u_\lambda\}_t)$.
This approximation also provides the time evolution at each scale $\lambda$.

\subsection{Time evolution at each scale as a path integral}
\label{subsec:path-integral}

The stochastic RG in space introduces randomness into the time evolution of the coarse-grained state.
From Eq.~(\ref{eq:ou-solution-spatiotemporal}), the temporal difference of $u_\lambda$ over a time step $\Delta t$ splits into two parts:
\begin{equation}
\label{eq:temporal-difference}
    \frac{u_\lambda(t + \Delta t) - u_\lambda(t)}{\Delta t} = \frac{\mathcal{C}_\lambda \left[u_0(t + \Delta t) - u_0(t)\right]}{\Delta t} + \frac{\sqrt{\Sigma_\lambda} \left[\epsilon(t + \Delta t) - \epsilon(t)\right]}{\Delta t}.
\end{equation}
On the right-hand side, the first term is the coarse-grained increment of the fine-resolution field $u_0$, and the second is the increment of the forward-process noise $\epsilon$.
When the fine-resolution dynamics are known, the increment $u_0(t + \Delta t) - u_0(t)$ is given, for example, by a discretized partial differential equation.
The noise increment in Eq.~(\ref{eq:temporal-difference}) is a random variable correlated across time (adjacent increments share $\epsilon(t)$) and across space (through $\Sigma_\lambda$).
The evolution at each $\lambda > 0$ is thus stochastic even when that at $\lambda = 0$ is deterministic.

To describe this stochastic process simply, we assume the following form:
\begin{equation}
\label{eq:governing-eq-along-t}
    \partial_t u_\lambda(t) = f_\lambda^\theta(u_\lambda(t)) + \sigma_\lambda \xi(t),
\end{equation}
where $f_\lambda^\theta$ is a drift that depends on the current state, $\theta$ parameterizes the assumed family of drifts, $\xi(t)$ is spatiotemporal white Gaussian noise, and $\sigma_\lambda$ is a state-independent amplitude that depends on $\lambda$ only.
We set $\sigma_\lambda$ from the forward process in Eq.~(\ref{eq:forward-rg-sde}) by matching the variance of the noise increment $\sigma_\lambda \xi(t)\,\Delta t$ to the mean diagonal element of $\Sigma_\lambda$ (\Cref{subsec:specification-of-the-noise-amplitude-sigma-lambda}).
Under the approximation of Eq.~(\ref{eq:governing-eq-along-t}), the time evolution at each $\lambda$ is determined once $f_\lambda^\theta$ is specified.
This approximation is revisited in \Cref{subsec:optimal-predictor}.

The stochastic equation~(\ref{eq:governing-eq-along-t}) determines the probability density of a path.
Under the It\^{o} convention, this density takes the Onsager--Machlup form \citep{Onsager+Machlup53a,Shiraishi23}:
\begin{align}
\label{eq:path-probability-density}
    p_\lambda(\{u_\lambda\}_t) &= \frac{r_\lambda}{Z_\lambda} \exp\!\left[- S_\lambda^\theta(\{u_\lambda\}_t)\right], \notag \\
    S_\lambda^\theta(\{u_\lambda\}_t) &:= \frac{1}{2\sigma_\lambda^2} \int \mathrm{d}^d x\,\mathrm{d}t \, \left\| \partial_t u_\lambda(t) - f_\lambda^\theta(u_\lambda(t)) \right\|^2,
\end{align}
where $r_\lambda$ denotes the initial density, $Z_\lambda$ a normalization constant independent of $f_\lambda^\theta$, and $\|\cdot\|$ the Euclidean norm over the components of its argument (i.e., the channels at each spatiotemporal grid point).
In discrete time, $p_\lambda(\{u_\lambda\}_t)$ is the likelihood of the path under the Euler--Maruyama discretization (\Cref{subsec:path-probability-measure-via-the-path-integral-representation}).
If $f_\lambda^\theta$ is chosen so that $p_\lambda(\{u_\lambda\}_t) \simeq q_\lambda(\{u_\lambda\}_t)$, then $S_\lambda^\theta$ approximates the effective action $S_\lambda$ of \Cref{subsec:rg-coarse-graining} up to the contributions of the initial density and the normalization.

The drift is determined by minimizing the Kullback--Leibler (KL) divergence between $q_\lambda(\{u_\lambda\}_t)$ and $p_\lambda(\{u_\lambda\}_t)$:
\begin{align}
\label{eq:kl-divergence}
    D_{\mathrm{KL}}(q_\lambda \| p_\lambda) &= \int \mathcal{D}u_\lambda\, q_\lambda(\{u_\lambda\}_t) \ln \frac{q_\lambda(\{u_\lambda\}_t)}{p_\lambda(\{u_\lambda\}_t)} \notag \\
    &= \frac{1}{2\sigma_\lambda^2} \mathbb{E}_{q_\lambda}\!\left[\int \mathrm{d}^d x\,\mathrm{d}t \, \left\| \partial_t u_\lambda(t) - f_\lambda^\theta(u_\lambda(t)) \right\|^2\right] + \mathrm{const},
\end{align}
where the terms independent of $f_\lambda^\theta$, including $\ln Z_\lambda$ and $\ln r_\lambda$, are collected in the constant.
The minimization becomes a least-squares fit of the time derivative, weighted by $1/(2\sigma_\lambda^2)$.
The expectation is taken over samples of $q_\lambda(\{u_\lambda\}_t)$, i.e., paths at $\lambda = 0$ passed through the solution in Eq.~(\ref{eq:ou-solution-spatiotemporal}).
To treat all scales at once, the KL divergence is averaged over $\lambda \sim \mathcal{U}(0, 1)$:
\begin{equation}
\label{eq:averaged-kl}
    \mathbb{E}_{\lambda \sim \mathcal{U}(0,1)}\!\left[ D_{\mathrm{KL}}(q_\lambda \| p_\lambda) \right].
\end{equation}
The minimizer $f_\lambda^*$ is the optimal drift (or predictor) at each scale, and its form is derived next.

\subsection{The optimal predictor \texorpdfstring{$f_\lambda^*$}{f\_lambda*}}
\label{subsec:optimal-predictor}

The minimizer of Eq.~(\ref{eq:kl-divergence}) is the conditional expectation of the temporal increment given the current state (\Cref{subsec:kl-minimization-and-optimal-predictor}):
\begin{equation}
\label{eq:optimal-drift}
    f_\lambda^*(u_\lambda(t)) = \mathbb{E}\!\left[\frac{u_\lambda(t + \Delta t) - u_\lambda(t)}{\Delta t} \;\middle|\; u_\lambda(t)\right],
\end{equation}
where the expectation is taken under the joint distribution of $u_0$ and $\epsilon$ through Eq.~(\ref{eq:ou-solution-spatiotemporal}).
Substituting the temporal difference in Eq.~(\ref{eq:temporal-difference}) splits $f_\lambda^*$ into two terms:
\begin{equation}
\label{eq:optimal-drift-two-terms}
    f_\lambda^*(u_\lambda(t)) = \mathbb{E}\!\left[\frac{\mathcal{C}_\lambda \left[u_0(t + \Delta t) - u_0(t)\right]}{\Delta t} \;\middle|\; u_\lambda(t)\right] - \mathbb{E}\!\left[\frac{\sqrt{\Sigma_\lambda}\, \epsilon(t)}{\Delta t} \;\middle|\; u_\lambda(t)\right].
\end{equation}
Since $\epsilon(t + \Delta t)$ is independent of $u_\lambda(t)$, its conditional expectation vanishes, whereas $\epsilon(t)$ is contained in $u_\lambda(t)$ and remains.
The first term is the evolution at $\lambda = 0$ followed by coarse-graining, so it carries the influence of the eliminated small scales on the large scales.
The second term corrects for the uncertainty in $u_\lambda(t)$ itself.

This second term can be represented by a score function (hereafter simply the score).
For the additive Gaussian noise of the forward process, Tweedie's formula \citep{Efron11} gives (\Cref{subsec:kl-minimization-and-optimal-predictor})
\begin{equation}
\label{eq:tweedie-single-time}
    -\mathbb{E}\!\left[\sqrt{\Sigma_\lambda}\, \epsilon(t) \;\middle|\; u_\lambda(t)\right] = \Sigma_\lambda \nabla_{u_\lambda(t)} \ln q_\lambda(u_\lambda(t)),
\end{equation}
where $q_\lambda(u_\lambda(t))$ is the marginal density of the state at time $t$.
From Eqs.~(\ref{eq:optimal-drift-two-terms}) and~(\ref{eq:tweedie-single-time}), the optimal predictor $f_\lambda^*(u_\lambda(t))$ contains the score $\nabla_{u_\lambda(t)} \ln q_\lambda(u_\lambda(t))$, and both quantities are derived from the same distribution.
This suggests that the predictor and the score may be computed from shared features.
This score, however, is that of the marginal at time $t$ (i.e., the single-time score), not the path score $\nabla_{u_\lambda(t)} \ln q_\lambda(\{u_\lambda\}_t)$ required by the reverse equation~(\ref{eq:reverse-rg-sde}).
The relation between the two scores is discussed in \Cref{subsec:sa-pdd}.

The white-noise assumption in Eq.~(\ref{eq:governing-eq-along-t}) can now be interpreted.
The noise in the increment from $t$ to $t + \Delta t$ is $\sqrt{\Sigma_\lambda}[\epsilon(t + \Delta t) - \epsilon(t)]$, and $\epsilon(t)$ also appears in the preceding increment from $t-\Delta t$ to $t$.
This shared term is the origin of the temporal correlation.
The current state $u_\lambda(t) = \mathcal{C}_\lambda u_0(t) + \sqrt{\Sigma_\lambda}\,\epsilon(t)$ contains $\epsilon(t)$, so the contribution $-\sqrt{\Sigma_\lambda}\,\epsilon(t)$ to the increment is partly predictable from $u_\lambda(t)$.
The second term of $f_\lambda^*$ in Eq.~(\ref{eq:optimal-drift-two-terms}) is exactly this predictable part, expressed as the single-time score in Eq.~(\ref{eq:tweedie-single-time}).
The remaining part is $\epsilon(t + \Delta t)$ and the unpredictable residual of $\epsilon(t)$.
Representing these by white noise is the assumption of Eq.~(\ref{eq:governing-eq-along-t}).

\subsection{Stochastic effective theory for various scales}
\label{subsec:effective-theories}

The optimal predictor $f_\lambda^*$ gives a stochastic description of the coarse-grained time evolution at every $\lambda \in [0, 1]$, a scale-indexed family of stochastic effective models.
These models are obtained from the data at $\lambda = 0$ alone, or from $q_0(\{u_0\}_t)$ when its analytical form is known.
Let $p_\lambda^*(\{u_\lambda\}_t)$ denote $p_\lambda(\{u_\lambda\}_t)$ with $f_\lambda^\theta = f_\lambda^*$.
Since $p_\lambda^*(\{u_\lambda\}_t)$ is given explicitly, it also provides the path score required by the reverse equation~(\ref{eq:reverse-rg-sde}),
\begin{equation}
\label{eq:score-approximation}
    \nabla_{u_\lambda(t)} \ln q_\lambda(\{u_\lambda\}_t) \approx \nabla_{u_\lambda(t)} \ln p_\lambda^*(\{u_\lambda\}_t).
\end{equation}

From the optimal predictor and the path score, three kinds of computations follow.
\emph{Simulation} integrates Eq.~(\ref{eq:governing-eq-along-t}) forward in $t$ at a fixed $\lambda$ from an initial state.
\emph{Generation} integrates the reverse equation~(\ref{eq:reverse-rg-sde}) from the Gaussian at $\lambda = 1$ down to $\lambda = 0$ and produces a spatiotemporal sample distributed approximately as $q_0(\{u_0\}_t)$.
\emph{Super-resolution} combines the two: a simulation at a coarse $\lambda > 0$ is followed by the reverse equation from that $\lambda$ down to $\lambda = 0$, restoring the eliminated components.
All three tasks are performed with the same $f_\lambda^*$.
\section{Computational realizations}
\label{sec:realizations}

The optimal drift $f_\lambda^*$ in Eq.~(\ref{eq:optimal-drift}) has no closed form, so we compute it by deep learning.
A single neural network $\hat f_\lambda^\theta$ that takes $\lambda$ as an input approximates $f_\lambda^*$ for all $\lambda$ at once.
Its parameters $\theta$ are obtained by minimizing the $\lambda$-averaged KL divergence in Eq.~(\ref{eq:averaged-kl}) over coarse-grained samples, a minimization referred to as training below.
The trained network is written $\hat f_\lambda^*$.
With $\hat f_\lambda^*$ in place of $f_\lambda^*$, simulation proceeds as in \Cref{subsec:effective-theories}.
Generation and super-resolution need the path score for the reverse equation~(\ref{eq:reverse-rg-sde}), and we present two methods to obtain it.
Predictor-Driven Diffusion (PDD) differentiates the path density built from $\hat f_\lambda^*$; Score-Augmented Predictor-Driven Diffusion (SA-PDD) adds a second output that estimates the noise $\epsilon$ of the forward process, from which the score is computed.

\subsection{Predictor-Driven Diffusion (PDD)}
\label{subsec:pdd}

The network $\hat f_\lambda^\theta$ takes as input $\lambda$ and the current and a finite number of past states,
\begin{equation}
\label{eq:definition-U}
    U_\lambda(t) := (u_\lambda(t), u_\lambda(t - \Delta t), \ldots, u_\lambda(t - (m-1)\Delta t)),
\end{equation}
where $m$ is the history length.
The history length is fixed in the experiments (\Cref{sec:experiments}), and its effect is discussed in supplementary material Section~S3.2.
For simplicity, the theory in \Cref{sec:framework} was written with the current state as the only argument (i.e., $m=1$), and it remains valid for a finite history once the state is augmented (\Cref{subsec:path-probability-measure-via-the-path-integral-representation}).
The network output is the increment to the next time step, $\hat f_\lambda^\theta(U_\lambda(t))\,\Delta t$.

The training loss is the criterion [Eq.~(\ref{eq:averaged-kl})] discretized in space and time:
\begin{equation}
\label{eq:loss-pdd}
    \mathcal{L}^{\mathrm{PDD}} = \mathbb{E}_{\lambda \sim \mathcal{U}(0,1),\, q_\lambda}\!\left[\frac{\Delta x^d}{2\sigma_\lambda^2 \Delta t} \sum_{n=0}^{N_t - 1} \left\| u_\lambda(t^{(n+1)}) - u_\lambda(t^{(n)}) - \hat f_\lambda^\theta(U_\lambda(t^{(n)}))\,\Delta t \right\|^2\right],
\end{equation}
where $t^{(n)} = n \Delta t$ ($n = 0, \ldots, N_t$) are the discrete time steps.
The argument of $\|\cdot\|$ is the discretized field at one time step, so the norm runs over the channels and spatial grid points.
The product $\Delta x^d \|\cdot\|^2$ is then the discrete form of $\int \mathrm{d}^d x\,\|\cdot\|^2$ in Eq.~(\ref{eq:path-probability-density}).
The training data are paths at $\lambda = 0$ (i.e., $\{u_0\}_t$) passed through the closed form in Eq.~(\ref{eq:ou-solution-spatiotemporal}), with $\lambda$ drawn for each sample and $\epsilon$ for each spatiotemporal grid point.

The score needed for generation and super-resolution is obtained by differentiating the path density $p_\lambda(\{u_\lambda\}_t)$ in Eq.~(\ref{eq:path-probability-density}), in which $\hat f_\lambda^*$ replaces $f_\lambda^\theta$.
Up to a constant, $\ln p_\lambda(\{u_\lambda\}_t) = -S_\lambda^\theta(\{u_\lambda\}_t) + \ln r_\lambda$.
The first term is a differentiable function of $\{u_\lambda\}_t$, so the score is computed by automatic differentiation \citep{Baydin+18} without a separate network.
The second term, $\ln r_\lambda$, enters only at the initial time step $t^{(0)}$ through $\nabla_{u_\lambda(t^{(0)})} \ln r_\lambda$.
Instead of estimating it, we obtain the mean of the update at $t^{(0)}$ by linear extrapolation in $t$ from the following time steps (supplementary material Section~S1.6).
Since $S_\lambda^\theta$ contains finite differences of neighboring time steps, the score at $t$ depends on the states at past and future times and is in general not causal in $t$.
With $\hat f_\lambda^*$ and the path score from it, the three tasks of \Cref{subsec:effective-theories} are performed by a single network.

The PDD formulation is computationally expensive for generation and super-resolution.
Each score evaluation requires a forward computation of $\hat f_\lambda^*$ over all time steps and a backward computation through $S_\lambda^\theta$.
This is repeated at every step of the reverse-$\lambda$ integration, so both computation time and memory grow with the temporal length of the path.
Estimating the score directly as a network output would avoid this cost, which motivates the second realization, SA-PDD.

\subsection{Score-Augmented Predictor-Driven Diffusion (SA-PDD)}
\label{subsec:sa-pdd}

Equations~(\ref{eq:optimal-drift-two-terms}) and~(\ref{eq:tweedie-single-time}) showed that $f_\lambda^*$ contains the score.
The score can therefore be estimated as a second output that shares the representation of the predictor, instead of running automatic differentiation or training a separate network.
SA-PDD is the realization built on this idea.

We propose a neural network with shared hidden layers and two outputs: the predictor and the score.
In the simple form used here, the hidden layers are followed by two pointwise linear layers, one for the increment $\hat f_\lambda^\theta(U_\lambda(t))\,\Delta t$ and the other for an estimate $\hat\epsilon_\lambda^\theta(U_\lambda(t))$ of the noise $\epsilon(t)$ in Eq.~(\ref{eq:ou-solution-spatiotemporal}).
The input is $\lambda$ and $U_\lambda(t)$, as in \Cref{subsec:pdd}.
The noise estimate $\hat\epsilon_\lambda^\theta$ gives a score,
\begin{equation}
\label{eq:sa-pdd-score}
    \hat s_\lambda^\theta(U_\lambda(t)) := -\Sigma_\lambda^{-1/2}\, \hat\epsilon_\lambda^\theta(U_\lambda(t)),
\end{equation}
by a relation similar to Eq.~(\ref{eq:tweedie-single-time}), which is extended to the history window $U_\lambda(t)$ (\Cref{subsec:path-score-and-sa-pdd-optimum}).
The loss adds a denoising term \citep{Vincent11,Ho+20} to the PDD loss:
\begin{equation}
\label{eq:loss-sa-pdd}
    \mathcal{L}^{\mathrm{SA\text{-}PDD}} = \mathbb{E}\!\left[\frac{1}{2}\sum_{n=0}^{N_t} \left\| \hat\epsilon_\lambda^\theta(U_\lambda(t^{(n)})) - \epsilon(t^{(n)}) \right\|^2\right] + w\, \mathcal{L}^{\mathrm{PDD}},
\end{equation}
where $w$ is a weight, and the expectation is taken over $\lambda \sim \mathcal{U}(0,1)$ and the joint distribution of $\{u_0\}_t$ and $\epsilon$.
We fix $w$ to 0.1 throughout the paper.
With this choice, the two terms in Eq.~(\ref{eq:loss-sa-pdd}) were of roughly the same order in the experiments (\Cref{sec:experiments}), and we confirmed that the results are not highly sensitive to $w$.

Simulation proceeds as in \Cref{subsec:pdd}, whereas generation and super-resolution use $\hat s_\lambda^*$ at every time step as the score of the reverse equation~(\ref{eq:reverse-rg-sde}).
Since the score follows from the noise estimate at each time step, neither $r_\lambda$ nor its gradient needs to be estimated, in contrast to PDD.
Furthermore, a score evaluation is one forward computation (i.e., without automatic differentiation), reducing the memory use in generation to about one ninth of that of PDD (supplementary material Section~S1.7).
The three tasks thus run on a single network, now with the two outputs $\hat f_\lambda^*\,\Delta t$ and $\hat\epsilon_\lambda^*$.

The score $\hat s_\lambda^*$ has the same form as the single-time score contained in $f_\lambda^*$ [Eqs.~(\ref{eq:optimal-drift-two-terms}) and~(\ref{eq:tweedie-single-time})] and as the path score required by the reverse equation~(\ref{eq:reverse-rg-sde}), but $\hat s_\lambda^*$ is a different object from both.
All three are conditional expectations of the forward-process noise $\epsilon(t)$ and differ only in the conditioning (\Cref{subsec:path-score-and-sa-pdd-optimum}):
\begin{align}
    \nabla_{u_\lambda(t)} \ln q_\lambda(u_\lambda(t)) &= -\Sigma_\lambda^{-1/2}\, \mathbb{E}\!\left[\epsilon(t) \mid u_\lambda(t)\right], \label{eq:score-single-time} \\
    \nabla_{u_\lambda(t)} \ln q_\lambda(\{u_\lambda\}_t) &= -\Sigma_\lambda^{-1/2}\, \mathbb{E}\!\left[\epsilon(t) \mid \{u_\lambda\}_t\right], \label{eq:score-path} \\
    \hat s_\lambda^*(U_\lambda(t)) &= -\Sigma_\lambda^{-1/2}\, \mathbb{E}\!\left[\epsilon(t) \mid U_\lambda(t)\right], \label{eq:score-window}
\end{align}
conditioned on the state $u_\lambda(t)$ at $t$, on the whole path $\{u_\lambda\}_t$, and on the current and past states $U_\lambda(t)$, respectively.
The third is the minimizer of the denoising term in Eq.~(\ref{eq:loss-sa-pdd}) over all functions of $U_\lambda(t)$.
By iterated conditioning, each is the conditional average of the score with more conditioning:
\begin{align}
    \hat s_\lambda^*(U_\lambda(t)) &= \mathbb{E}\!\left[\nabla_{u_\lambda(t)} \ln q_\lambda(\{u_\lambda\}_t) \;\middle|\; U_\lambda(t)\right],\label{eq:score-tower1}\\
    \nabla_{u_\lambda(t)} \ln q_\lambda(u_\lambda(t)) &= \mathbb{E}\!\left[\hat s_\lambda^*(U_\lambda(t)) \;\middle|\; u_\lambda(t)\right].\label{eq:score-tower2}
\end{align}

\subsection{Two approximations to the path score \texorpdfstring{$\nabla_{u_\lambda(t)} \ln q_\lambda(\{u_\lambda\}_t)$}{}}
\label{subsec:two-approximations}

The reverse equation~(\ref{eq:reverse-rg-sde}) integrates all time steps of the path at once and requires the path-score component at each $t$ [Eq.~(\ref{eq:score-path})].
In Eq.~(\ref{eq:reverse-rg-sde}), the time steps are coupled only through this score.
PDD and SA-PDD approximate the path score in different ways.

PDD approximates the density itself, replacing $q_\lambda(\{u_\lambda\}_t)$ by $p_\lambda(\{u_\lambda\}_t)$.
Automatic differentiation of $p_\lambda(\{u_\lambda\}_t)$ then gives its score exactly, up to numerical error.
Thus, the error in the score is determined by the accuracy of $\hat f_\lambda^*$ and by the assumptions of \Cref{subsec:path-integral}.
Extending the model family of $p_\lambda(\{u_\lambda\}_t)$ can reduce this error (\Cref{sec:discussion}).
The score at $t$ is computed from a finite number of past and future steps, since $\hat f_\lambda^\theta$ takes $U_\lambda(t)$ as its input and $S_\lambda^\theta$ in Eq.~(\ref{eq:path-probability-density}) includes $\partial_t u_\lambda(t)$ and $\hat f_\lambda^\theta$.

SA-PDD estimates the path-score component conditioned on the window $U_\lambda(t)$ [Eq.~(\ref{eq:score-window})] rather than on the whole path.
Its error comes from the finite extent of the window, most importantly from the exclusion of future states.
In practice, the whole path is available during generation, so the input can in principle be extended to future states (\Cref{sec:discussion}).
\section{Experiments}
\label{sec:experiments}

We evaluate the framework on two representative chaotic systems with multiscale structure: the 1D Lorenz-96 model with two timescales \citep{Lorenz96} and the 2D Kolmogorov flow \citep{Obukhov+83}.
Both are commonly used as testbeds in machine learning \citep[e.g.,][]{Kochkov+21,Kuhbacher+26}.
For each system, PDD and SA-PDD perform simulation, unconditional generation, and super-resolution, each with a single trained network (\Cref{subsec:pdd,subsec:sa-pdd}).
The goal is to validate the framework rather than to pursue benchmark performance.
Implementation details and extended results are provided in the supplementary material (Sections~S1--S3).
Code is available in the repositories listed under \nameref{sec:data-availability}.

\subsection{Experimental setup}
\label{subsec:experimental-setup}

\paragraph{Training data.}
For both systems, spatiotemporal samples are generated from physics-based simulations in a statistically steady regime.
Each Lorenz-96 sample has two channels, the slow variable $X$ and the fast variable $Y$, with a temporal length of 64 snapshots and a spatial length of 128 grid points.
We use 3,000 and 100 samples for training and testing, respectively.
Each Kolmogorov-flow sample has one vorticity channel with a temporal length of 40 snapshots.
The flow is simulated at $256 \times 256$ and downsampled to $40 \times 40$ or $128 \times 128$ (supplementary material Section~S1.1).
At each resolution, we use 6,000 and 100 samples for training and testing, respectively.

\paragraph{Neural network, training, and sampling.}
We use a U-Net \citep{Saharia+22b} for $\hat f_\lambda^\theta$.
The input is $\lambda$ and $U_\lambda(t)$ with $m = 5$, whose states are concatenated along the channel dimension, and the output is the increment to the next time step.
For SA-PDD with $w = 0.1$, a second linear layer outputs $\hat\epsilon_\lambda^\theta$ (\Cref{subsec:sa-pdd}).
Training minimizes Eq.~(\ref{eq:loss-pdd}) or~(\ref{eq:loss-sa-pdd}) on local time differences, without rollout in $t$.
The diffusion scale $\lambda \in [0, 1]$ is discretized into 1,000 steps.
The reverse-$\lambda$ integration uses a predictor--corrector scheme \citep{Song+21}, with three corrector steps per predictor step, where the signal-to-noise ratio parameter is $\kappa = 0.7$ for the Lorenz-96 model and $0.3$ for the Kolmogorov flow.
The same settings are used for PDD, SA-PDD, and DDPM (supplementary material Section~S1.6).
The coarse-graining parameters are selected by grid search: $(\alpha, \beta) = (0.1, \sqrt{2})$ for the Lorenz-96 model and $(0.3, \sqrt{6})$ for the Kolmogorov flow (supplementary material Section~S1.3).
In the main text, we report results at $\lambda = 0$ (fine) and $\lambda = 0.2$ (coarse-grained), where the cutoff wavenumber $k_{\mathrm{cut}} \sim 1/\sqrt{\alpha\lambda}$ (\Cref{subsec:rg-coarse-graining}) is about 7 for the Lorenz-96 model and about 4 for the Kolmogorov flow.
Each network is trained five times with different initial weights, to confirm that the results do not depend strongly on the initialization and to report their uncertainty as standard deviations.
We also confirmed with a Fourier neural operator \citep{Li+21} that the results are not highly sensitive to the network architecture (supplementary material Section~S3.5).

\paragraph{DDPM.}
For comparison at $\lambda = 0$, two denoising diffusion probabilistic models (DDPMs) \citep{Ho+20} are trained separately with the same U-Net: one for unconditional generation and the other for simulation in a conditional setup \citep[e.g.,][]{Price+25}.
Both are referred to as DDPM (supplementary material Section~S1.4).
A DDPM has no spatial coarse-graining hierarchy and is undefined at $\lambda > 0$, so the comparison is made only at $\lambda = 0$.
Each DDPM is specialized to one task (i.e., simulation or generation), so it provides a reference for the validity of the results at a fixed $\lambda$ rather than a controlled comparison (\Cref{subsec:extensions}).

\paragraph{Higher-resolution demonstration.}
The $128 \times 128$ Kolmogorov flow is run with SA-PDD only, to show that the framework works at a higher resolution rather than to compare with PDD or DDPM.
Note that PDD uses about nine times the memory of SA-PDD at $40 \times 40$ (\Cref{subsec:sa-pdd}; supplementary material Section~S1.7), suggesting that PDD is difficult to apply at higher resolution.
All SA-PDD settings are the same as at $40 \times 40$.

\paragraph{Simulation in the evaluation phase.}
The evaluation of simulation is deterministic.
We run the rollout with $\hat f_\lambda^*$ and $\sigma_\lambda = 0$, i.e., along the most probable path (\Cref{subsec:path-probability-measure-via-the-path-integral-representation}).
The result is compared with the reference, the coarse-grained physics-based solution $\mathcal{C}_\lambda u_0$, using the same initial condition.
With a noisy $u_\lambda$ as the reference (and $\sigma_\lambda > 0$) instead, the following evaluation metrics in Eqs.~(\ref{eq:l2-err-definition}) and~(\ref{eq:spec-err-definition}) would compare two random realizations, making the results difficult to interpret.
Simulations with $\sigma_\lambda > 0$ remain stable and are shown in supplementary material Section~S3.1.

\paragraph{Evaluation metrics.}
We assess the short-term accuracy of simulation by the $L^2$ error and its long-run statistical consistency by the spatiotemporal spectral error \citep{Jiang+23}:
\begin{align}
    L^2 \text{ error}(t) &:= \frac{\| u(t) - \hat{u}(t)\|_2}{\| u(t) \|_2},  \label{eq:l2-err-definition} \\
    \text{spectral error} &:= \frac{\| \; |\mathcal{G}[u]|^2 - |\mathcal{G}[\hat{u}]|^2 \; \|_1}{\| \; |\mathcal{G}[u]|^2 \; \|_1}, \label{eq:spec-err-definition}
\end{align}
where $u$ and $\hat{u}$ are the reference and the model output, $\|\cdot\|_2$ is taken over the spatial grid and $\|\cdot\|_1$ over all spatiotemporal modes, and $\mathcal{G}$ is the Fourier transform over $x$ and $t$.
In chaotic systems, small differences between trajectories grow rapidly, so the $L^2$ error is reported at the sixth time step.
Since $m = 5$, this is the first step at which all the inputs are neural network outputs.
For the spectral error, the first 10 segments (each spanning the temporal length of one sample) of a long rollout are discarded and the power spectral densities (PSDs) are computed from the 11th segment.
We confirmed that a much longer burn-in (e.g., 100 segments) gives similar values.
Further details are provided in supplementary material Section~S1.5.

\subsection{Experimental results}
\label{subsec:experimental-results}

The figures in this section show the Kolmogorov flow, and the Lorenz-96 figures are collected in supplementary material Section~S2.
The tables report both systems.

\begin{figure}[tbp!]
    \begin{center}
        \includegraphics[width=\textwidth]{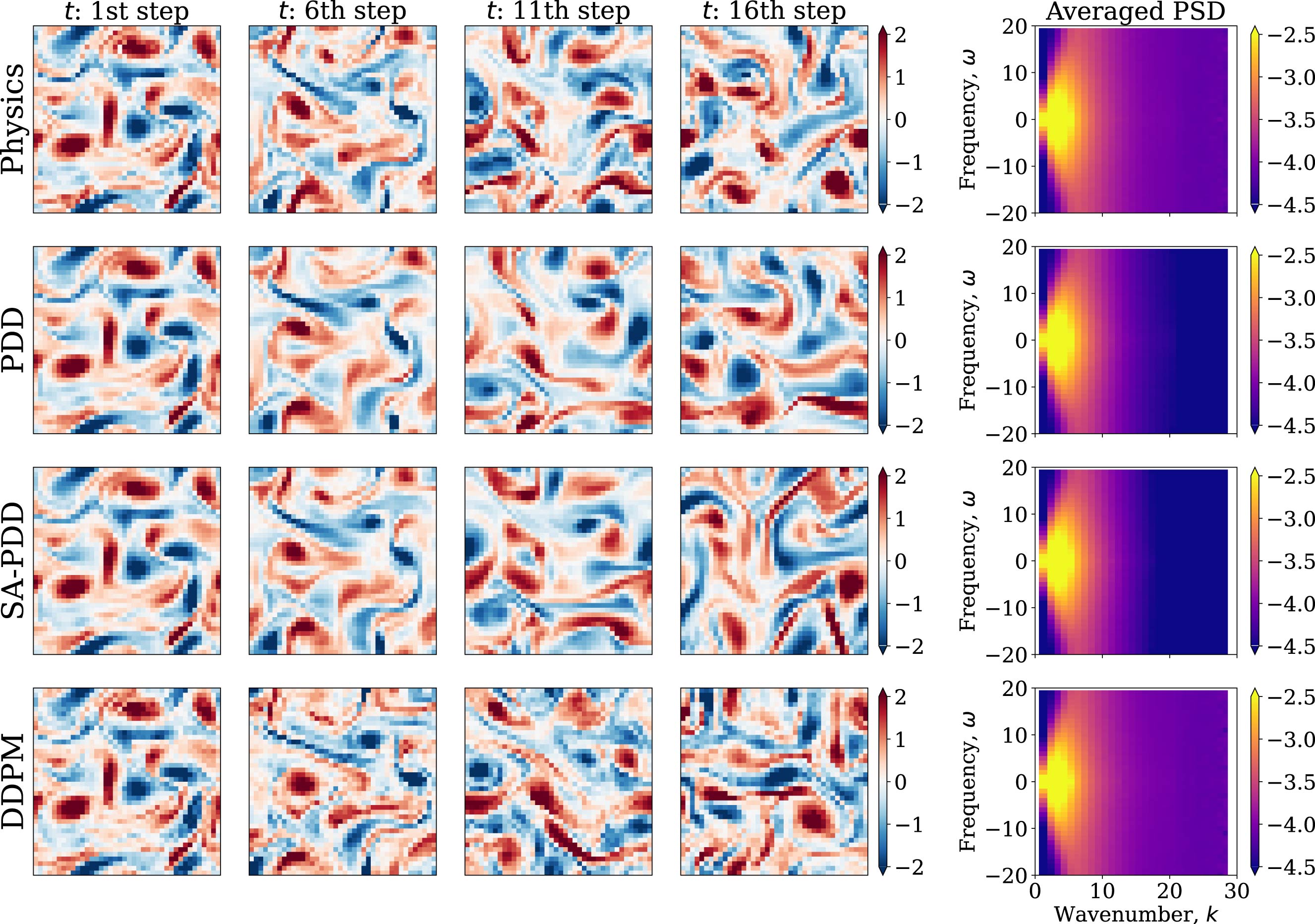} 
        \caption{
        Simulation of the $40 \times 40$ Kolmogorov flow (vorticity) at fine resolution ($\lambda = 0$).
        Rows: physics-based simulation (Physics), PDD, SA-PDD, and DDPM, all started from the same initial condition, the history window at time steps $-4$ to $0$.
        Left: vorticity at the 1st, 6th, 11th, and 16th time steps of the rollout; right: spatiotemporal power spectral density (PSD) as a function of wavenumber $k$ and frequency $\omega$.
        PSDs are averaged over the 100 samples, and $\log_{10}$ is applied to the average.
        All quantities are non-dimensionalized.
        }
        \label{fig:kf040_simulation_lambda0}
    \end{center}
\end{figure}

\begin{figure}[tbp!]
    \begin{center}
        \includegraphics[width=\textwidth]{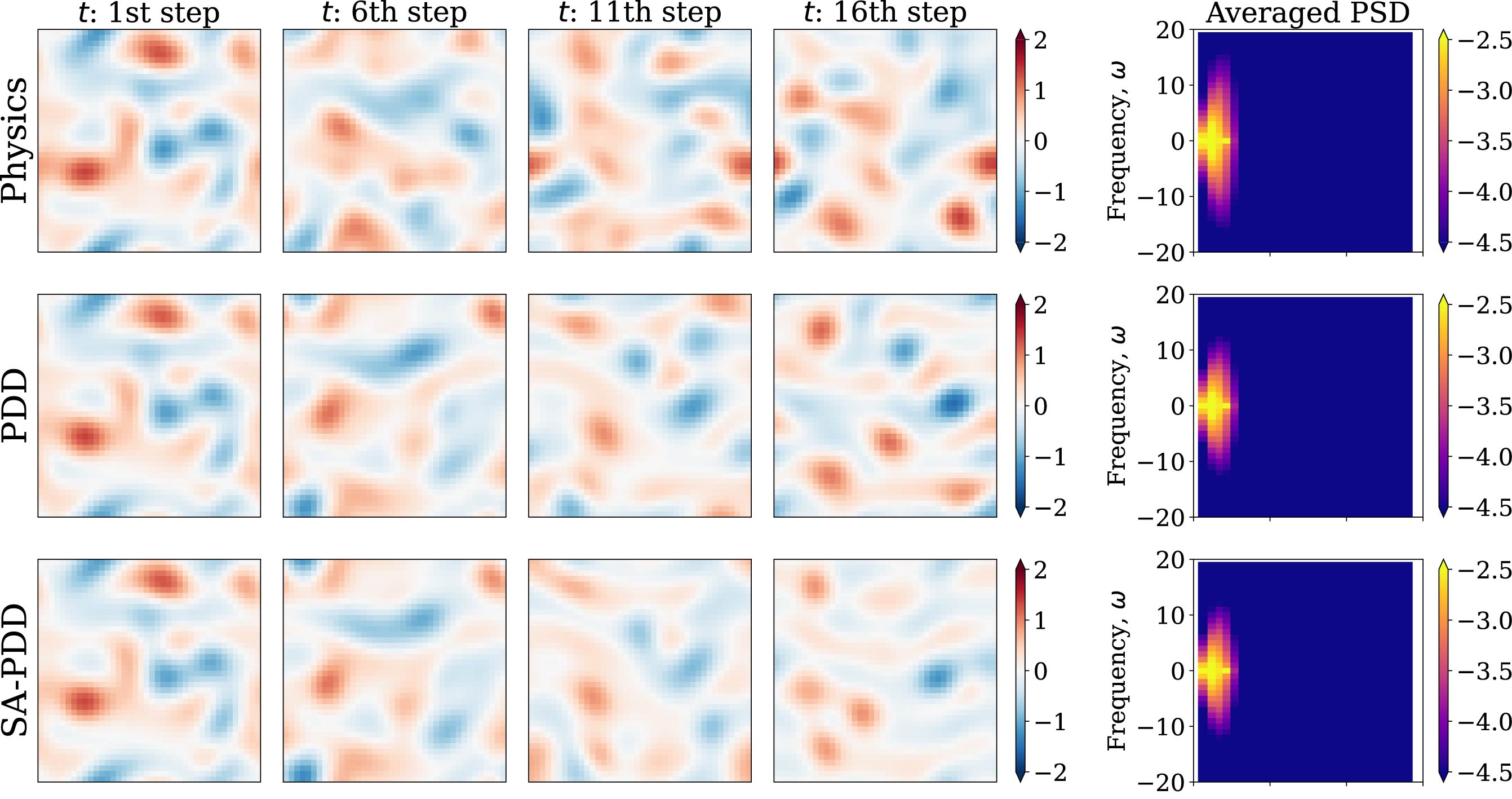} 
        \caption{
        Simulation of the $40 \times 40$ Kolmogorov flow (vorticity) at coarse-grained resolution ($\lambda = 0.2$), in the same layout as Fig.~\ref{fig:kf040_simulation_lambda0}.
        Rows: coarse-grained physics-based simulation $\mathcal{C}_\lambda u_0$ (Physics), PDD, and SA-PDD; note that DDPM is undefined at $\lambda > 0$.
        Coarse-graining removes small-scale structure while the large-scale dynamics are preserved.
        PSDs are averaged over the 100 samples, and $\log_{10}$ is applied to the average.
        }
        \label{fig:kf040_simulation_lambda02}
    \end{center}
\end{figure}

\paragraph{Simulation.}
Figures~\ref{fig:kf040_simulation_lambda0} and~\ref{fig:kf040_simulation_lambda02} show simulations of the $40 \times 40$ Kolmogorov flow at $\lambda = 0$ and $0.2$, respectively.
The learned predictor, for both PDD and SA-PDD, emulates the fine and the coarse-grained dynamics well.
As the rollout proceeds, the vorticity fields of Physics, PDD, SA-PDD, and DDPM diverge from one another, as expected in a chaotic system.
Their spatiotemporal PSDs are nevertheless close, indicating good agreement of the spatiotemporal patterns.
At $\lambda = 0$, the three models give similar errors, except for the $L^2$ error of the $40 \times 40$ Kolmogorov flow and the spectral error of the Lorenz-96 model, where DDPM gives a larger error than both PDD and SA-PDD (Table~\ref{table:main_text_error_simulation}).
As $\lambda$ increases, small-scale components are progressively removed while large-scale structures are preserved.
The retained large scales still fluctuate, and this variability likely reflects the influence of the eliminated small scales (supplementary material Section~S3.2).

\begin{table}[tbp!]
    \caption{
    Simulation errors at $\lambda = 0$ for the Lorenz-96 model (L96) and the Kolmogorov flow (KF) at $40 \times 40$ and $128 \times 128$.
    The $L^2$ error in Eq.~(\ref{eq:l2-err-definition}) measures the short-term simulation accuracy (at the sixth time step) and the spectral error in Eq.~(\ref{eq:spec-err-definition}) measures the long-run statistical consistency.
    Values are means $\pm$ standard deviations over five training runs (100 test samples each).
    DDPM is the model trained for simulation, separate from the DDPM for unconditional generation in Table~\ref{table:main_text_error_generation} (\Cref{subsec:experimental-setup}).
    At $128 \times 128$, only SA-PDD was run.
    }
    \label{table:main_text_error_simulation}
    \centering
    \small
    \begin{tabular}{lccc}
        \toprule
                                     & PDD                 & SA-PDD              & DDPM                \\
        \midrule
        L96 $L^2$ error              & 0.503 ($\pm$ 0.011) & 0.553 ($\pm$ 0.045) & 0.557 ($\pm$ 0.031) \\
        KF $40 \times 40$ $L^2$ error   & 0.691 ($\pm$ 0.057) & 0.648 ($\pm$ 0.040) & 0.861 ($\pm$ 0.023) \\
        KF $128 \times 128$ $L^2$ error & --                  & 0.717 ($\pm$ 0.193) & --                  \\
        \midrule
        L96 spectral error           & 0.120 ($\pm$ 0.014) & 0.149 ($\pm$ 0.077) & 0.226 ($\pm$ 0.077) \\
        KF $40 \times 40$ spectral error   & 0.182 ($\pm$ 0.029) & 0.160 ($\pm$ 0.003) & 0.139 ($\pm$ 0.021) \\
        KF $128 \times 128$ spectral error & --                  & 0.172 ($\pm$ 0.041) & --                  \\
        \bottomrule
    \end{tabular}
\end{table}

\FloatBarrier

\begin{figure}[tbp!]
    \begin{center}
        \includegraphics[width=0.9\textwidth]{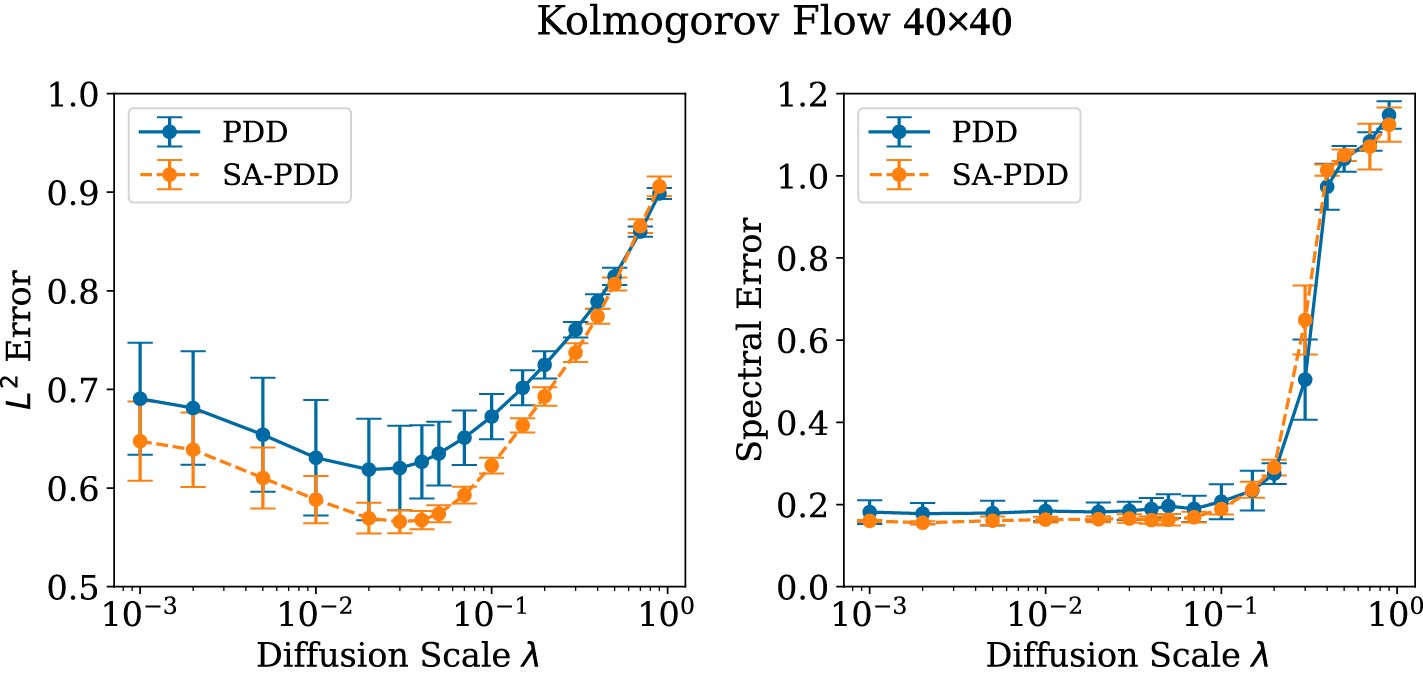} 
        \caption{
        Dependence of the simulation errors on the diffusion scale $\lambda$ for the $40 \times 40$ Kolmogorov flow.
        (Left) $L^2$ error in Eq.~(\ref{eq:l2-err-definition}) at the sixth time step; (right) spectral error in Eq.~(\ref{eq:spec-err-definition}).
        Blue solid lines: PDD; orange dashed lines: SA-PDD.
        The reference at each $\lambda$ is the coarse-grained physics-based simulation $\mathcal{C}_\lambda u_0$ (\Cref{subsec:experimental-setup}).
        To use a log scale, a small offset of $10^{-3}$ is added to $\lambda$ (e.g., $\lambda = 0$ is displayed as $\lambda = 10^{-3}$).
        Error bars indicate the standard deviation over five training runs.
        }
        \label{fig:kf040_error_vs_lambda}
    \end{center}
\end{figure}

Figure~\ref{fig:kf040_error_vs_lambda} shows the $\lambda$ dependence of the simulation accuracy for the $40 \times 40$ Kolmogorov flow.
The $L^2$ error is non-monotonic in $\lambda$ and takes a minimum at intermediate $\lambda$, likely because coarse-graining first removes the small scales that are hard to predict.
The steep rise of the spectral error for $\lambda > 0.2$ is likely a normalization effect: both errors are ratios to the amplitude of the reference state, and this amplitude decreases under coarse-graining, so the ratio grows even if the absolute discrepancy does not.
SA-PDD gives a slightly smaller $L^2$ error than PDD over most of the range, while the spectral errors nearly coincide for $\lambda \lesssim 0.2$.
The relative accuracy of the two models is reversed on the Lorenz-96 model (Table~\ref{table:main_text_error_simulation}), and a detailed comparison of their accuracy requires theoretical analysis or experiments on more systems.

\begin{figure}[tbp!]
    \begin{center}
        \includegraphics[width=\textwidth]{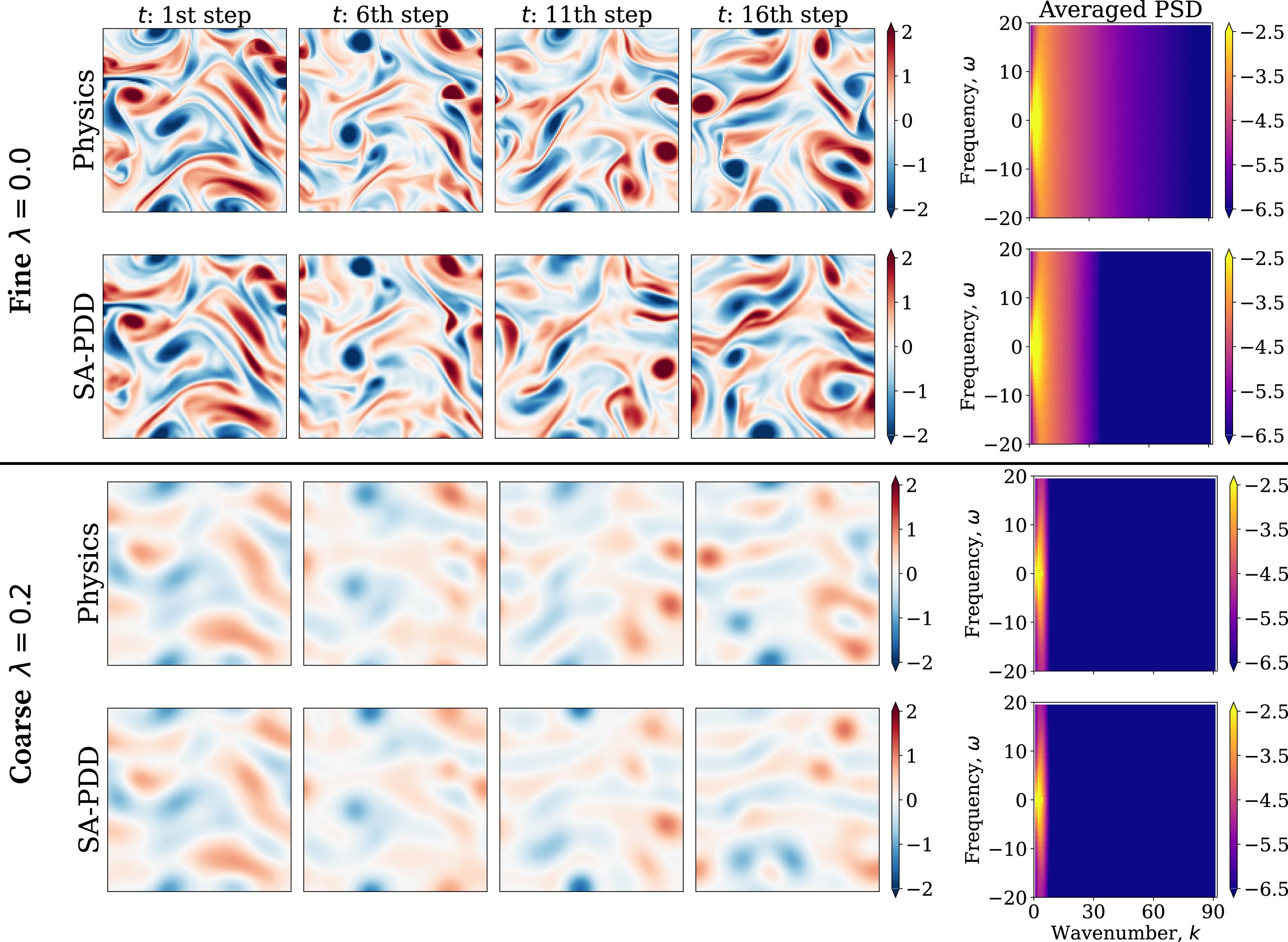} 
        \caption{
        Simulation of the $128 \times 128$ Kolmogorov flow (vorticity) with SA-PDD at $\lambda = 0$ (top block) and $\lambda = 0.2$ (bottom block), in the same layout as Fig.~\ref{fig:kf040_simulation_lambda0}.
        Rows in each block: physics-based simulation (Physics) and SA-PDD.
        All settings, including $\alpha$ and $\beta$, are the same as at $40 \times 40$.
        PSDs are averaged over the 100 samples, and $\log_{10}$ is applied to the average.
        }
        \label{fig:kf128_simulation}
    \end{center}
\end{figure}

At $128 \times 128$, SA-PDD also reproduces the fine and the coarse-grained dynamics well (Fig.~\ref{fig:kf128_simulation}; Table~\ref{table:main_text_error_simulation}).
At $\lambda = 0$, however, SA-PDD underestimates the spectral power at wavenumbers $k \gtrsim 30$, and further improving the small-scale accuracy at higher resolution is one direction for future work (\Cref{sec:discussion}).

\FloatBarrier

\paragraph{Generation.}
With the network trained for simulation, spatiotemporal samples are generated by integrating the reverse equation~(\ref{eq:reverse-rg-sde}) from the Gaussian at $\lambda = 1$ down to $\lambda = 0$ (\Cref{subsec:pdd,subsec:sa-pdd}).
The ranking of the three models by spectral error differs between the two systems (Table~\ref{table:main_text_error_generation}).
On the Lorenz-96 model, DDPM gives the smallest spectral error, whereas on the Kolmogorov flow, PDD is the smallest and SA-PDD the largest.
Supplementary material Section~S2.2 shows the generated samples for all settings, including the case of SA-PDD at $128 \times 128$.
These results indicate that the path density defined by the predictor (PDD) and the window-conditioned noise estimate (SA-PDD) both provide a score adequate for generation.

\begin{table}[tbp!]
    \caption{
    Spectral errors of unconditional generation for the Lorenz-96 model (L96) and the Kolmogorov flow (KF) at $40 \times 40$ and $128 \times 128$.
    The spectral error in Eq.~(\ref{eq:spec-err-definition}) measures the statistical consistency of the spatiotemporal patterns between generated and physics-based samples.
    Values are means $\pm$ standard deviations over five training runs (100 generated samples each).
    DDPM is the model trained for unconditional generation, separate from the DDPM for simulation in Table~\ref{table:main_text_error_simulation} (\Cref{subsec:experimental-setup}).
    At $128 \times 128$, only SA-PDD was run.
    }
    \label{table:main_text_error_generation}
    \centering
    \small
    \begin{tabular}{lccc}
        \toprule
                                     & PDD                 & SA-PDD              & DDPM                \\
        \midrule
        L96 spectral error           & 0.343 ($\pm$ 0.028) & 0.367 ($\pm$ 0.016) & 0.267 ($\pm$ 0.020) \\
        KF $40 \times 40$ spectral error   & 0.457 ($\pm$ 0.025) & 0.654 ($\pm$ 0.045) & 0.615 ($\pm$ 0.121) \\
        KF $128 \times 128$ spectral error & --                  & 0.492 ($\pm$ 0.026) & --                  \\
        \bottomrule
    \end{tabular}
\end{table}

\FloatBarrier

\begin{figure}[tbp!]
    \begin{center}
        \includegraphics[width=\textwidth]{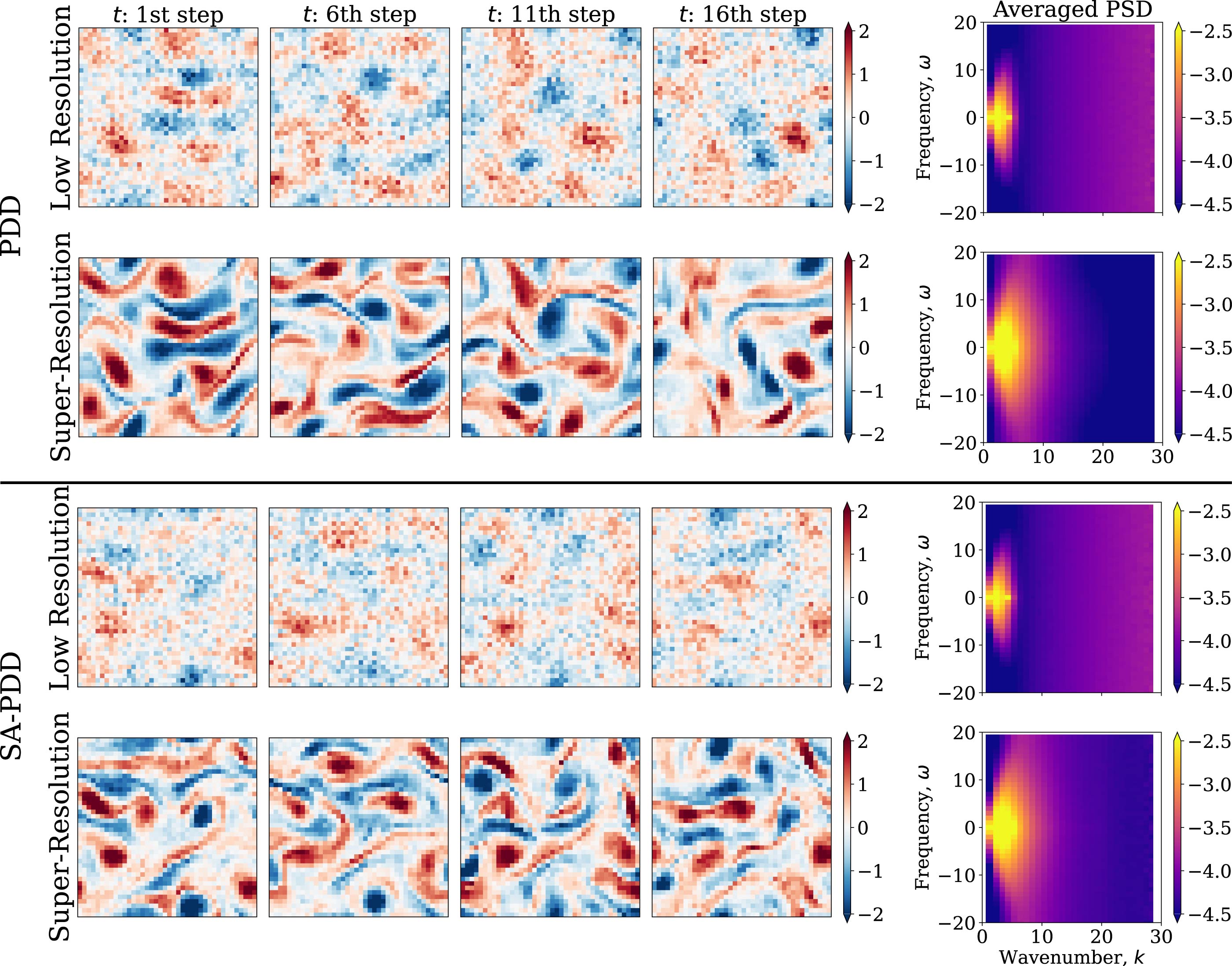} 
        \caption{
        Super-resolution of the $40 \times 40$ Kolmogorov flow (vorticity) with PDD (top block) and SA-PDD (bottom block).
        Rows in each block: low-resolution input, which is simulated at $\lambda = 0.2$ with $\sigma_\lambda > 0$ (i.e., a sample of $q_\lambda$), and the super-resolved output at $\lambda = 0$.
        Left: vorticity at the 1st, 6th, 11th, and 16th time steps; right: spatiotemporal PSD.
        PSDs are averaged over the 100 samples, and $\log_{10}$ is applied to the average.
        The reverse-$\lambda$ integration restores the small-scale structure absent from the input.
        }
        \label{fig:kf040_super_resolution}
    \end{center}
\end{figure}

\begin{figure}[tbp!]
    \begin{center}
        \includegraphics[width=\textwidth]{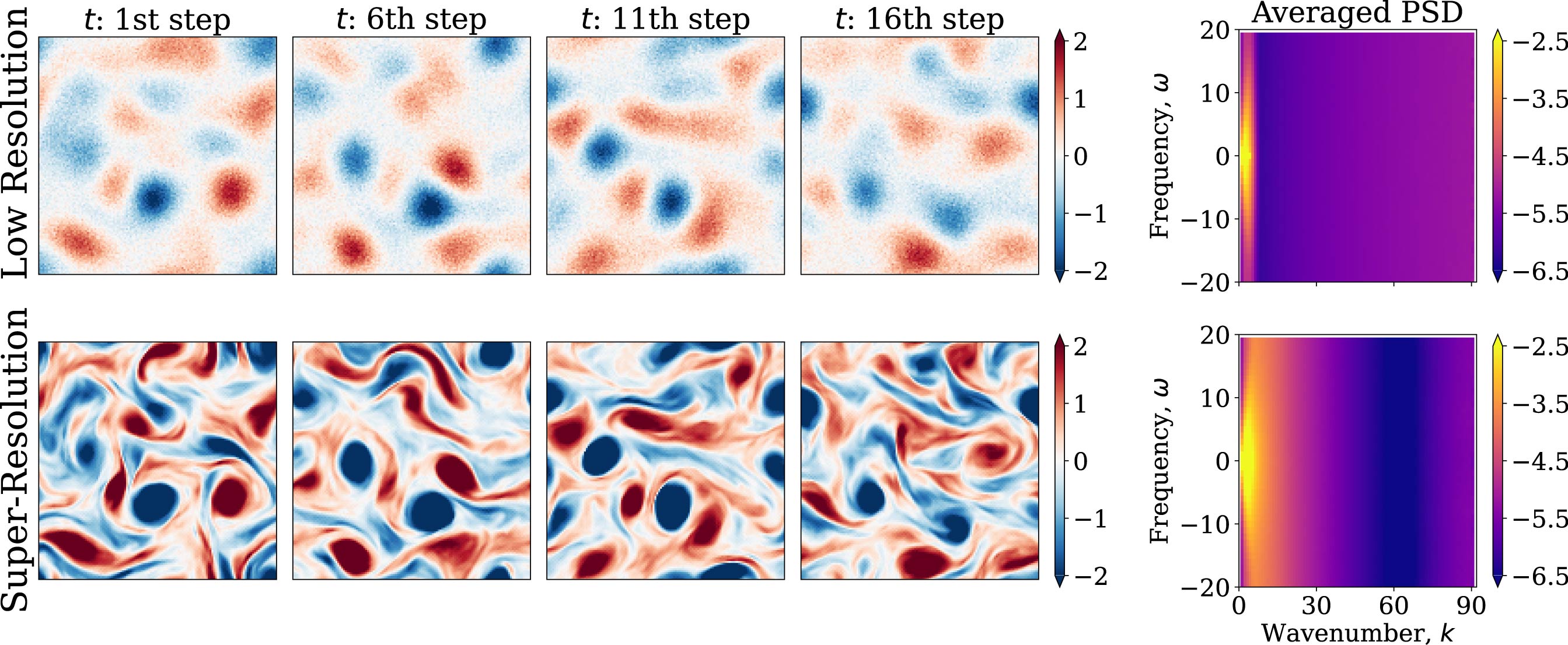} 
        \caption{
        Super-resolution of the $128 \times 128$ Kolmogorov flow (vorticity) with SA-PDD, in the same layout as Fig.~\ref{fig:kf040_super_resolution}.
        Rows: low-resolution input, which is simulated at $\lambda = 0.2$ with $\sigma_\lambda > 0$, and the super-resolved output at $\lambda = 0$.
        PSDs are averaged over the 100 samples, and $\log_{10}$ is applied to the average.
        }
        \label{fig:kf128_super_resolution}
    \end{center}
\end{figure}

\paragraph{Super-resolution.}
In the super-resolution task, the input is a coarse-grained simulation at $\lambda = 0.2$, and the reverse-$\lambda$ integration is performed down to $\lambda = 0$ (Figs.~\ref{fig:kf040_super_resolution} and~\ref{fig:kf128_super_resolution}), using the same trained model as in the simulation and unconditional generation tasks.
Unlike in the evaluation above, the input is simulated with $\sigma_\lambda > 0$ instead of $\sigma_\lambda = 0$, so that the simulation output is noisy and regarded as a sample of $q_\lambda$.

Table~\ref{table:main_text_error_super_resolution} compares the low-resolution (LR) error of the input at $\lambda = 0.2$ with the super-resolution (SR) error of the output at $\lambda = 0$, both against the $\lambda = 0$ reference.
The SR error is much smaller than the LR error for PDD and SA-PDD.
Their SR errors are comparable on the Lorenz-96 model (supplementary material Section~S2.3), whereas on the Kolmogorov flow SA-PDD gives the larger error.
At $128 \times 128$, SA-PDD reduces the error from LR to SR as well.
The spectral error is insensitive to phase, so we also confirmed in a control experiment that the super-resolved output retains the large-scale structure of its own low-resolution input and is therefore distinct from an unconditional sample (supplementary material Section~S2.3).
These results indicate that the super-resolution procedure restores the small-scale components absent from the input.

\begin{table}[tbp!]
    \caption{
    Spectral errors of super-resolution, defined in Eq.~(\ref{eq:spec-err-definition}), for the Lorenz-96 model (L96) and the Kolmogorov flow (KF) at $40 \times 40$ and $128 \times 128$.
    LR: low-resolution input, which is simulated at $\lambda = 0.2$ with $\sigma_\lambda > 0$, evaluated against the $\lambda = 0$ physics-based simulation; SR: super-resolved output at $\lambda = 0$, evaluated against the same reference.
    Values are means $\pm$ standard deviations over five training runs (100 test samples each).
    At $128 \times 128$, only SA-PDD was run.
    }
    \label{table:main_text_error_super_resolution}
    \centering
    \small
    \begin{tabular}{lcccc}
        \toprule
                            & \multicolumn{2}{c}{PDD}                   & \multicolumn{2}{c}{SA-PDD}                \\
                            & LR                  & SR                  & LR                  & SR                  \\
        \midrule
        L96                 & 0.873 ($\pm$ 0.005) & 0.345 ($\pm$ 0.027) & 0.867 ($\pm$ 0.003) & 0.355 ($\pm$ 0.023) \\
        KF $40 \times 40$   & 0.743 ($\pm$ 0.005) & 0.303 ($\pm$ 0.011) & 0.757 ($\pm$ 0.004) & 0.372 ($\pm$ 0.024) \\
        KF $128 \times 128$ & --                  & --                  & 0.804 ($\pm$ 0.020) & 0.202 ($\pm$ 0.031) \\
        \bottomrule
    \end{tabular}
\end{table}

\FloatBarrier
\section{Related work}
\label{sec:related-work}

\paragraph{Effective dynamics at a fixed scale.}
The effect of unresolved degrees of freedom has long been modeled as a time evolution closed in terms of the resolved scales.
A major framework is the Mori--Zwanzig formalism, which represents this effect as memory and fluctuation terms \citep{Zwanzig61,Mori65}.
The memory term has been approximated analytically and from data \citep{Chorin+00,Parish+Duraisamy17,Lin+Lu21}.
More recently, coarse-grid dynamics have been closed by a learned correction \citep{Kochkov+21}, by sampling the unresolved term with a generative model \citep{Perezhogin+23,Wang+24,Dong+25}, or by identifying effective stochastic dynamics from data \citep{Kaltenbach+Koutsourelakis20,Dietrich+23}.
These previous studies close the dynamics at a fixed scale, whereas our framework formulates the effective dynamics as a family indexed by the scale parameter $\lambda$.

The finite-history input of the predictor $f_\lambda^*$ plays the role of the memory term, as has been argued for autoregressive generative rollouts \citep{Armegioiu26}.
Furthermore, a recent study reports that data-driven weather models trained on coarse-grained data learn the effect of the removed small scales implicitly \citep{Hassanzadeh+26}.
This corresponds to the first term of $f_\lambda^*$ in Eq.~(\ref{eq:optimal-drift-two-terms}), the evolution at $\lambda = 0$ followed by coarse-graining.

\paragraph{Path probability and action-based learning.}
Coarse-grained dynamics have also been described as a distribution over paths.
Onsager--Machlup actions were derived from microscopic dynamics \citep{Uneyama20}, and the corresponding variational principle was combined with deep learning \citep{Li+25}.
Objectives of this kind have been used to find transition paths \citep{Raja+25} and to train neural stochastic differential equations \citep{Shen+Chen25}.
For static configurations, the counterpart of the action is an energy, and energy-based models learn $E$ in $p(x) \propto \exp(-E(x))$, using the differentiable log-density for both sampling and learning \citep{Xie+16,Gao+21,Zhu+24}.
Our formulation is analogous, with $S_\lambda^\theta$ in Eq.~(\ref{eq:path-probability-density}) as the action over paths and its gradient as the path score for reverse-$\lambda$ sampling.
None of these previous works carries a hierarchy of spatial scales, whereas our action $S_\lambda^\theta$ is indexed by $\lambda$.

\paragraph{Diffusion-like generative models.}
Diffusion and flow-based generative models have been extensively studied \citep{Sohl-Dickstein+15,Ho+20,Song+21,Lipman+23,Albergo+23}.
These models treat a diffusion scale as a time-like parameter, often called diffusion time, with a neural network modeling the score or drift.
Standard formulations apply scale-uniform damping and noise across Fourier modes, so intermediate states are organized by noise level rather than by spatial scale \citep{Kingma+21,Rissanen+23,Falck+25}.
The coarse-to-fine appearance of generation arises from differences in signal-to-noise ratio across modes, not from an explicit spatial hierarchy \citep{Rissanen+23,Falck+25,Masuki+Ashida25}.
Applied to physical dynamics, some diffusion models identify physical time with diffusion time \citep{Ruhling+23,Ruhling+24}.
In contrast, we introduce the diffusion scale $\lambda$, which corresponds to diffusion time, as an axis separate from the physical time $t$.
The scale hierarchy is indexed by $\lambda$ and the drift $f_\lambda^\theta$ is modeled in $t$, leading to the KL-divergence objective between path densities in physical time rather than in diffusion time \citep{Ho+20,Kingma+21}.

\paragraph{RG-based generative models.}
Before diffusion-based approaches, the inverse RG flow was used as a generative model with normalizing flows \citep{Li+Wang18}.
The RG transformation was then formulated theoretically as a stochastic process \citep{Carosso20,Monthus25}, and this kind of stochastic formulation has been used to build diffusion-like generative models \citep{Cotler+Rezchikov23a,Sheshmani+25,Masuki+Ashida25,Masuki+Ashida26}.
These models employ damping and noise levels that depend on both wavenumber $k$ and diffusion scale $\lambda$, and the Laplacian-based formulation in our framework is one instance of this broader class.
Some diffusion models also combine Laplacian damping or advection--diffusion with Gaussian noise, but are not formulated as RG transformations \citep{Rissanen+23,Hoogeboom+Salimans23,Gruszczynski+25,Chen+26}.
All these previous works treat static data and define no time evolution at each scale.

\paragraph{Other scale-aware generative models.}
Several generative models introduce multiscale structure without modifying the forward process.
Some use hierarchical networks or multiresolution representations \citep{Ryu+Ye22,Fan+23,Phung+23}, while others factorize the data distribution across scales \citep{Guth+22}.
RG-based diffusion differs from all these studies in that the scale dependence is incorporated into the forward process through explicit coarse-graining.
Despite this distinction, RG-based diffusion and these approaches share the principle of coarse-to-fine generation, suggesting avenues for integration.

\paragraph{Super-resolution.}
Diffusion-like generative models have been widely applied to super-resolution \citep{Ho+22a,Saharia+22a,Mardani+25,Fotiadis+25}.
Some approaches interpret or exploit intermediate diffusion states as coarse-grained representations, and perform super-resolution by integrating from these states along the diffusion-scale axis \citep{Shu+23,Bischoff+Deck24}.
With a scale-ordered forward process, a single unconditional network performs both generation and super-resolution of static data \citep{Chen+26}; our framework shares this principle and adds simulation in $t$.

When super-resolution is combined with simulation, a common strategy is to simulate at a low resolution and then super-resolve the result \citep{Chen+23,Brenowitz+25,Xue+26}, but the two models are typically trained separately.
Our framework unifies these two steps: the predictor defines the path density whose score drives reverse-$\lambda$ super-resolution.
\section{Discussion}
\label{sec:discussion}

\subsection{Simulation and generation from one effective action}
\label{subsec:generation-as-flow}

For generation, the score in Eq.~(\ref{eq:reverse-rg-sde}) is the negative gradient of the effective action $S_\lambda^\theta$, a functional of the whole path, so the score at $t$ depends on past and future states.
Simulation with Eq.~(\ref{eq:governing-eq-along-t}) is causal in $t$, whereas generation is not.
Our framework realizes both in one theory with a single neural network.
From this viewpoint, autoregressive simulation \citep{Price+25} and non-autoregressive generation of a whole time series \citep{Brenowitz+25} may be described in one framework.
The former advances the state with the drift that appears in $S_\lambda^\theta$, and the latter follows the negative gradient of $S_\lambda^\theta$ over the whole path.

This viewpoint may connect to a geometric picture.
In the space of probability densities for static fields, the change of the density under RG has been formulated as a Wasserstein gradient flow \citep{Cotler+Rezchikov23b,Sheshmani+25}.
If the same formulation holds for path densities, the deformation of the spatiotemporal landscape $S_\lambda^\theta$ along $\lambda$ is governed by a gradient system for some potential.
The stochastic dynamics at each $\lambda$ realize paths near the minima of $S_\lambda^\theta$, i.e., the valleys of the landscape in $t$, which are the paths of high probability.
Establishing this picture is one direction for future work.

\subsection{Roles of the noise and the Laplacian in coarse-graining}
\label{subsec:intermediate-states}

Our framework extends the existing description of scale-dependent generation with diffusion models \citep{Cotler+Rezchikov23a,Falck+25,Masuki+Ashida25,Sheshmani+25} from static fields to spatiotemporal paths.
The intermediate state of the diffusion is the coarse-grained path $\{u_\lambda\}_t$, which follows $q_\lambda(\{u_\lambda\}_t)$, obtained by the RG transformation of $q_0(\{u_0\}_t)$.
Concretely, at each time, $u_\lambda(t)$ is the coarse-grained field $\mathcal{C}_\lambda u_0(t)$ ($\mathcal{C}_\lambda = e^{\alpha\lambda\nabla_x^2}$) plus the noise $\sqrt{\Sigma_\lambda}\,\epsilon(t)$ in Eq.~(\ref{eq:ou-solution-spatiotemporal}).
The Laplacian $\nabla_x^2$ and the noise $\epsilon$ play different roles.
Owing to the noise, $q_\lambda(\{u_\lambda\}_t)$ is the convolution of $q_0(\{u_0\}_t)$ with a Gaussian kernel (\Cref{subsec:rg-coarse-graining}), an operation regarded as integrating out the small-scale components \citep{Carosso20,Cotler+Rezchikov23a}.
Generation in our framework restores the small-scale components integrated out in $q_\lambda(\{u_\lambda\}_t)$.

This coarse-to-fine order of restoration requires both the Laplacian and the noise.
Without the noise, nothing is integrated out, and generation and super-resolution failed in our experiments (supplementary material Section~S3.4).
Without the Laplacian, all Fourier modes are damped at the same rate, and the order is set by the signal-to-noise ratio of the data spectrum rather than by an explicit scale \citep{Rissanen+23,Falck+25,Masuki+Ashida25} (supplementary material Section~S3.3).
Standard diffusion models add noise and damp all modes uniformly; a probability density is still defined for each intermediate state, but that state is not a physical state coarse-grained at a definite scale.
Even if a time evolution were defined at every $\lambda$, it would be the evolution of a uniformly damped, noisy copy of the original field.

\subsection{Limitations and future work}
\label{subsec:extensions}

Our theory proceeds from RG to stochastic time evolution, then to path probability, and finally to computation by deep learning (\Cref{sec:framework,sec:realizations}).
Each element is based on a simplification that can be relaxed.

In the forward process [Eq.~(\ref{eq:forward-rg-sde})], the damping $\alpha$ and the noise amplitude $\beta$ were held constant in $\lambda$, so a single schedule sets the damping and the noise level of the mode at each wavenumber $k$.
Damping and noise that depend on $k$ and $\lambda$ have been reported to improve efficiency and accuracy in RG-based diffusion models \citep{Sheshmani+25,Masuki+Ashida25}.
This extension may improve the spectrum at high wavenumbers, which is underestimated for the $128 \times 128$ Kolmogorov flow (\Cref{subsec:experimental-results}).
In the time evolution [Eq.~(\ref{eq:governing-eq-along-t})], the residual not explained by the predictor $f_\lambda^\theta$ was taken as spatiotemporally white, isotropic Gaussian noise with the amplitude $\sigma_\lambda$, so correlations of the residual are neglected.
Including the correlations explicitly in the path integral is expected to give a more precise effective model.
The amplitude $\sigma_\lambda$ can also be optimized as a learned parameter instead of being fixed by RG (\Cref{subsec:specification-of-the-noise-amplitude-sigma-lambda}).
In the path probability, the initial density $r_\lambda$ was given empirically, assuming stationarity, and PDD approximates the effect of its score by linear extrapolation (supplementary material Section~S1.6).
Modeling $r_\lambda$ would improve the accuracy of the score and is necessary for transient systems.

Several limitations also remain in the numerical experiments.
For PDD, the score is computed by automatic differentiation of $p_\lambda(\{u_\lambda\}_t)$, so its accuracy is limited by the assumed form of that density.
The training criterion is the KL divergence [Eq.~(\ref{eq:averaged-kl})], not a score-matching objective, so the accuracy of the score is not controlled directly, and the score error is not evaluated in this study.
For SA-PDD, the network takes only the current and past states as input, although the whole path (including the future states) is available when a spatiotemporal sample is generated.
We do not settle which of PDD and SA-PDD is more accurate, since the ordering can depend on the system and on the backbone network.
The history length $m$ was fixed in the experiments, so how much history each scale $\lambda$ requires remains unresolved (the present dependence on $m$ is discussed in supplementary material Section~S3.2).
A network architecture that accepts a variable $m$, such as a transformer \citep{Vaswani+17}, is a natural first step, and would allow the memory effect at each $\lambda$ to be measured directly.

The evaluation used the $L^2$ error and the PSD, so the higher-order structure of the fields remains untested even though the framework gives the distribution over paths at each scale.
The DDPMs used for comparison are specialized to one task (i.e., simulation at $\lambda = 0$ or generation), whereas our single network covers three tasks (simulation at various $\lambda$, generation, and super-resolution).
The comparison therefore does not separate the quality of a model from the range of tasks.
A comparison under equal conditions is left for future work.
The experiments also cover only two systems in a statistically steady regime at resolutions up to $128 \times 128$.
Further validation on other systems and at higher resolutions is necessary.
\section{Conclusions}
\label{sec:conclusions}

This study combines RG-based stochastic coarse-graining in space with a path-integral description of the time evolution.
Our theory gives a spatial hierarchy of coarse-grained dynamics together with the distribution over spatiotemporal paths.
The predictor of these dynamics contains the score that connects the scales, so the predictor and the score are two aspects of the same path formulation.
One trained network evolves the state in time at a fixed spatial scale and generates spatiotemporal samples from coarse to fine scales, unifying simulation, generation, and super-resolution.
The predictor is computed by deep learning in two realizations, PDD and SA-PDD.
Both are validated by numerical experiments on the Kolmogorov flow and the two-timescale Lorenz-96 model.

\section*{Acknowledgements}
The deep learning experiments were performed on the Earth Simulator at the Japan Agency for Marine-Earth Science and Technology (JAMSTEC).

\section*{Funding}
Y. Yasuda was supported by the Japan Society for the Promotion of Science (JSPS) KAKENHI (JP26K07197).

\section*{Data availability}\label{sec:data-availability}
The source code that supports the findings of this study is preserved in the Zenodo repository (\url{https://doi.org/10.5281/zenodo.22979347}) and developed openly in the GitHub repository (\url{https://github.com/YukiYasuda2718/predictor-driven-diffusion}).

\section*{Supplementary material}
The supplementary material contains details of experimental methods (S1), details of experimental results (S2), and additional experimental results (S3).

\appendix
\crefalias{subsection}{subappendix}
\renewcommand{\thesubsection}{A.\arabic{subsection}}
\renewcommand{\theequation}{A.\arabic{equation}}
\setcounter{equation}{0}

\makeatletter
\@namedef{theHequation}{A.\arabic{equation}}
\makeatother

\section*{Appendix}
\label{sec:appendix}

This appendix derives the theoretical results used in the main text: the coarse-graining operator (\Cref{subsec:coarse-graining-operator-and-covariance-matrix}), the path density defined by the predictor (\Cref{subsec:path-probability-measure-via-the-path-integral-representation}), the noise amplitude $\sigma_\lambda$ (\Cref{subsec:specification-of-the-noise-amplitude-sigma-lambda}), the optimal predictor (\Cref{subsec:kl-minimization-and-optimal-predictor}), and the relations among the scores (\Cref{subsec:path-score-and-sa-pdd-optimum}).

We adopt spatiotemporal discretization throughout, a standard convention in statistical physics \citep[e.g.,][]{Shiraishi23,Hirono+24}.
All quantities are defined at finite $\Delta x$ and $\Delta t$.
Continuum notation is a compact shorthand for discrete sums.
For instance, $\int \mathrm{d}^d x\,\mathrm{d}t = \sum_{i,n} \Delta x^d \, \Delta t$, where $d$ is the spatial dimension and $i$ and $n$ index the spatial grid points and the time steps.
Time is discretized as $t^{(n)} = n\Delta t$ with $n = 0, \ldots, N_t$.
We write $u_\lambda^{(n)} := u_\lambda(t^{(n)})$, and likewise for $\epsilon^{(n)}$ and $u_0^{(n)}$.
The spatial argument $x^{(i)}$ is omitted, as in the main text.
The path $\{u_\lambda\}_t$ is the collection of field values over all spatiotemporal grid points, $\big(u_\lambda(x^{(i)},t^{(n)})\big)_{i,n} \in \mathbb{R}^{N_x \times (N_t+1) \times N_c}$, where $N_x$ is the number of spatial grid points and $N_c$ the number of channels.
The path measure is $\mathcal{D}u_\lambda := \prod_{i,n} \mathrm{d}u_\lambda(x^{(i)},t^{(n)})$.
For a scalar function $h(\{u_\lambda\}_t)$, the gradient with respect to the field values at one time step acts componentwise, $\nabla_{u_\lambda^{(n)}} h := \partial h / \partial u_\lambda(x^{(i)},t^{(n)}) \in \mathbb{R}^{N_x \times N_c}$.
We use the It\^{o} convention, in which the drift of an increment is evaluated at the earlier time $t^{(n)}$.
Accordingly, we follow the Euler--Maruyama discretization: $(\partial_t u_\lambda)(t^{(n)}) = (u_\lambda^{(n+1)} - u_\lambda^{(n)})/\Delta t$.

\subsection{Coarse-graining operator and covariance matrix}
\label{subsec:coarse-graining-operator-and-covariance-matrix}

We derive explicit expressions for the coarse-graining operator $\mathcal{C}_\lambda$ and the covariance matrix $\Sigma_\lambda$ in Eqs.~(\ref{eq:ou-solution-spatiotemporal}) and~(\ref{eq:conditional-density}).
To obtain diagonal representations in Fourier space, we consider a finite spatial domain with periodic boundary conditions, $x \in [0, 2\pi)^d$:
\begin{align}
    \widetilde{u}_\lambda(k,t) &= \mathcal{F}[u_\lambda(x,t)] = \frac{1}{\sqrt{N_x}} \sum_{x} e^{-i k \cdot x}\, u_\lambda(x,t),  \\
    {u}_\lambda(x,t) &= \mathcal{F}^{-1}[\widetilde{u}_\lambda(k,t)] = \frac{1}{\sqrt{N_x}} \sum_k e^{i k \cdot x}\, \widetilde{u}_\lambda(k,t),
\end{align}
where $x$ runs over the $N_x$ grid points and $k$ over the integer wavenumbers up to the Nyquist limit of the grid in each direction.
The main text uses the real-space representation, while the operators are computed in the Fourier representation below.

In Fourier space, the forward process in Eq.~(\ref{eq:forward-rg-sde}) takes the form
\begin{align}
    \partial_\lambda \widetilde{u}_\lambda(k,t) &= -\alpha |k|^2 \widetilde{u}_\lambda(k,t) +\beta \widetilde{\eta}_\lambda(k,t),
\end{align}
where $\widetilde{\eta}_\lambda$ denotes the Fourier transform of the white Gaussian noise $\eta_\lambda$.
In discrete form, $\eta_\lambda$ is sampled at each diffusion scale as $\eta_\lambda \sim \mathcal{N}(0, \frac{1}{\Delta\lambda} I_{D'})$ with $D' := N_x \times (N_t+1) \times N_c$.
The increment $\eta_\lambda \Delta\lambda$ has covariance $\Delta\lambda \, I_{D'}$, consistent with the standard Wiener process convention \citep[e.g.,][]{Shiraishi23}.
Each Fourier mode evolves independently.
For $k \ne 0$ (the $k = 0$ mode is discussed below), each mode follows an Ornstein--Uhlenbeck process, for which an analytical solution can be obtained \citep[e.g.,][]{Gardiner09}:
\begin{align}
    \widetilde{u}_\lambda(k,t) = e^{-\alpha |k|^2 \lambda} \widetilde{u}_0(k,t) + \beta \int_0^\lambda \mathrm{d}\lambda'\, e^{-\alpha |k|^2\left(\lambda - \lambda'\right)} \widetilde{\eta}_{\lambda'}(k,t).
\end{align}
The conditional density at each $t$ is given by
\begin{align}
    \widetilde{q}_\lambda(\widetilde{u}_\lambda(t)\mid\widetilde{u}_0(t)) \propto \prod_{k} \exp \left(-\frac{1}{2}\frac{2 \alpha|k|^2}{\beta^2\left(1-e^{-2\alpha |k|^2 \lambda}\right)}\left|\widetilde{u}_\lambda(k,t)-\widetilde{u}_0(k,t)\, e^{-\alpha |k|^2 \lambda}\right|^2\right). \label{eq:cond-q-lambda-in-fourier}
\end{align}
The conditional density used in the main text, $q_\lambda(\{u_\lambda\}_t \mid \{u_0\}_t)$ in Eq.~(\ref{eq:conditional-density}), is obtained by expressing this Gaussian density in real space.
In Fourier space, both the coarse-graining operator and the covariance matrix are diagonal, and are given for each $k$ by
\begin{align}
    \widetilde{\mathcal{C}}_\lambda(k) &= e^{-\alpha |k|^2 \lambda}, \label{eq:coarse-graining-operator-in-fourier} \\
    \widetilde{\Sigma}_\lambda(k) &=\beta^2\frac{1-e^{-2\alpha |k|^2 \lambda}} {2\alpha |k|^2}.\label{eq:coarse-graining-covariance-in-fourier}
\end{align}
These coefficients satisfy $\widetilde{\mathcal{C}}_{\lambda'}(k)\,\widetilde{\mathcal{C}}_\lambda(k) = \widetilde{\mathcal{C}}_{\lambda + \lambda'}(k)$ and $\widetilde{\mathcal{C}}_{\lambda'}(k)^2\,\widetilde{\Sigma}_\lambda(k) + \widetilde{\Sigma}_{\lambda'}(k) = \widetilde{\Sigma}_{\lambda + \lambda'}(k)$ for each $k$.
Thus, coarse-graining to $\lambda$ and then to $\lambda'$ equals coarse-graining to $\lambda + \lambda'$.
This is the semigroup property of the convolution in Eq.~(\ref{eq:joint-probability-by-RG}) stated in \Cref{subsec:rg-coarse-graining}.
In practice, $\mathcal{C}_\lambda$ is applied by (i) transforming the field $u_\lambda(x,t)$ to Fourier space, (ii) multiplying by the diagonal coefficients $\widetilde{\mathcal{C}}_\lambda(k)$, and (iii) transforming back to real space.
The same procedure applies to $\sqrt{\Sigma_\lambda}$ and $\epsilon$ in Eq.~(\ref{eq:ou-solution-spatiotemporal}).

This Fourier representation provides a direct interpretation of the diffusion scale $\lambda$ in terms of spatial scales \citep{Carosso20,Cotler+Rezchikov23a}.
Since $k$ is dimensionless, the Fourier mode at wavenumber $k$ has spatial wavelength $2\pi/|k|$.
In terms of the wavelength $\ell_k:=2\pi/|k|$, the exponent of $\widetilde{\mathcal{C}}_\lambda$ in Eq.~(\ref{eq:coarse-graining-operator-in-fourier}) can be rewritten as $\alpha|k|^2\lambda=\alpha\lambda(2\pi/\ell_k)^2$, so $\sqrt{\alpha\lambda}$ corresponds to a wavelength (i.e., length scale) up to the factor $2\pi$.
Equivalently, modes with $|k|\gg k_{\mathrm{cut}}=1/\sqrt{\alpha\lambda}$ are exponentially suppressed, while modes with $|k|\lesssim k_{\mathrm{cut}}$ remain.

Throughout the paper, we retain the zero-wavenumber mode ($k=0$), for which $\widetilde{\Sigma}_\lambda(0) = \beta^2\lambda$ is the $k \to 0$ limit of Eq.~(\ref{eq:coarse-graining-covariance-in-fourier}).
Accordingly, products over Fourier modes [e.g., in Eq.~(\ref{eq:cond-q-lambda-in-fourier})] run over all $k$, and $\Sigma_\lambda$ is positive definite for every $\lambda > 0$.
The $k=0$ mode is never damped, since $\widetilde{\mathcal{C}}_\lambda(0) = 1$ in Eq.~(\ref{eq:coarse-graining-operator-in-fourier}).
To damp this mode, a scale-independent damping term $-\gamma u_\lambda$ can be added to the forward equation~(\ref{eq:forward-rg-sde}) \citep{Cotler+Rezchikov23a}.
We confirmed that our method also works under this variant (supplementary material Section~S3.3).

\subsection{Path probability density by the predictor}
\label{subsec:path-probability-measure-via-the-path-integral-representation}

We derive the path density $p_\lambda(\{u_\lambda\}_t)$ defined by the predictor in discrete time [Eq.~(\ref{eq:path-probability-density})], assuming that the dynamics in $t$ are Markov \citep{Onsager+Machlup53a,Shiraishi23}.
Under the It\^{o} convention, the governing equation~(\ref{eq:governing-eq-along-t}) takes the form
\begin{align}
    u_\lambda^{(n+1)} = u_\lambda^{(n)} + f_\lambda^\theta(u_\lambda^{(n)}) \Delta t + \sigma_\lambda\xi^{(n)} \Delta t. \label{eq:discrete-time-governing-equation}
\end{align}
Let $D := N_x \times N_c$ denote the dimension of the discretized state $u_\lambda^{(n)}$ at a fixed time step.
The white Gaussian noise is sampled at each time step as $\xi^{(n)} \sim \mathcal{N}(0, \frac{1}{\Delta x^d \Delta t} I_D)$, so that the increment $\xi^{(n)} \Delta t$ has covariance $(\Delta t/\Delta x^d) \, I_D$ \citep[e.g.,][]{Shiraishi23}.

From the governing equation~(\ref{eq:discrete-time-governing-equation}), the transition probability density of $u_\lambda^{(n+1)}$ is \citep[e.g.,][]{Shiraishi23}
\begin{align}
    p_\lambda(u_\lambda^{(n+1)} \mid u_\lambda^{(n)}) = \left(\frac{2\pi \sigma_\lambda^2\Delta t}{\Delta x^d}\right)^{-D/2} \exp\left[-\frac{\Delta x^d}{2\sigma_\lambda^2\Delta t}\left\| u_\lambda^{(n+1)} - u_\lambda^{(n)} - f_\lambda^\theta(u_\lambda^{(n)}) \Delta t \right\|^2 \right], \label{eq:transition-probability-density}
\end{align}
where $\|v\|^2 := \sum_i |v_i|^2$ for a field $v$ at a fixed time step, the sum running over all spatial grid points and channels, and $\Delta x^d$ is the grid-cell volume.
Together with the factor $\Delta x^d$, this norm is the discrete form of $\int \mathrm{d}^d x\,\|\cdot\|^2$ in the main text.
Since $\sigma_\lambda$ is state independent, the Jacobian associated with the change of variables from $\xi^{(n)}$ to $u_\lambda^{(n+1)}$ is the constant $(\sigma_\lambda \Delta t)^{-D}$, and the normalization does not depend on $f_\lambda^\theta$.

The path density is the product of the transition density over successive time steps,
\begin{align}
    p_\lambda(\{u_\lambda\}_t) &= \frac{r_\lambda(u_\lambda^{(0)})}{Z_\lambda} \exp \left[ - S_\lambda^\theta(\{u_\lambda\}_t)\right], \notag \\
    S_\lambda^\theta(\{u_\lambda\}_t) &= \frac{\Delta x^d}{2 \sigma_\lambda^2} \sum_{n=0}^{N_t-1} \Delta t \left\| \frac{u_\lambda^{(n+1)} - u_\lambda^{(n)}}{\Delta t} - f_\lambda^\theta(u_\lambda^{(n)}) \right\|^2, \label{eq:path-integral-formal-representation}
\end{align}
where $r_\lambda(u_\lambda^{(0)})$ denotes the probability density of the initial condition and $Z_\lambda$ the normalization constant, $Z_\lambda := (2\pi \sigma_\lambda^2 \Delta t/\Delta x^d)^{D N_t/2}$.
The continuum notation of Eq.~(\ref{eq:path-probability-density}) is shorthand for this sum.
For a given initial state, the path that maximizes $p_\lambda(\{u_\lambda\}_t)$ satisfies $u_\lambda^{(n+1)} = u_\lambda^{(n)} + f_\lambda^\theta(u_\lambda^{(n)})\,\Delta t$ for every $n$.
The deterministic rollout is therefore the most probable path, which is used to evaluate simulation in \Cref{subsec:experimental-setup}.
As $\sigma_\lambda \to 0$, $p_\lambda(\{u_\lambda\}_t)$ concentrates on this path.
The numerical treatment of small $\sigma_\lambda$ is given in \Cref{subsec:specification-of-the-noise-amplitude-sigma-lambda}.

In practice, the drift takes a history window of $m$ steps as input \citep{Tashiro+21,Yuan+24,Price+25}, which we write as $U_\lambda^{(n)} := (u_\lambda^{(n)}, u_\lambda^{(n-1)}, \ldots, u_\lambda^{(n-m+1)})$ in Eq.~(\ref{eq:definition-U}).
This does not violate the Markov formulation, since the process augmented with the past $m - 1$ time steps is Markov.
Its transition adds the noise $\sigma_\lambda \xi^{(n)} \Delta t$ to the new component $u_\lambda^{(n+1)}$ alone, the remaining components of $U_\lambda^{(n+1)}$ being copies of states already in the window.
Following Eq.~(\ref{eq:transition-probability-density}), the density of the new component $u_\lambda^{(n+1)}$ is
\begin{align}
    p_\lambda\!\left(u_\lambda^{(n+1)} \mid U_\lambda^{(n)}\right) = \mathcal{N}\!\Big(u_\lambda^{(n+1)}\,;\, u_\lambda^{(n)} + f_\lambda^\theta(U_\lambda^{(n)})\,\Delta t,\;\frac{\sigma_\lambda^2 \Delta t}{\Delta x^d}\, I_D\Big). \label{eq:history-transition-density}
\end{align}
Let $r_\lambda(U_\lambda^{(0)})$ denote the initial density of the history window, where the missing past states $u_\lambda^{(-1)}, \ldots, u_\lambda^{(-m+1)}$ are filled by zero padding (supplementary material Section~S1.2).
The path density then factorizes as
\begin{align}
    p_\lambda(\{u_\lambda\}_t) = r_\lambda(U_\lambda^{(0)}) \prod_{n=0}^{N_t-1} p_\lambda\!\left(u_\lambda^{(n+1)} \mid U_\lambda^{(n)}\right). \label{eq:history-factorized-path-density}
\end{align}
In the PDD formalism, the path score $\nabla_{u_\lambda^{(n)}}\ln p_\lambda(\{u_\lambda\}_t)$ is computed by automatic differentiation of this log-density (\Cref{subsec:pdd}).

\subsection{Specification of the noise amplitude \texorpdfstring{$\sigma_\lambda$}{sigma\_lambda}}
\label{subsec:specification-of-the-noise-amplitude-sigma-lambda}

The noise amplitude $\sigma_\lambda$ characterizes the magnitude of the uncertainty arising from the elimination of small-scale components through coarse-graining \citep{Carosso20,Cotler+Rezchikov23a}.
Although $\sigma_\lambda$ could in principle be treated as a learnable parameter, we determine $\sigma_\lambda$ from the solution of the forward process in Eq.~(\ref{eq:forward-rg-sde}).
The coarse-grained state at each time step is $u_\lambda^{(n)} = \mathcal{C}_\lambda u_0^{(n)} + \sqrt{\Sigma_\lambda} \epsilon^{(n)}$ [Eq.~(\ref{eq:ou-solution-spatiotemporal})] with $\epsilon^{(n)} \sim \mathcal{N}(0,I_D)$ independent across $n$, and its temporal difference is
\begin{align}
    u_\lambda^{(n+1)} - u_\lambda^{(n)} = \mathcal{C}_\lambda \left(u_0^{(n+1)} - u_0^{(n)}\right) + \sqrt{\Sigma_\lambda} \left(\epsilon^{(n+1)} - \epsilon^{(n)}\right), \label{eq:temporal-diff-coarse-grained-u}
\end{align}
which is Eq.~(\ref{eq:temporal-difference}) of the main text up to the division by $\Delta t$.
Compared with the governing equation~(\ref{eq:discrete-time-governing-equation}), the first term on the right-hand side corresponds to the drift term $f_\lambda^\theta \Delta t$ and the second to the noise term $\sigma_\lambda \xi^{(n)} \Delta t$.
Coarse-graining through a stochastic process thus induces a noise term in the temporal dynamics.

To obtain a concrete value for $\sigma_\lambda$, we match the conditional covariance of the dynamics in Eq.~(\ref{eq:discrete-time-governing-equation}) to that of $u_\lambda^{(n+1)}$ given $u_\lambda^{(n)}$.
This covariance of $u_\lambda^{(n+1)}$ in Eq.~(\ref{eq:temporal-diff-coarse-grained-u}) consists of the uncertainty in predicting $\mathcal{C}_\lambda u_0^{(n+1)}$ and the irreducible contribution $\Sigma_\lambda$ from the noise $\epsilon^{(n+1)}$ at the next time step.
The noise $\epsilon^{(n)}$ at the current step is not considered here, since the optimal drift in \Cref{subsec:kl-minimization-and-optimal-predictor} absorbs the part of $\epsilon^{(n)}$ predictable from the current state (\Cref{subsec:optimal-predictor}).
Equating the traces of $(\sigma_\lambda^2 \Delta t/\Delta x^d) I_D$ and $\Sigma_\lambda$, i.e., matching the average variance, yields
\begin{align}
  \sigma_\lambda^2 = \frac{\Delta x^d}{\Delta t \cdot D} \mathrm{tr}(\Sigma_\lambda).
  \label{eq:sigma-trace-matching}
\end{align}
This indicates that $\sigma_\lambda$ characterizes the typical scale of coarse-graining noise.
We use this setting of $\sigma_\lambda$ for all the experiments.
Equation~(\ref{eq:sigma-trace-matching}) provides a conservative estimate of $\sigma_\lambda$ because it counts only the contribution from $\sqrt{\Sigma_\lambda}\,\epsilon^{(n+1)}$ in Eq.~(\ref{eq:temporal-diff-coarse-grained-u}).
We consider $\sqrt{\Sigma_\lambda}\,\epsilon^{(n+1)}$ alone because it is available in closed form, whereas the prediction uncertainty of $\mathcal{C}_\lambda u_0^{(n+1)}$ is state dependent and would have to be learned (\Cref{subsec:extensions}).
We confirmed with PDD on the Lorenz-96 model and the $40 \times 40$ Kolmogorov flow that moderate rescaling of $\sigma_\lambda$ (by a factor of 1.5) does not significantly affect the results.

We now consider the limit of $\sigma_\lambda$ as $\lambda \to 0$.
From Eq.~(\ref{eq:coarse-graining-covariance-in-fourier}), $\Sigma_\lambda \to 0$ as $\lambda \to 0$, which implies $\sigma_\lambda \to 0$ in Eq.~(\ref{eq:sigma-trace-matching}).
Mathematically, no singularity arises in the path probability itself: the path density concentrates on the most probable path as a well-defined distributional limit (\Cref{subsec:path-probability-measure-via-the-path-integral-representation}).
However, the path score $\nabla_{u_\lambda^{(n)}} \ln p_\lambda(\{u_\lambda\}_t)$ diverges in this limit.
For numerical stability, we lower-bound $\Sigma_\lambda$ by $\Sigma_{\Delta\lambda}$, the covariance at the smallest non-zero diffusion scale ($\lambda\in[0,1]$ is discretized into 1,000 steps, so $\Delta\lambda = 10^{-3}$; supplementary material Section~S1.3).
This regularization ensures $\sigma_\lambda > 0$ in Eq.~(\ref{eq:sigma-trace-matching}), preventing divergence of the path score.
The same bound caps the weight $1/(2\sigma_\lambda^2)$ of the losses in Eqs.~(\ref{eq:loss-pdd}) and~(\ref{eq:loss-sa-pdd}) at small $\lambda$; without it, the $\lambda$-average in Eq.~(\ref{eq:averaged-kl}) can diverge.

\subsection{KL divergence minimization and the optimal predictor}
\label{subsec:kl-minimization-and-optimal-predictor}

We derive the optimal predictor $f_\lambda^*$ as the minimizer of the KL divergence $D_{\mathrm{KL}}(q_\lambda \| p_\lambda)$ in Eq.~(\ref{eq:kl-divergence}) using the discrete-time path density.
We first define the joint density by
\begin{align}
    q_\lambda^{\rm joint}(\{u_\lambda\}_t,\{u_0\}_t) := q_\lambda(\{u_\lambda\}_t \mid \{u_0\}_t)\,q_0(\{u_0\}_t). \label{eq:q-joint}
\end{align}
Here, $q_\lambda(\{u_\lambda\}_t \mid \{u_0\}_t)$ in Eq.~(\ref{eq:conditional-density}) factorizes over time steps into the Gaussian factors $\mathcal{N}(u_\lambda^{(n)}\,;\, \mathcal{C}_\lambda u_0^{(n)}, \Sigma_\lambda)$.
The marginal of $q_\lambda^{\rm joint}$ with respect to $\{u_0\}_t$ gives $q_\lambda(\{u_\lambda\}_t)$, as in Eq.~(\ref{eq:joint-probability-by-RG}).
We assume stationarity, so that $r_\lambda$ does not depend on time and the optimal predictor (or drift) $f_\lambda^*$ does not depend on $n$.

The KL divergence is
\begin{align}
    D_{\mathrm{KL}}(q_\lambda \| p_\lambda) &= \int \mathcal{D}u_\lambda\, q_\lambda(\{u_\lambda\}_t)\, \ln\frac{q_\lambda(\{u_\lambda\}_t)}{p_\lambda(\{u_\lambda\}_t)} \notag \\
    &= \mathbb{E}_{(\{u_\lambda\}_t,\{u_0\}_t)\sim q_\lambda^{\rm joint}} \left[-\ln p_\lambda(\{u_\lambda\}_t)\right] + \mathrm{const}, \label{eq:kl-reduction}
\end{align}
where $q_\lambda(\{u_\lambda\}_t)$ does not depend on $\theta$, and $\mathrm{const}$ collects all terms independent of $\theta$.
From Eq.~(\ref{eq:path-integral-formal-representation}), the $\theta$-dependent part of $D_{\mathrm{KL}}(q_\lambda \| p_\lambda)$ is
\begin{align}
    \mathbb{E}_{(\{u_\lambda\}_t,\{u_0\}_t)\sim q_\lambda^{\rm joint}} \left[ \frac{\Delta x^d}{2\sigma_\lambda^2\Delta t} \sum_{n=0}^{N_t-1} \left\|u_\lambda^{(n+1)} - u_\lambda^{(n)} - f_\lambda^\theta(u_\lambda^{(n)})\Delta t\right\|^2 \right], \label{eq:kl-to-increments-mse}
\end{align}
which is the weighted least-squares form of Eq.~(\ref{eq:kl-divergence}).
Since the minimizer of a squared loss over all functions is the conditional expectation, the optimal drift satisfies
\begin{align}
    f_\lambda^*(u)\Delta t = \mathbb{E}\!\left[u_\lambda^{(n+1)}-u_\lambda^{(n)}\;\middle|\;u_\lambda^{(n)}=u\right], \label{eq:optimal-drift-conditional-mean-increment}
\end{align}
where the conditional expectation is taken under $q_\lambda^{\rm joint}$.

Substituting the temporal difference in Eq.~(\ref{eq:temporal-diff-coarse-grained-u}) and taking the conditional expectation yields
\begin{align}
    f_\lambda^*(u_\lambda^{(n)})\Delta t
    &= \mathbb{E}\!\left[\mathcal{C}_\lambda\left(u_0^{(n+1)} - u_0^{(n)}\right) \mid u_\lambda^{(n)}\right] - \sqrt{\Sigma_\lambda}\,\mathbb{E}\!\left[\epsilon^{(n)} \mid u_\lambda^{(n)}\right]. \label{eq:optimal-drift-noise-form}
\end{align}
Since $\epsilon^{(n+1)}$ is independent of $(u_0^{(n)}, \epsilon^{(n)})$ and hence of $u_\lambda^{(n)}$, its conditional expectation vanishes.
That of $\epsilon^{(n)}$ does not, because $\epsilon^{(n)}$ is contained in $u_\lambda^{(n)} = \mathcal{C}_\lambda u_0^{(n)} + \sqrt{\Sigma_\lambda}\epsilon^{(n)}$.
From Eq.~(\ref{eq:ou-solution-spatiotemporal}), the second term is written equivalently as
\begin{equation}
    - \sqrt{\Sigma_\lambda}\,\mathbb{E}\!\left[\epsilon^{(n)} \mid u_\lambda^{(n)}\right] = \mathbb{E}[\mathcal{C}_\lambda u_0^{(n)} \mid u_\lambda^{(n)}] - u_\lambda^{(n)}. \label{eq:denoising-correction}
\end{equation}
This is a denoising correction induced by conditioning on the noisy coarse-grained state.

The correction depends on the $q_0$-induced posterior distribution of $u_0^{(n)}$ given $u_\lambda^{(n)}$, and hence is not available in closed form without further assumptions.
Nevertheless, the correction admits a general identity \citep[Tweedie's formula;][]{Efron11} due to the Gaussianity of the conditional density in Eq.~(\ref{eq:conditional-density}).
The marginal density of $u_\lambda^{(n)}$ under $q_\lambda^{\rm joint}$ is
\begin{align}
    q_\lambda(u_\lambda^{(n)}) &= \int q_\lambda^{\rm joint}(\{u_\lambda\}_t,\{u_0\}_t)\, \mathrm{d}u_0^{(n)}\prod_{j \ne n} \left(\mathrm{d}u_0^{(j)} \mathrm{d}u_\lambda^{(j)}\right) \notag\\
    &= \int \mathcal{N}(u_\lambda^{(n)}; \mathcal{C}_\lambda u_0^{(n)}, \Sigma_\lambda)\, q_0(u_0^{(n)})\, \mathrm{d}u_0^{(n)},
\end{align}
where $q_0(u_0^{(n)})$ is the marginal of $q_0(\{u_0\}_t)$ at $t^{(n)}$.
Differentiating under the integral with $\nabla_u \mathcal{N}(u; \mu, \Sigma) = -\Sigma^{-1}(u - \mu)\,\mathcal{N}(u; \mu, \Sigma)$ gives
\begin{align}
    \Sigma_\lambda \nabla_{u_\lambda^{(n)}}\ln q_\lambda(u_\lambda^{(n)}) &= -\int \left( u_\lambda^{(n)} - \mathcal{C}_\lambda u_0^{(n)} \right) q_\lambda(u_0^{(n)} \mid u_\lambda^{(n)})\, \mathrm{d}u_0^{(n)} \notag\\ 
    &= -\sqrt{\Sigma_\lambda}\,\mathbb{E}\!\left[\epsilon^{(n)} \mid u_\lambda^{(n)}\right], \label{eq:tweedie-correction}
\end{align}
where Eq.~(\ref{eq:denoising-correction}) is used in the second line.
The identity is well defined because $\Sigma_\lambda$ is positive definite for $\lambda > 0$ with the $k = 0$ mode treated as in \Cref{subsec:coarse-graining-operator-and-covariance-matrix}.
As $\lambda \to 0$, $\Sigma_\lambda \to 0$ and the posterior of $u_0^{(n)}$ concentrates on $u_\lambda^{(n)}$, so the correction vanishes.

\subsection{Path score, single-time score, and the SA-PDD optimum}
\label{subsec:path-score-and-sa-pdd-optimum}

We derive the component of the path score that the reverse equation~(\ref{eq:reverse-rg-sde}) needs.
The forward process in Eq.~(\ref{eq:forward-rg-sde}) acts independently at each time, so
\begin{align}
    q_\lambda(\{u_\lambda\}_t) = \int \mathcal{D}u_0\, \prod_{j} \mathcal{N}\!\left(u_\lambda^{(j)}; \mathcal{C}_\lambda u_0^{(j)}, \Sigma_\lambda\right)\, q_0(\{u_0\}_t), \label{eq:path-density-as-integral}
\end{align}
in which only the factor $j = n$ depends on $u_\lambda^{(n)}$.
The same computation as in Eq.~(\ref{eq:tweedie-correction}) gives
\begin{align}
    \nabla_{u_\lambda^{(n)}}\ln q_\lambda(\{u_\lambda\}_t) &= -\Sigma_\lambda^{-1}\left(u_\lambda^{(n)} - \mathbb{E}\!\left[\mathcal{C}_\lambda u_0^{(n)} \mid \{u_\lambda\}_t\right]\right) \notag\\
    &= -\Sigma_\lambda^{-1/2}\,\mathbb{E}\!\left[\epsilon^{(n)} \mid \{u_\lambda\}_t\right], \label{eq:path-score-component}
\end{align}
which is Eq.~(\ref{eq:score-path}) of the main text.
The expectation is taken under the posterior $q_\lambda(\{u_0\}_t \mid \{u_\lambda\}_t) \propto q_\lambda(\{u_\lambda\}_t \mid \{u_0\}_t)\, q_0(\{u_0\}_t) = q_\lambda^{\rm joint}(\{u_\lambda\}_t,\{u_0\}_t)$.
It depends on all times, past and future, through the temporal correlation of $q_0(\{u_0\}_t)$.

The single-time score, $\nabla_{u_\lambda^{(n)}}\ln q_\lambda(u_\lambda^{(n)})$ in Eq.~(\ref{eq:tweedie-correction}), is the conditional average of Eq.~(\ref{eq:path-score-component}) given $u_\lambda^{(n)}$:
\begin{align}
    \nabla_{u_\lambda^{(n)}}\ln q_\lambda(u_\lambda^{(n)}) = \mathbb{E}\!\left[\nabla_{u_\lambda^{(n)}}\ln q_\lambda(\{u_\lambda\}_t) \;\middle|\; u_\lambda^{(n)}\right]. \label{eq:marginal-score-as-average}
\end{align}
This follows by differentiating $q_\lambda(u_\lambda^{(n)}) = \int q_\lambda(\{u_\lambda\}_t) \prod_{j \ne n} \mathrm{d}u_\lambda^{(j)}$ under the integral and dividing by $q_\lambda(u_\lambda^{(n)})$.
The ratio $q_\lambda(\{u_\lambda\}_t)/q_\lambda(u_\lambda^{(n)})$ is the conditional density of the other times given $u_\lambda^{(n)}$.
Substituting Eqs.~(\ref{eq:tweedie-correction}) and~(\ref{eq:path-score-component}) into~(\ref{eq:marginal-score-as-average}) gives $\mathbb{E}[\epsilon^{(n)} \mid u_\lambda^{(n)}] = \mathbb{E}[\mathbb{E}[\epsilon^{(n)} \mid \{u_\lambda\}_t] \mid u_\lambda^{(n)}]$, consistent with the tower property.
The single-time score thus averages out the dependence of the path score on the other times.

The same argument applies to any subset of times.
For the history window $U_\lambda^{(n)}$,
\begin{align}
    \nabla_{u_\lambda^{(n)}}\ln q_\lambda(U_\lambda^{(n)}) &= \mathbb{E}\!\left[\nabla_{u_\lambda^{(n)}}\ln q_\lambda(\{u_\lambda\}_t) \;\middle|\; U_\lambda^{(n)}\right] \notag\\
    &= -\Sigma_\lambda^{-1/2}\,\mathbb{E}\!\left[\epsilon^{(n)} \mid U_\lambda^{(n)}\right], \label{eq:window-score}
\end{align}
where $q_\lambda(U_\lambda^{(n)})$ is the marginal for the window $U_\lambda^{(n)}$.
The case $m = 1$ recovers Eq.~(\ref{eq:marginal-score-as-average}), and a window covering all times (i.e., the whole path) recovers Eq.~(\ref{eq:path-score-component}).
The denoising term of the SA-PDD loss [the first term in Eq.~(\ref{eq:loss-sa-pdd})] is minimized over all functions of $U_\lambda^{(n)}$ by $\hat\epsilon_\lambda^*(U_\lambda^{(n)}) = \mathbb{E}[\epsilon^{(n)} \mid U_\lambda^{(n)}]$.
The corresponding score $\hat s_\lambda^* = -\Sigma_\lambda^{-1/2}\hat\epsilon_\lambda^*$ coincides with Eq.~(\ref{eq:window-score}), i.e., Eq.~(\ref{eq:score-window}) of the main text.
What SA-PDD estimates is thus the path-score component averaged over the times outside the window $U_\lambda^{(n)}$.

The squared error of $\hat\epsilon_\lambda^*$ decomposes orthogonally as
\begin{align}
    \mathbb{E}\!\left[\left\|\epsilon^{(n)} - \hat\epsilon_\lambda^*(U_\lambda^{(n)})\right\|^2\right]
    = \mathbb{E}\!\left[\left\|\epsilon^{(n)} - \mathbb{E}\!\left[\epsilon^{(n)} \mid \{u_\lambda\}_t\right]\right\|^2\right]
    + \mathbb{E}\!\left[\mathrm{Var}\!\left(\mathbb{E}\!\left[\epsilon^{(n)} \mid \{u_\lambda\}_t\right] \;\middle|\; U_\lambda^{(n)}\right)\right], \label{eq:sa-pdd-error-decomposition}
\end{align}
where $\mathrm{Var}$ denotes the trace of the conditional covariance.
The first term is the part of $\epsilon^{(n)}$ not recoverable even from the whole path, while the second is the part of $\epsilon^{(n)}$ that the times outside the window, mostly the future, would explain.
The difference of the two scores is $\hat s_\lambda^*(U_\lambda^{(n)}) - \nabla_{u_\lambda^{(n)}}\ln q_\lambda(\{u_\lambda\}_t) = -\Sigma_\lambda^{-1/2}\big(\hat\epsilon_\lambda^*(U_\lambda^{(n)}) - \mathbb{E}[\epsilon^{(n)} \mid \{u_\lambda\}_t]\big)$.
The second term in Eq.~(\ref{eq:sa-pdd-error-decomposition}) weighted by $\Sigma_\lambda^{-1}$ is therefore the mean squared deviation of $\hat s_\lambda^*$ from the path score.
This deviation is the error of SA-PDD discussed in \Cref{subsec:two-approximations,subsec:extensions}.

\bibliographystyle{unsrtnat}
\bibliography{references}


\makeatletter
%
\let\all@ref\ref
\let\all@Cref\Cref

\def\all@stop{}
\def\all@stripM M-#1\all@stop{#1}
\def\all@ifM#1{\all@ifM@#1M-\all@stop}
\def\all@ifM@#1M-#2\all@stop{%
  \ifx\relax#1\relax
    \expandafter\@firstoftwo
  \else
    \expandafter\@secondoftwo
  \fi}
\newcommand{\all@name}[1]{\all@ifM{#1}{\all@stripM #1\all@stop}{#1}}

\newcommand{\all@apply}[2]{%
  \begingroup\edef\all@next{\endgroup\noexpand#1{\all@name{#2}}}\all@next}

\protected\def\ref#1{\all@apply{\all@ref}{#1}}
\protected\def\Cref#1{\all@apply{\all@Cref}{#1}}
\makeatother

\setcounter{section}{0}
\setcounter{equation}{0}
\setcounter{figure}{0}
\setcounter{table}{0}
\setcounter{algorithm}{0}
\renewcommand{\thesection}{S\arabic{section}}
\renewcommand{\thesubsection}{\thesection.\arabic{subsection}}
\renewcommand{\theequation}{S\arabic{equation}}
\renewcommand{\thefigure}{S\arabic{figure}}
\renewcommand{\thetable}{S\arabic{table}}
\renewcommand{\thealgorithm}{S\arabic{algorithm}}

\makeatletter
\@namedef{theHsection}{S\arabic{section}}
\@namedef{theHsubsection}{S\arabic{section}.\arabic{subsection}}
\@namedef{theHequation}{S\arabic{equation}}
\@namedef{theHfigure}{S\arabic{figure}}
\@namedef{theHtable}{S\arabic{table}}
\@namedef{theHalgorithm}{S\arabic{algorithm}}
\makeatother

\crefalias{appendix}{section}
\crefalias{subappendix}{subsection}
\crefname{subsection}{Section}{Sections}
\Crefname{subsection}{Section}{Sections}
\crefname{algorithm}{Algorithm}{Algorithms}
\Crefname{algorithm}{Algorithm}{Algorithms}

\clearpage
\phantomsection
\addcontentsline{toc}{section}{Supplementary material}

\begin{center}
    {\LARGE\bfseries Supplementary material}\\[1.5ex]
    {\large for ``A renormalization-group hierarchy of stochastic effective dynamics learned through path integrals''}
\end{center}

\bigskip

\noindent
This supplementary material provides details of experimental methods (\Cref{sec:details-of-experimental-methods}), details of experimental results (\Cref{supp:experimental-results}), and additional experimental results (\Cref{supp:additional-experimental-results}), supplementing the main article.
Numbers without the prefix S, such as Section~4, Eq.~(5), Table~1, and Appendix~A.2, refer to the main article.

\section{Details of experimental methods}
\label{sec:details-of-experimental-methods}

This section provides details of the experimental methods in main article Section~\ref{M-sec:experiments}.
Notation follows the main article and its Appendix.
In particular, $u_\lambda^{(n)} := u_\lambda(t^{(n)})$ denotes the coarse-grained state at time step $t^{(n)} = n\Delta t$, and $U_\lambda^{(n)} := (u_\lambda^{(n)}, u_\lambda^{(n-1)}, \ldots, u_\lambda^{(n-m+1)})$ the history window of length $m$ that is input to the neural network (Section~\ref{M-subsec:pdd} and Appendix~\ref{M-subsec:path-probability-measure-via-the-path-integral-representation}).
The neural network is written $\hat f_\lambda^\theta$ during training and $\hat f_\lambda^*$ after training (Section~\ref{M-sec:realizations}).

\subsection{Construction of training data}
\label{subsec:construction-of-training-data}

\paragraph{Lorenz-96 model.}
The Lorenz-96 model is a one-dimensional idealized model that represents zonal atmospheric dynamics \citep{Lorenz96}.
Periodic boundary conditions are assumed, analogous to the longitudinal direction of the Earth, with $x \in [0, 2\pi)$.
The state variables $X$ and $Y$ represent slow and fast variables in physical time $t$, respectively, and evolve according to
\begin{align}
    \frac{\mathrm{d}}{\mathrm{d}t} X_i &= -X_{i-1}\left(X_{i-2}-X_{i+1}\right)-X_i+F-\left(\frac{h c}{b}\right) \sum_{j=1}^{J} Y_{j,i}, \\
    \frac{\mathrm{d}}{\mathrm{d}t} Y_{j,i} &= -c b Y_{j+1,i}\left(Y_{j+2, i}-Y_{j-1,i}\right)-c Y_{j,i}+\frac{h c}{b} X_i.
\end{align}
The indices $i$ and $j$ denote grid-point locations on the one-dimensional spatial domain.
For each of the $I$ slow variables ($i = 1, \ldots, I$), $J$ fast variables $Y_{j,i}$ interact with the corresponding slow variable $X_i$ for $j = 1, \ldots, J$.
To match the spatial resolution, each $X_i$ is replicated $J$ times so that both $X$ and $Y$ are defined on $IJ$ grid points.
The ordering of grid points follows Russell et al.\ \citep{Russell+17}, where $X_i=X_{i+I}$, $Y_{j,i}=Y_{j,i+I}$, and $Y_{j,i}=Y_{j-J,i+1}$.
We take $h = 1$ and $F = b = c = 10$ from Lorenz \citep{Lorenz96}, and set $I = 32$ and $J = 4$, so the grid has $IJ = 128$ points.
The parameter $b$ characterizes the amplitude ratio between $X$ and $Y$, while $c$ characterizes the timescale separation.
With $b = c = 10$, the slow variable $X$ has amplitudes and timescales approximately one order of magnitude larger than those of the fast variable $Y$.

Training data for the Lorenz-96 model are generated by numerical integration using the Euler method with a time step of $5 \times 10^{-4}$.
Initial conditions are randomly varied, and samples are collected after the system reaches a statistically stationary state, with a burn-in period of 30 time units.
Each sample consists of two channels $(X, Y)$, with 64 snapshots in time (i.e., $N_t=63$ increments) and a spatial length of 128 grid points.
The temporal sampling interval is set to 0.05 time units, corresponding to 100 integration steps.
We verify that varying this sampling interval changes the temporal power spectrum only slightly.
A total of 3,000 samples are used for training and 100 samples with distinct initial conditions are used for testing.
We confirm that the test results are not strongly dependent on the sample size.

\paragraph{Kolmogorov flow.}
The two-dimensional Kolmogorov flow is an incompressible flow system driven by sinusoidal forcing \citep[e.g.,][]{Obukhov+83} and is widely used in turbulence research and machine learning applications \citep[e.g.,][]{Lucas+Kerswell14,Kochkov+21,Shu+23}.
The governing equations are
\begin{align}
    \frac{\partial \zeta}{\partial t} + \mathbf{V} \cdot \nabla_x \zeta = - \mu \zeta + \nu \nabla_x^2 \zeta - k_{\rm forcing} \cos (k_{\rm forcing} y), \\
    \zeta = \nabla_x^2 \psi \quad\text{and}\quad\mathbf{V} = \left(-\frac{\partial \psi}{\partial y}, \frac{\partial \psi}{\partial x}\right)^{\top}.
\end{align}
Here, $\nabla_x$ and $\nabla_x^2$ denote the gradient and Laplacian with respect to the spatial coordinates $(x,y)$, $\zeta$ the vorticity field, $\mathbf{V}$ the velocity field, and $\psi$ the stream function.
Doubly periodic boundary conditions are assumed on $(x,y) \in [0, 2\pi)^2$.
Model parameters are chosen following previous studies \citep{Kochkov+21,Shu+23}, with $\mu = 0.1$, $\nu = 1 \times 10^{-3}$, and $k_{\rm forcing} = 4$.

Spatial derivatives and nonlinear terms are computed using a pseudospectral method \citep[e.g.,][]{Canuto+88}, and time integration is performed using the explicit midpoint method, also known as the modified Euler method.
The spatial grid resolution is set to $256 \times 256$, and the integration time step is $1 \times 10^{-3}$.
Initial conditions are randomly varied, and samples are collected after the system reaches a statistically stationary state, with a burn-in period of 50 time units.
Each sample consists of a single channel corresponding to $\zeta$, with 40 snapshots in time (i.e., $N_t=39$ increments) and a spatial resolution of $40 \times 40$ or $128 \times 128$.
The spatial fields are resized from the $256 \times 256$ grid using bicubic interpolation, and the temporal dimension is subsampled at an interval of 0.5 time units.
We verify that changing the temporal sampling interval affects the temporal power spectrum only slightly.
Both resolutions are obtained from the same $256 \times 256$ simulations, differing only in the target size of the resizing.
At each resolution, a total of 6,000 samples are used for training and 100 samples with distinct initial conditions are used for testing.
We confirm that the test results are not strongly dependent on the sample size.

\paragraph{Data preprocessing.}
All state variables (i.e., $X, Y, \zeta$) are standardized to have zero mean and unit standard deviation.

\subsection{Neural network architecture}
\label{subsec:neural-network-architecture}

\paragraph{U-Net.}
We use a U-Net-based neural network architecture \citep{Ronneberger+15,Ho+20,Ho+22b,Saharia+22b,Saharia+22a}.
The number of channels is first increased to 32 by convolution, after which the spatial resolution is progressively reduced.
Self-attention mechanisms are then applied \citep{Vaswani+17}, and finally the spatial resolution is restored.
For the Lorenz-96 model, one-dimensional convolutions are used to reduce the spatial resolution to 1/16 ($=1/2^4$) of the original size, while the number of channels is gradually increased to 64, 128, 128, and 192.
For the Kolmogorov flow, two-dimensional convolutions are used to reduce the spatial resolution to 1/8 ($=1/2^3$) of the original size, with the number of channels increased to 64, 128, and 192.
Downsampling is performed using convolutions with stride 2, and upsampling is performed using transposed convolutions.
The Swish activation function is used throughout the network \citep{Ramachandran+17}.
The diffusion scale $\lambda$ is provided via feature-wise linear modulation (FiLM) conditioning \citep{Perez+18}, which modulates intermediate features through learned scale and shift parameters.

The input to the neural network consists of the state vectors at the current time and the four most recent past time steps, i.e., the history window $U_\lambda^{(n)}$ with $m = 5$ (Section~\ref{M-subsec:pdd}).
For the Lorenz-96 model, this corresponds to 10 input channels, representing $X$ and $Y$ over five time steps, while for the Kolmogorov flow the input consists of 5 channels, representing $\zeta$ over five time steps.
All variables are concatenated along the channel dimension before being provided to the network.
The network output is the increment to the next time step, $\hat f_\lambda^\theta(U_\lambda^{(n)})\,\Delta t$, i.e., a finite-difference approximation of the time derivative in physical time $t$ multiplied by $\Delta t$.
In \Cref{subsec:dependence-on-the-number-of-input-time-steps}, we vary the number of input time steps; the network architecture is kept fixed throughout.

\paragraph{Output layers of PDD and SA-PDD.}
Predictor-Driven Diffusion (PDD; Section~\ref{M-subsec:pdd}) and Score-Augmented Predictor-Driven Diffusion (SA-PDD; Section~\ref{M-subsec:sa-pdd}) use the same U-Net.
In PDD, the network ends with a single pointwise linear layer (i.e., a $1 \times 1$ convolution) that outputs $\hat f_\lambda^\theta(U_\lambda^{(n)})\,\Delta t$.
In SA-PDD, this final layer is duplicated: two pointwise linear layers are applied to the same hidden features, one outputting $\hat f_\lambda^\theta(U_\lambda^{(n)})\,\Delta t$ and the other the noise estimate $\hat\epsilon_\lambda^\theta(U_\lambda^{(n)})$.
The number of parameters therefore differs from PDD only by the additional output layer.
The same network is used for the $40 \times 40$ and $128 \times 128$ Kolmogorov flows.

\paragraph{FNO.}
As an alternative architecture, we also employ a Fourier neural operator (FNO) \citep{Li+21}, which we use only with PDD.
The FNO applies spectral convolutions in Fourier space with mode truncation, followed by pointwise nonlinearities.
We use 8 Fourier layers for the Lorenz-96 model and 6 layers for the Kolmogorov flow, with a hidden width of 64 in both cases.
The number of retained Fourier modes is 56 (out of 64 Nyquist modes) for the Lorenz-96 model and 16 (out of 20 Nyquist modes) for the Kolmogorov flow, for which only the $40 \times 40$ resolution is considered.
The Swish activation and FiLM conditioning for $\lambda$ are the same as in the U-Net.
The input and output formats are also identical.
Unlike the U-Net, the FNO uses neither convolutions nor self-attention, providing a test of whether our proposed method works across different network architectures.
Results with the FNO-based PDD are presented in \Cref{subsec:experimental-results-with-fno}.

\paragraph{Temporal boundary handling.}
Each sample starts at $t^{(0)}$, so the history window $U_\lambda^{(n)}$ is not fully available for the first $m - 1$ time steps.
The missing past states $u_\lambda^{(-1)}, \ldots, u_\lambda^{(-m+1)}$ are filled by zero padding.
This padding is used in training, generation, and super-resolution, for PDD, SA-PDD, and DDPM alike.
In simulation, the history window at the initial time is taken from the reference data (\Cref{alg:prediction}), so no padding is needed.
PDD additionally overwrites the initial time step during reverse-$\lambda$ integration by linear extrapolation in $t$; this treatment concerns the initial density $r_\lambda$ rather than the padding and is described in \Cref{subsec:detailed-implementation}.

\subsection{Hyperparameters}
\label{subsec:hyperparameters}

\paragraph{Scope of tuning.}
For hyperparameter tuning, we use 100 samples with distinct initial conditions as the validation set for both the Lorenz-96 model and the Kolmogorov flow, separate from both the training and test sets, with the Kolmogorov flow validation set prepared only at $40 \times 40$.
The $128 \times 128$ Kolmogorov flow is run with SA-PDD only (Section~\ref{M-subsec:experimental-setup}), and all values given below, including $\alpha$, $\beta$, $m$, $w$, the network, and the optimization, are the same as at $40 \times 40$; no tuning was performed for the higher resolution.

\begin{figure}[tb!]
    \begin{center}
        \includegraphics[width=\textwidth]{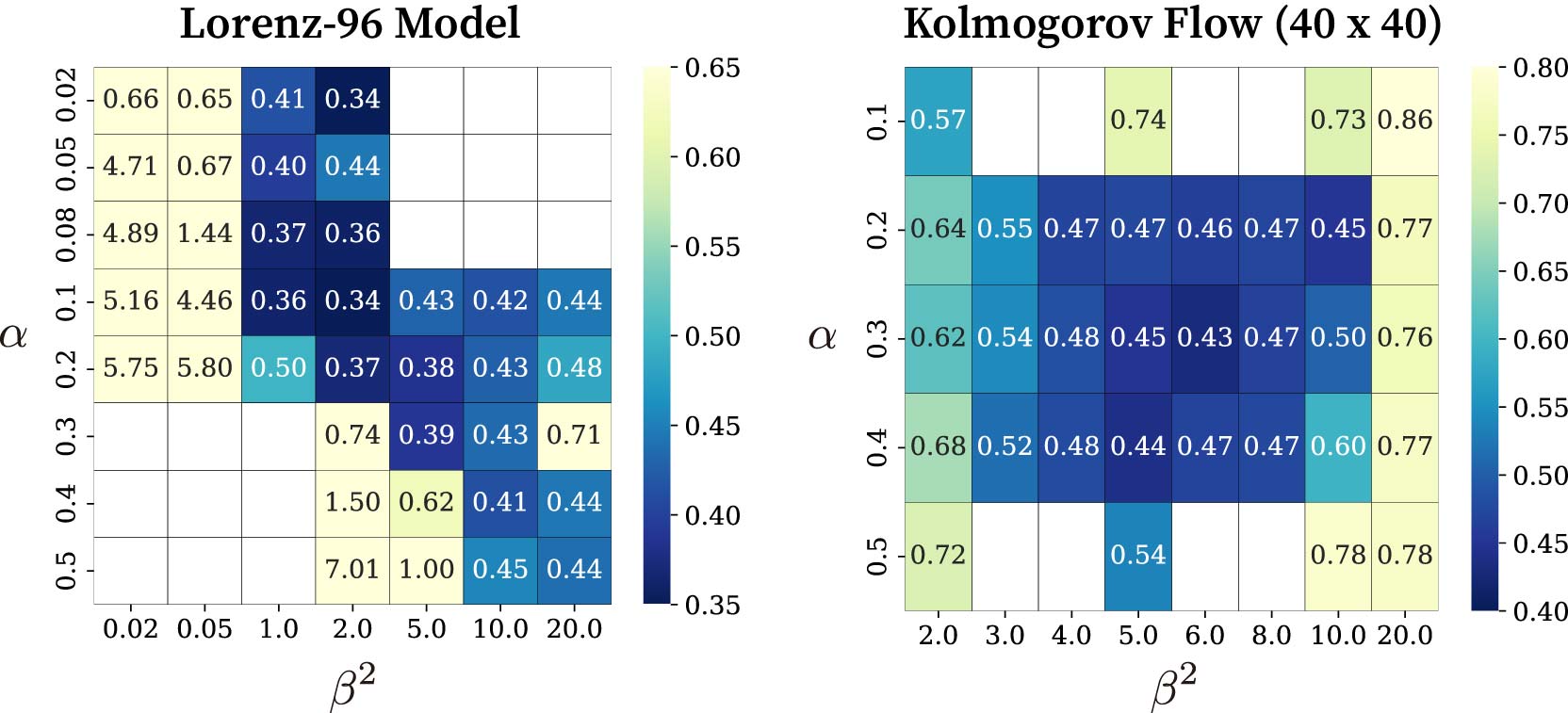} 
        \caption{
        Grid search over the coarse-graining parameters $(\alpha, \beta^2)$ for the Lorenz-96 model (left) and the $40 \times 40$ Kolmogorov flow (right).
        Each cell gives the spectral error in Eq.~(\ref{eq:supp-spec-err-definition}) between unconditionally generated samples and the validation set, for the damping $\alpha$ (rows) and the noise variance $\beta^2$ (columns).
        The horizontal axis is $\beta^2$, not $\beta$.
        Each cell is a single training run of PDD, separate from the five runs used for the reported results.
        Blank cells are settings that were not run, since the grid was refined around the minimum.
        The color scale saturates at both ends, so the annotated numbers give the values outside its range.
        The adopted values are $(\alpha, \beta^2) = (0.1, 2)$ for the Lorenz-96 model and $(0.3, 6)$ for the Kolmogorov flow, i.e., $\beta = \sqrt{2}$ and $\sqrt{6}$, respectively.
        }
        \label{fig:supp_L96_KF_grid_search}
    \end{center}
\end{figure}

\paragraph{Coarse-graining.}
The diffusion scale is taken as $\lambda \in [0,1]$ ($\lambda_{\min}=0$ and $\lambda_{\max}=1$), and a discretization with 1,000 steps is used (i.e., $\Delta \lambda = 10^{-3}$).
For the Lorenz-96 model, we set $\alpha = 0.1$ and $\beta = \sqrt{2}$, while for the Kolmogorov flow we set $\alpha = 0.3$ and $\beta = \sqrt{6}$.
These values are selected by a grid search over $(\alpha, \beta^2)$ with PDD, and SA-PDD reuses them without further tuning.
We perform unconditional generation for each pair $(\alpha, \beta^2)$ and adopt the pair that minimizes the spectral error in Eq.~(\ref{eq:supp-spec-err-definition}) against the validation set (Fig.~\ref{fig:supp_L96_KF_grid_search}).
The grid was refined around the minimum, so the blank cells are settings that were not run.
Near the minimum, the error surface is broad and shallow.
For the Lorenz-96 model, the error stays within 0.34--0.50 at every explored pair with $\alpha \le 0.2$ and $\beta^2 \ge 1$.
For the Kolmogorov flow, it stays within 0.43--0.48 for $\alpha \in [0.2, 0.4]$ and $\beta^2 \in [4, 8]$.
The error tends to grow steeply where $\beta^2$ is small relative to $\alpha$.
The results are therefore not overly sensitive to the exact values.
The two selected pairs share the ratio $\beta^2/(2\alpha) = 10$, even though $\alpha$ and $\beta^2$ were varied independently in the search.
By Eq.~(\ref{M-eq:coarse-graining-covariance-in-fourier}), the variance injected into the mode at wavenumber $k$ approaches $10/|k|^2$ once $2\alpha|k|^2\lambda \gg 1$, so both settings reach the same noise level relative to the unit-variance data.

\paragraph{Loss weight of SA-PDD.}
The weight $w$ of the PDD term in the SA-PDD loss~(\ref{M-eq:loss-sa-pdd}) is fixed to $0.1$ in all experiments (Section~\ref{M-subsec:sa-pdd}).
This value makes the two terms of the SA-PDD loss of roughly the same order.
As noted in the main article, the results are not highly sensitive to $w$.

\paragraph{Optimization.}
We train all models using Adam \citep{Kingma+15} with learning rate $2\times 10^{-4}$ and batch size 50 for the Lorenz-96 model and 40 for the Kolmogorov flow.
We train for $3\times 10^4$ optimization iterations (no epoch-based schedule and no early stopping).
We use the default Adam hyperparameters, with no weight decay and no gradient clipping.
We use automatic mixed precision (AMP) with FP16.
The DDPMs in \Cref{subsec:ddpm-models} use the same optimization settings.

\FloatBarrier

\subsection{DDPMs used for comparison}
\label{subsec:ddpm-models}

For comparison at $\lambda = 0$, we train denoising diffusion probabilistic models (DDPMs) using the denoising score matching objective \citep{Ho+20,Vincent11}.
We use the same neural network architecture as described in \Cref{subsec:neural-network-architecture} and the same training hyperparameters as in \Cref{subsec:hyperparameters}.
The forward diffusion process is variance-preserving \citep{Ho+20,Song+21} with $\alpha=0$ and $\gamma=3$, i.e., $\partial_\lambda u_\lambda(x,t) = -\gamma u_\lambda(x,t) + \sqrt{2\gamma}\,\eta_\lambda(x,t)$, and we discretize $\lambda \in [0,1]$ into 1,000 steps.
As in PDD and SA-PDD, we do not use a $\lambda$-dependent noise schedule.
Input time series are concatenated along the channel dimension, and we use the same zero padding at temporal boundaries as in \Cref{subsec:neural-network-architecture}.
Two DDPMs are trained separately, one for simulation and the other for unconditional generation; both are referred to as DDPM in the main article (Section~\ref{M-subsec:experimental-setup}).
Since these models use $\alpha=0$, they have no spatial coarse-graining hierarchy and are undefined at $\lambda>0$; we therefore compare with them only at $\lambda=0$.

\paragraph{DDPM for simulation.}
For simulation, we follow the conditional diffusion setup used in GenCast \citep{Price+25}.
The denoiser estimates the score of the standardized residual, the difference between the states at $t^{(n)}$ and $t^{(n+1)}$, conditioned on the five most recent clean states $(u_0^{(n)}, u_0^{(n-1)}, \ldots, u_0^{(n-4)})$.
At inference, we draw one residual per step and roll forward in $t$ by sliding the conditioning window.

\paragraph{DDPM for unconditional generation.}
For unconditional generation, we model the joint distribution over full spatiotemporal fields and draw samples via DDPM-style reverse-$\lambda$ integration with 1,000 steps \citep{Ho+22b}, using the same reverse-$\lambda$ discretization as PDD and SA-PDD (\Cref{subsec:detailed-implementation}).
To evaluate the score at each time step $t^{(n)}$, the denoiser takes a local temporal window $(u_\lambda^{(n)}, u_\lambda^{(n-1)}, \ldots, u_\lambda^{(n-4)})$ from the noisy sample, with zero padding where $n < 4$.
Unlike the DDPM for simulation, no clean conditioning context is used---all five inputs are noisy.
During reverse-$\lambda$ integration, the score is evaluated at every $t^{(n)}$ and the full spatiotemporal field is updated simultaneously.

\subsection{Evaluation metrics}
\label{subsec:evaluation-metrics}

We use the relative $L^2$ error for the short-term accuracy of simulation and the relative spectral error for long-term statistical consistency \citep{Jiang+23}, defined as:
\begin{align}
    \text{Relative } L^2 \text{ error}(t) &:= \frac{\| u(t) - \hat{u}(t)\|_2}{\| u(t) \|_2},  \label{eq:supp-l2-err-definition} \\
    \text{Relative spectral error} &:= \frac{\| \; |\mathcal{G}[u]|^2 - |\mathcal{G}[\hat{u}]|^2 \; \|_1}{\| \; |\mathcal{G}[u]|^2 \; \|_1}, \label{eq:supp-spec-err-definition}
\end{align}
where $u$ and $\hat{u}$ denote the reference from the physics-based simulation and the model output, respectively, $\|\cdot\|_2$ is taken over the spatial grid and $\|\cdot\|_1$ over all spatiotemporal modes, and $\mathcal{G}$ denotes the Fourier transform over $x$ and $t$.
Unless otherwise stated, we compare $u$ and $\hat{u}$ at the same diffusion scale $\lambda$.
The only exception in the main article is the low-resolution (LR) column in Table~\ref{M-table:main_text_error_super_resolution}, which compares the super-resolution input (i.e., an LR simulation at $\lambda=0.2$) against the $\lambda=0$ physics-based reference.
A large LR error is therefore a statement about the difference between the two resolutions, not about the accuracy of the coarse-grained simulation, which is evaluated at equal $\lambda$ (Section~\ref{M-subsec:experimental-results} and \Cref{subsec:details-of-experimental-results-simulation}).
We omit the term ``relative'' [in Eqs.~(\ref{eq:supp-l2-err-definition}) and~(\ref{eq:supp-spec-err-definition})] below and in the main article.

Both metrics are computed after standardizing $u$ and $\hat{u}$ with the same normalization parameters.
The $L^2$ error is evaluated at each $t$ and averaged over test samples; in chaotic systems, small differences between trajectories grow rapidly, so the main article reports the $L^2$ error at the sixth time step.
For the spectral error, we average the spatiotemporal power spectral densities (PSDs) over test samples and compute the relative difference.
To estimate statistical uncertainty, we train the neural network five times with different random initializations and compute standard deviations across these runs.

For the $L^2$ error, simulations use the same initial condition as the reference, namely the history window at time steps $-(m-1)$ to $0$.
For the spectral error, we run a long rollout, discard the first 10 segments (each spanning the temporal length of one sample), and compute PSDs from the 11th segment; discarding 100 segments instead of 10 yields similar spectral errors and confirms long-rollout stability.

For $\lambda > 0$, the reference $u_\lambda$ is constructed by applying $\mathcal{C}_\lambda$ to $u_0$.
Since this reference excludes noise, we set $\sigma_\lambda = 0$ in Eq.~(\ref{M-eq:governing-eq-along-t}) for evaluation, so that the rollout follows the most probable path (Section~\ref{M-subsec:experimental-setup}).
Although simulations with $\sigma_\lambda \ne 0$ are stable (\Cref{subsec:simulation-with-noise}), we omit noise to allow clear comparison with the reference.
The input of the super-resolution task is the exception: it is simulated with $\sigma_\lambda > 0$ so that the resulting path is an approximate sample of $q_\lambda(\{u_\lambda\}_t)$ (Section~\ref{M-subsec:experimental-results}).

\subsection{Detailed implementation}
\label{subsec:detailed-implementation}

This subsection provides implementation details for training (\Cref{alg:training}), forward simulation (\Cref{alg:prediction}), and generation/super-resolution (\Cref{alg:generation}), including numerical techniques for stable integration.
PDD and SA-PDD share the three algorithms; steps that apply to only one of them are marked (PDD) or (SA-PDD).

For the discretized stochastic equations in Algorithms~\ref{alg:prediction} and~\ref{alg:generation}, we adopt the white Gaussian noise convention \citep[e.g.,][]{Shiraishi23} of the Appendix: the noise variables $\xi^{(n)}$ and $\eta_\lambda$ are sampled from $\mathcal{N}(0, \frac{1}{\Delta x^d \Delta t}I_D)$ and $\mathcal{N}(0, \frac{1}{\Delta\lambda}I_{D'})$, respectively, so that the increments $\xi^{(n)}\,\Delta t$ and $\eta_\lambda\,\Delta\lambda$ have covariance matrices $(\Delta t/\Delta x^d) I_D$ and $\Delta\lambda\, I_{D'}$ (Appendices~\ref{M-subsec:coarse-graining-operator-and-covariance-matrix} and~\ref{M-subsec:path-probability-measure-via-the-path-integral-representation}).
Here $D = N_x \times N_c$ is the dimension of the state at one time step and $D' = N_x \times (N_t + 1) \times N_c$ that of a path.
Our source code sets the grid-cell volume to $\Delta x^d = 1$, which loses no generality: with $\sigma_\lambda$ given by Eq.~(\ref{M-eq:sigma-trace-matching}), the grid-cell volume $\Delta x^d$ cancels wherever it appears.
The prefactor of the loss~(\ref{M-eq:loss-pdd}) becomes $\Delta x^d/(2\sigma_\lambda^2\Delta t) = D/(2\,\mathrm{tr}\,\Sigma_\lambda)$, and the covariance of the noise increment $\sigma_\lambda \xi^{(n)} \Delta t$ in Eq.~(\ref{M-eq:governing-eq-along-t}) becomes $(\sigma_\lambda^2 \Delta t/\Delta x^d) I_D = (\mathrm{tr}\,\Sigma_\lambda/D)\, I_D$, both independent of $\Delta x$.

The Laplacian is discretized by second-order finite differences on the periodic domain $[0,2\pi)$ with $N$ grid points per direction, so that $|k|^2$ in Appendix~\ref{M-subsec:coarse-graining-operator-and-covariance-matrix} and Eq.~(\ref{eq:exponential-integrator}) is replaced by the eigenvalue $\sum_{i=1}^{d}(N/\pi)^2\sin^2(\pi k_i/N)$, which agrees with $|k|^2$ at low wavenumbers.

In \Cref{alg:training}, the superscript $(b,n)$ denotes time step $n$ of sample $b$, and the history window $U_{\lambda}^{(b,n)}$ is built from the coarse-grained states of the same sample with zero padding for $n < m - 1$ (\Cref{subsec:neural-network-architecture}).
The noise $\epsilon^{(b,n)}$ sampled in line 3 is the target of the denoising term of SA-PDD in line 8.

\begin{algorithm}[tb]
    \caption{Training of PDD and SA-PDD}
    \label{alg:training}
    \textbf{Input:} training dataset of sample paths $\{u_0\}_t$; functions $\mathcal{C}_\lambda$, $\Sigma_\lambda$, $\sigma_\lambda$; neural network $\hat f_\lambda^\theta$ (PDD) or $(\hat f_\lambda^\theta, \hat\epsilon_\lambda^\theta)$ (SA-PDD) with parameters $\theta$; weight $w$ (SA-PDD) \\
    \textbf{Output:} trained parameters $\theta^*$, i.e., $\hat f_\lambda^*$ (and $\hat\epsilon_\lambda^*$)
    \begin{algorithmic}[1]
        \REPEAT
        \STATE Sample a minibatch of spatiotemporal paths $\{\{u_0\}_t^{(b)}\}_{b=1}^{B}$ from the training dataset
        \STATE Sample $\lambda^{(b)} \sim \mathcal{U}(0,1)$ for each $b$, and $\epsilon^{(b,n)} \sim \mathcal{N}(0,I_{D})$ for each $b$ and $n = 0, \ldots, N_t$
        \STATE Compute $\mathcal{C}_{\lambda^{(b)}}$, $\Sigma_{\lambda^{(b)}}$, and $\sigma_{\lambda^{(b)}}$ for each $b$
        \STATE $u_{\lambda^{(b)}}^{(b,n)} \leftarrow \mathcal{C}_{\lambda^{(b)}} u_0^{(b,n)} + \sqrt{\Sigma_{\lambda^{(b)}}}\,\epsilon^{(b,n)}$ for each $b$ and $n$ {\footnotesize in Eq.~(\ref{M-eq:ou-solution-spatiotemporal})}
        \STATE $\mathcal{L}^{\mathrm{PDD}} \leftarrow \frac{1}{B}\sum_{b=1}^{B}\frac{\Delta x^d}{2\sigma_{\lambda^{(b)}}^2 \Delta t}\sum_{n=0}^{N_t-1}\left\| u_{\lambda^{(b)}}^{(b,n+1)} - u_{\lambda^{(b)}}^{(b,n)} - \hat f_{\lambda^{(b)}}^\theta(U_{\lambda^{(b)}}^{(b,n)})\,\Delta t\right\|^2$ {\footnotesize in Eq.~(\ref{M-eq:loss-pdd})}
        \STATE \textbf{(PDD)} $\mathcal{L} \leftarrow \mathcal{L}^{\mathrm{PDD}}$
        \STATE \textbf{(SA-PDD)} $\mathcal{L} \leftarrow \frac{1}{B}\sum_{b=1}^{B}\frac{1}{2}\sum_{n=0}^{N_t}\left\| \hat\epsilon_{\lambda^{(b)}}^\theta(U_{\lambda^{(b)}}^{(b,n)}) - \epsilon^{(b,n)}\right\|^2 + w\,\mathcal{L}^{\mathrm{PDD}}$ {\footnotesize in Eq.~(\ref{M-eq:loss-sa-pdd})}
        \STATE Update $\theta$ using $\nabla_\theta \mathcal{L}$
        \UNTIL{convergence}
    \end{algorithmic}
\end{algorithm}

\Cref{alg:prediction} is identical for PDD and SA-PDD, since simulation uses only $\hat f_\lambda^*$.
To start a simulation at a given $\lambda$, we take the history window $U_0^{(0)}$ of the fine-resolution reference and apply $\mathcal{C}_\lambda$ to each of its states, adding $\sqrt{\Sigma_\lambda}\,\epsilon$ as in Eq.~(\ref{M-eq:ou-solution-spatiotemporal}) when $\sigma_\lambda > 0$; the resulting $U_\lambda^{(0)}$ is the initial condition.
The deterministic setting $\sigma_\lambda = 0$ is used for the evaluation in the main article, and $\sigma_\lambda > 0$ for the input of super-resolution (\Cref{subsec:evaluation-metrics}) and in \Cref{subsec:simulation-with-noise}.

\begin{algorithm}[tb]
    \caption{Forward simulation in $t$ at fixed $\lambda$}
    \label{alg:prediction}
    \textbf{Input:} fine-resolution initial condition, the history window $U_0^{(0)} = (u_0^{(0)}, u_0^{(-1)}, \ldots, u_0^{(-m+1)})$; diffusion scale $\lambda$; functions $\mathcal{C}_\lambda$, $\Sigma_\lambda$, $\sigma_\lambda$; trained network $\hat f_\lambda^*$; time step $\Delta t$; number of increments $N_t$ \\
    \textbf{Output:} simulated path $\{u_\lambda\}_t = (u_\lambda^{(0)}, \ldots, u_\lambda^{(N_t)})$, where $t^{(0)} = 0$ and $t^{(N_t)}=N_t \Delta t$
    \begin{algorithmic}[1]
        \STATE Set $\sigma_\lambda \leftarrow 0$ for deterministic simulation, otherwise compute $\sigma_\lambda$ for the given $\lambda$
        \STATE Initialize $u_\lambda^{(n)} \leftarrow \mathcal{C}_\lambda u_0^{(n)}$ for $n = -m+1, \ldots, 0$, adding $\sqrt{\Sigma_\lambda}\,\epsilon^{(n)}$ with $\epsilon^{(n)} \sim \mathcal{N}(0, I_D)$ if $\sigma_\lambda > 0$
        \FOR{$n=0$ {\bfseries to} $N_t - 1$}
        \STATE Sample $\xi^{(n)} \sim \mathcal{N}(0, \frac{1}{\Delta x^d \Delta t}I_D)$ if $\sigma_\lambda > 0$; otherwise set $\xi^{(n)} \leftarrow 0$
        \STATE $u_\lambda^{(n+1)} \leftarrow u_\lambda^{(n)} + \hat f_\lambda^*(U_\lambda^{(n)})\,\Delta t + \sigma_\lambda\,\xi^{(n)} \Delta t$ {\footnotesize in Eq.~(\ref{M-eq:governing-eq-along-t})}
        \ENDFOR
    \end{algorithmic}
\end{algorithm}

For generation and super-resolution (\Cref{alg:generation}), we adopt predictor--corrector sampling \citep{Song+21}.
Each iteration consists of a predictor step that integrates the reverse-$\lambda$ equation~(\ref{M-eq:reverse-rg-sde}), followed by Langevin corrector steps that refine the sample toward the target distribution.
Both steps require the score at every time step of the path, written $s_\lambda = (s_\lambda^{(0)}, \ldots, s_\lambda^{(N_t)})$, and the two realizations differ only in how it is evaluated (the score evaluation at the top of \Cref{alg:generation}).
In PDD, $s_\lambda^{(n)}$ is the path score $\nabla_{u_\lambda^{(n)}} \ln p_\lambda(\{u_\lambda\}_t)$ computed by automatic differentiation of the log-density~(\ref{M-eq:history-factorized-path-density}), in which $\hat f_\lambda^*$ replaces $f_\lambda^\theta$ (Section~\ref{M-subsec:pdd}).
In SA-PDD, $s_\lambda^{(n)}$ is $\hat s_\lambda^*(U_\lambda^{(n)}) = -\Sigma_\lambda^{-1/2}\hat\epsilon_\lambda^*(U_\lambda^{(n)})$ in Eq.~(\ref{M-eq:sa-pdd-score}), one forward computation without automatic differentiation (Section~\ref{M-subsec:sa-pdd}).
Following \citep[Algorithm 4]{Song+21}, each corrector step draws $z \sim \mathcal{N}(0,I_{D'})$, recomputes the score $s_\lambda$, defines $\chi := 2\left(\kappa\,\|z\|_2/\|s_\lambda\|_2\right)^2$ ($\kappa$: signal-to-noise ratio), and updates
\begin{equation}
    u_\lambda' = u_\lambda + \chi s_\lambda + \sqrt{2\chi}\, z,
    \label{eq:langevin-corrector}
\end{equation}
where $u_\lambda$ here denotes the whole path $\{u_\lambda\}_t$ regarded as a single vector of dimension $D'$.
We use three corrector steps per predictor step; in our experiments, results were not sensitive to this choice (one or two steps yielded similar performance).
We select the signal-to-noise ratio parameter $\kappa$ on the validation set by grid search over $[0,2]$ with step size 0.05 to minimize the spectral error in Eq.~(\ref{eq:supp-spec-err-definition}).
The search is run separately for PDD and for DDPM, and $\kappa$ is then set to the average of the two best values: $\kappa=0.7$ for the Lorenz-96 model and $\kappa=0.3$ for the Kolmogorov flow.
SA-PDD takes no part in this tuning, and all three models use these values in the test phase, so that the sampler is not tuned in favor of any one of them.

The predictor step employs an exponential time differencing (ETD) scheme \citep{Hochbruck+Ostermann10,Lord+Tambue19}, a standard approach for stiff equations.
In reverse-$\lambda$ integration, the Laplacian coefficient $\alpha|k|^2$ varies by orders of magnitude across Fourier modes, requiring small step sizes for high-wavenumber modes under explicit methods.
Let $\widetilde{u}_\lambda(k,t)=\mathcal{F}[u_\lambda(x,t)](k)$, $\widetilde{s}_\lambda(k,t)=\mathcal{F}[s_\lambda(x,t)](k)$, and $\widetilde{\eta}_\lambda(k,t)=\mathcal{F}[\eta_\lambda(x,t)](k)$ denote spatial Fourier transforms at each fixed $t$ (Appendix~\ref{M-subsec:coarse-graining-operator-and-covariance-matrix}), where our convention is $\mathcal{F}[\nabla_x^2 u_\lambda](k,t)=-|k|^2\widetilde{u}_\lambda(k,t)$.
The update proceeds as
\begin{align}
    \widetilde{u}_{\lambda-\Delta\lambda}(k,t) = e^{\alpha|k|^2\Delta\lambda}\widetilde{u}_\lambda(k,t) + \varphi(\alpha|k|^2\Delta\lambda)\,\beta^2 \widetilde{s}_\lambda(k,t)\Delta\lambda + \beta\,\widetilde{\eta}_\lambda(k,t)\Delta\lambda, \label{eq:exponential-integrator}
\end{align}
where $\varphi(z) = (e^z - 1)/z$ with $\varphi(0)=1$ by continuity.
Since integration proceeds backward in $\lambda$, the Laplacian term $\alpha \nabla_x^2 u_\lambda$ acts as anti-diffusion; the exponential factor $e^{\alpha|k|^2\Delta\lambda} > 1$ assigns a larger amplification rate to high-wavenumber modes, counteracting the scale-dependent damping in the forward coarse-graining.
The update is performed in Fourier space and the updated field $\widetilde{u}_{\lambda-\Delta\lambda}$ is then transformed back to real space.
Inspired by Lord and Tambue \citep{Lord+Tambue19}, we treat the deterministic and stochastic terms separately, with the $\varphi$-function applied only to the deterministic component (i.e., the score term).
For the noise term, we use the Euler--Maruyama scheme.
For the DDPM, we apply the same ETD scheme with $\alpha|k|^2$ replaced by a constant $\gamma$ (\Cref{subsec:ddpm-models}), which corresponds to scale-independent amplification.

The path density in Eq.~(\ref{M-eq:path-probability-density}) contains the initial density $r_\lambda$.
This $r_\lambda$ is not needed during training, where it drops out of the KL divergence as a constant in Eq.~(\ref{M-eq:kl-divergence}), nor during forward-$t$ simulation, which requires only the drift.
However, generation with PDD via Eq.~(\ref{M-eq:reverse-rg-sde}) requires the path score $\nabla_{u_\lambda^{(n)}} \ln p_\lambda(\{u_\lambda\}_t)$, and at $n = 0$ this score includes the boundary term $\nabla \ln r_\lambda(U_\lambda^{(0)})$ [Eq.~(\ref{M-eq:history-factorized-path-density})].
We do not model this term explicitly; instead, in each predictor step, we overwrite the noise-free part of the update at $t^{(0)}$ by linear extrapolation from $t^{(1)}$ and $t^{(2)}$ before the noise is added, and apply the same extrapolation within the Langevin corrector steps.
This extrapolation supplies the mean of the update at the initial time step and stabilizes reverse-$\lambda$ sampling near the temporal boundary.
Even when the drift depends on the past four time steps (\Cref{subsec:neural-network-architecture}), overwriting only $t^{(0)}$ was sufficient in our experiments.
SA-PDD needs no such treatment: its score follows from the noise estimate at each time step, and $r_\lambda$ does not appear (Section~\ref{M-subsec:sa-pdd}); the same holds for the DDPM, whose denoising objective defines a target at every time step.
Explicitly modeling $r_\lambda$ (or its score) is one direction for future work, particularly for statistically unsteady systems (Section~\ref{M-subsec:extensions}).

\begin{algorithm}[tb]
    \caption{Generation / super-resolution (reverse-$\lambda$ integration) with PDD and SA-PDD}
    \label{alg:generation}
    \textbf{Input:} $\lambda_{\max}>\lambda_{\min}$; $\alpha,\beta,\kappa$; function $\Sigma_\lambda$; trained network $\hat f_\lambda^*$ (PDD) or $(\hat f_\lambda^*, \hat\epsilon_\lambda^*)$ (SA-PDD); step size $\Delta\lambda>0$; coarse-grained path at $\lambda_{\max}$, $\left\{u_{\lambda_{\max}}^{\rm LR}\right\}_{t}$ (for super-resolution) \\
    \textbf{Output:} generated / super-resolved path $\left\{u_{\lambda_{\min}}\right\}_{t}$ \\
    \textbf{Score evaluation} $s_\lambda \leftarrow \mathrm{Score}(\{u_\lambda\}_t)$, for $n = 0, \ldots, N_t$: \\
    \hspace*{1em}\textbf{(PDD)} $s_\lambda^{(n)} \leftarrow \nabla_{u_\lambda^{(n)}} \ln p_\lambda(\{u_\lambda\}_t)$ by automatic differentiation {\footnotesize of Eq.~(\ref{M-eq:history-factorized-path-density}) with $\hat f_\lambda^*$} \\
    \hspace*{1em}\textbf{(SA-PDD)} $s_\lambda^{(n)} \leftarrow -\Sigma_\lambda^{-1/2}\,\hat\epsilon_\lambda^*(U_\lambda^{(n)})$ {\footnotesize in Eq.~(\ref{M-eq:sa-pdd-score})}
    \begin{algorithmic}[1]
        \STATE \textbf{(Generation)} Sample $u_{\lambda_{\max}}^{(n)} {\sim} \mathcal{N}(0,\Sigma_{\lambda_{\max}})$ for $n=0,\ldots,N_t$
        \STATE \textbf{(Super-resolution)} Initialize $\left\{u_{\lambda_{\max}}\right\}_{t} \leftarrow \left\{u_{\lambda_{\max}}^{\rm LR}\right\}_{t}$
        \FOR{$\lambda=\lambda_{\max},\,\lambda_{\max}-\Delta\lambda,\,\dots,\,\lambda_{\min}+\Delta\lambda$}
        \STATE $s_\lambda \leftarrow \mathrm{Score}(\{u_\lambda\}_t)$
        \STATE Sample $\eta_\lambda \sim \mathcal{N}(0, \frac{1}{\Delta\lambda}I_{D'})$
        \STATE \textbf{Predictor:} Update $u_{\lambda-\Delta\lambda}$ via ETD in Eq.~(\ref{eq:exponential-integrator}); \textbf{(PDD)} extrapolate at $t^{(0)}$
        \STATE \textbf{Corrector:} (with $\kappa$ as specified)
        \FOR{$j=1$ {\bfseries to} $3$}
        \STATE $s_{\lambda-\Delta\lambda} \leftarrow \mathrm{Score}(\{u_{\lambda-\Delta\lambda}\}_t)$
        \STATE Sample $z \sim \mathcal{N}(0, I_{D'})$
        \STATE Update $u_{\lambda-\Delta\lambda}$ via Langevin step in Eq.~(\ref{eq:langevin-corrector}); \textbf{(PDD)} extrapolate at $t^{(0)}$
        \ENDFOR
        \ENDFOR
    \end{algorithmic}
\end{algorithm}

\FloatBarrier

\subsection{Computational cost}
\label{subsec:computational-cost}

Each training iteration samples $\lambda$ and computes a loss on local time differences at that $\lambda$, without rollout in $t$ (\Cref{alg:training}); the per-iteration cost is comparable (up to constants) to DDPM-style training when using the same backbone and similar batching strategies \citep{Ho+20}.
Simulation at a fixed $\lambda$ is inexpensive for both PDD and SA-PDD: the trained drift $\hat f_\lambda^*$ is evaluated once per physical-time step to advance the state, with no backpropagation required (\Cref{alg:prediction}).
Generation and super-resolution are more costly, and here PDD and SA-PDD differ (\Cref{alg:generation}).

In PDD, a spatiotemporal path $\{u_\lambda\}_t$ is regarded as a single tensor, analogous to a multi-channel image, and the score is obtained as the input gradient of the scalar log-density with respect to this tensor.
This is a vector--Jacobian product computed by reverse-mode automatic differentiation \citep{Baydin+18}; we do \emph{not} backpropagate through a physical-time integrator that repeatedly applies $\hat f_\lambda^*$ to produce a rollout.
Each score evaluation is therefore a forward computation of $\hat f_\lambda^*$ over all time steps followed by a backward computation, and the activations of the whole path must be kept until the backward computation finishes, so the memory use grows with the temporal length $N_t$ and the spatial resolution.
Standard checkpointing strategies can trade additional computation for reduced activation memory \citep{Chen+16}, but we did not use them.

In SA-PDD, the score is a network output rather than a derivative of one.
A score evaluation is therefore a single forward computation, as in a DDPM.
Neither a backward computation nor automatic differentiation is involved, and no activations need to be stored.

\Cref{tab:generation-cost} reports the measured cost of unconditional generation, the task with the largest cost, since the reverse-$\lambda$ integration runs all 1,000 steps over the whole path (super-resolution is the same sampler over a shorter $\lambda$ interval).
With three corrector steps per predictor step, each reverse-$\lambda$ step evaluates the score four times.
PDD takes about 2.9 times the computation time of SA-PDD per reverse-$\lambda$ step on the Lorenz-96 model and about 4.7 times on the $40 \times 40$ Kolmogorov flow, and uses about 8.9 and 8.6 times the GPU memory, respectively.
These ratios are measured with our implementation and hardware.
The $128 \times 128$ Kolmogorov flow was run with SA-PDD only (Section~\ref{M-subsec:experimental-setup}), and its cost was not measured against PDD.

\begin{table}[tb]
    \caption{
    Measured cost of unconditional generation with PDD and SA-PDD for the Lorenz-96 model (L96) and the $40 \times 40$ Kolmogorov flow (KF).
    Time is the wall-clock time per reverse-$\lambda$ step, including one predictor and three corrector steps (i.e., four score evaluations), measured as the median of three timed runs after one warm-up run; a full generation of 1,000 steps takes 1,000 times this value to within 3\%.
    Memory is the GPU memory allocated during generation (maximum over the run).
    Both are measured on one NVIDIA L40S GPU with a batch of 10 paths in float32.
    Ratio: PDD divided by SA-PDD.
    }
    \label{tab:generation-cost}
    \centering
    \small
    \begin{tabular}{lrrrrrr}
        \toprule
                          & \multicolumn{3}{c}{Time per step (ms)} & \multicolumn{3}{c}{Memory (MiB)} \\
                          & PDD   & SA-PDD & Ratio & PDD  & SA-PDD & Ratio \\
        \midrule
        L96               & 186.2 & 65.0   & 2.9   & 1,923 & 217    & 8.9 \\
        KF $40 \times 40$ & 988.3 & 209.7  & 4.7   & 7,466 & 870    & 8.6 \\
        \bottomrule
    \end{tabular}
\end{table}

Overall, the sampling cost of both PDD and SA-PDD scales approximately linearly with the number of reverse-$\lambda$ steps and with the temporal length $N_t$ (since the tensor size grows with $N_t$).
We have not optimized the sampler for wall-clock performance, so there is room for improvement.
For instance, reducing the number of reverse-$\lambda$ steps via sampler or noise scheduling is a direct way to lower cost \citep{Karras+22,Masuki+Ashida25}; we leave such optimization for future work.

\FloatBarrier
\section{Details of experimental results}
\label{supp:experimental-results}

This section shows the details of the experimental results from main article Section~\ref{M-sec:experiments}: the Lorenz-96 results, for which the main article gives only the tables, and the generated samples for the Lorenz-96 model and the Kolmogorov flow.
In all figures, all quantities are non-dimensionalized.

\subsection{Simulation}
\label{subsec:details-of-experimental-results-simulation}

\begin{figure}[tb!]
    \begin{center}
        \includegraphics[width=\textwidth]{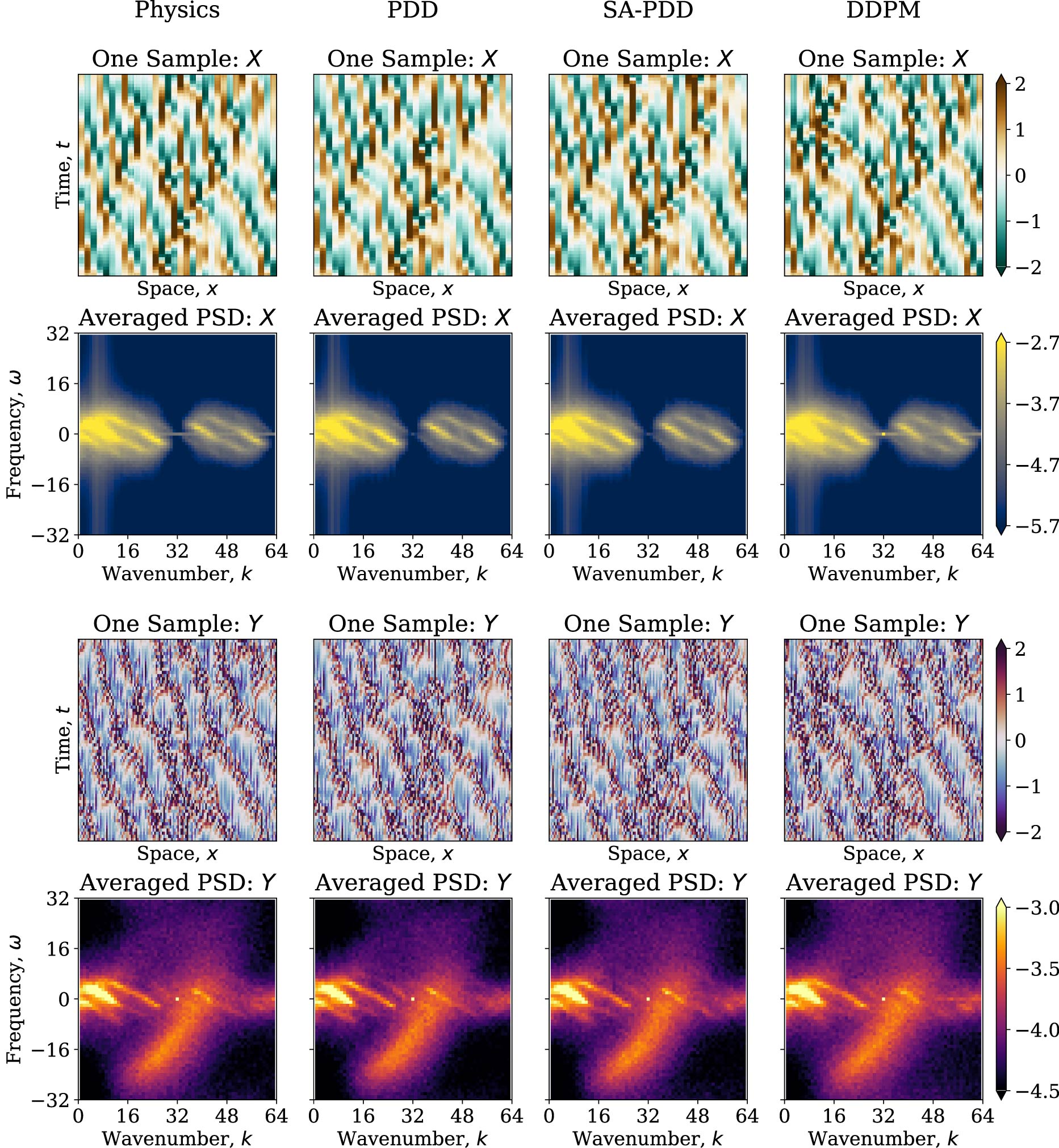} 
        \caption{
        Simulation of the Lorenz-96 model at fine resolution ($\lambda = 0$).
        Columns: physics-based simulation (Physics), PDD, SA-PDD, and DDPM, all started from the same initial condition, the history window at time steps $-4$ to $0$.
        Rows: one sample of the slow variable $X$ (space $x$ horizontal, physical time $t$ vertical), its spatiotemporal PSD as a function of wavenumber $k$ and frequency $\omega$, and the same two rows for the fast variable $Y$.
        PSDs are averaged over the 100 samples, and $\log_{10}$ is applied to the average.
        }
        \label{fig:supp_L96_simulation_lambda0}
    \end{center}
\end{figure}

\begin{figure}[tb!]
    \begin{center}
        \includegraphics[width=0.8\textwidth]{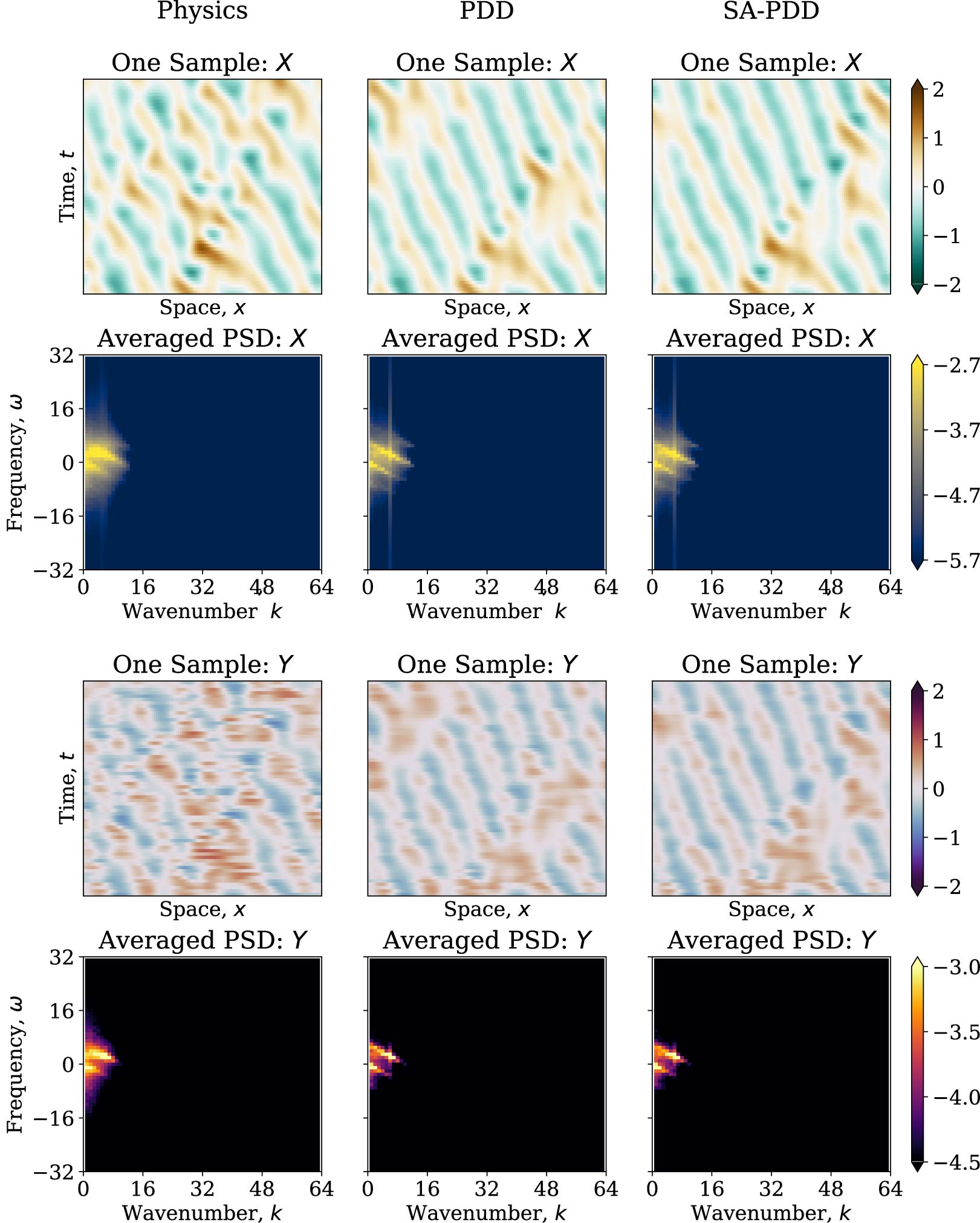} 
        \caption{
        Simulation of the Lorenz-96 model at coarse-grained resolution ($\lambda = 0.2$), in the same layout as Fig.~\ref{fig:supp_L96_simulation_lambda0}.
        Columns: coarse-grained physics-based simulation $\mathcal{C}_\lambda u_0$ (Physics), PDD, and SA-PDD.
        Here, DDPM is undefined at $\lambda > 0$ (\Cref{subsec:ddpm-models}).
        Coarse-graining removes small-scale structure while the large-scale dynamics are preserved.
        PSDs are averaged over the 100 samples, and $\log_{10}$ is applied to the average.
        }
        \label{fig:supp_L96_simulation_lambda02}
    \end{center}
\end{figure}

Figures~\ref{fig:supp_L96_simulation_lambda0} and~\ref{fig:supp_L96_simulation_lambda02} show simulation results for the Lorenz-96 model at diffusion scales $\lambda = 0$ and $0.2$, respectively, the counterparts of main article Figs.~\ref{M-fig:kf040_simulation_lambda0} and~\ref{M-fig:kf040_simulation_lambda02} for the Kolmogorov flow.
At $\lambda = 0$, we compare with the DDPM trained for simulation (\Cref{subsec:ddpm-models}).
PDD, SA-PDD, and DDPM all reproduce the spatiotemporal patterns of the physics-based simulation for both the slow variable $X$ and the fast variable $Y$, and their PSDs agree with the reference.
At $\lambda = 0.2$, the small-scale components, in particular those of $Y$, are removed by coarse-graining, and PDD and SA-PDD reproduce the remaining large-scale patterns.
These results support the quantitative evaluation in main article Table~\ref{M-table:main_text_error_simulation}.

\begin{figure}[tb!]
    \begin{center}
        \includegraphics[width=0.9\textwidth]{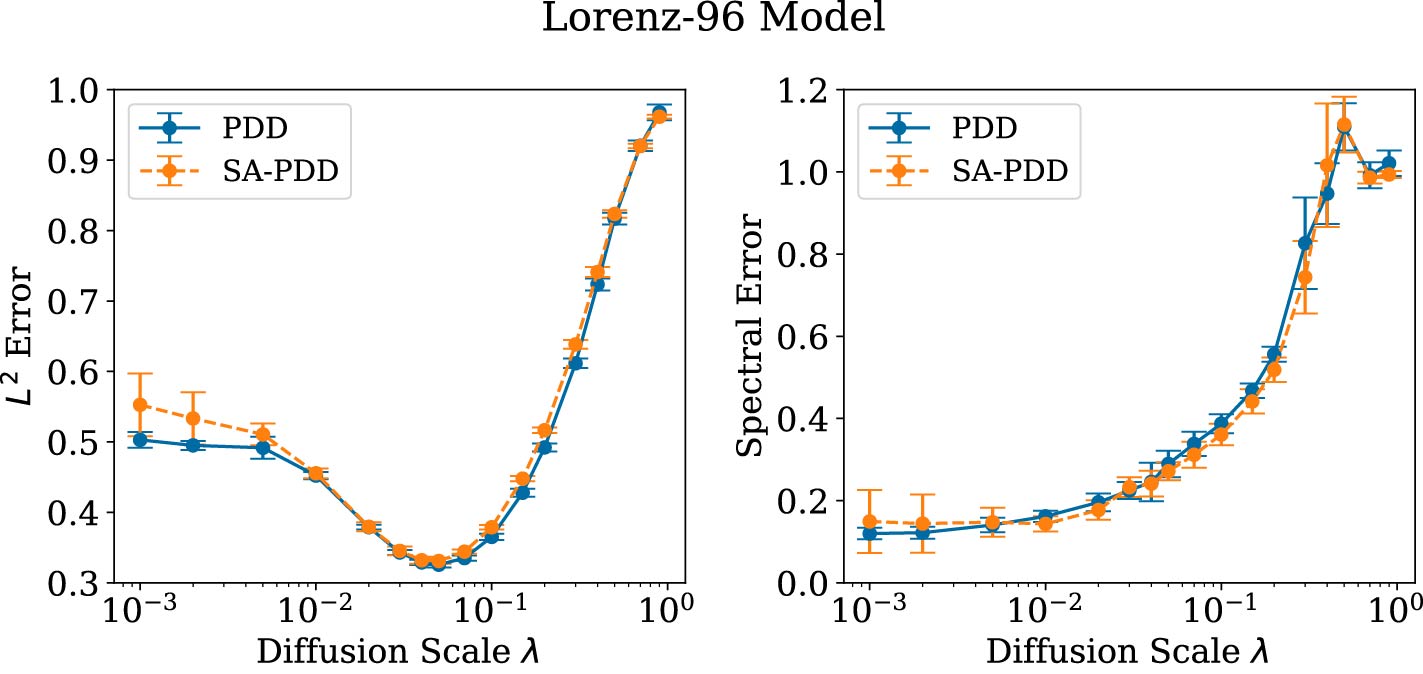} 
        \caption{
        Dependence of the simulation errors on the diffusion scale $\lambda$ for the Lorenz-96 model.
        (Left) $L^2$ error in Eq.~(\ref{eq:supp-l2-err-definition}) at the sixth time step; (right) spectral error in Eq.~(\ref{eq:supp-spec-err-definition}).
        Blue solid lines: PDD; orange dashed lines: SA-PDD.
        The reference at each $\lambda$ is the coarse-grained physics-based simulation $\mathcal{C}_\lambda u_0$ (\Cref{subsec:evaluation-metrics}).
        To use a log scale, a small offset of $10^{-3}$ is added to $\lambda$ (e.g., $\lambda = 0$ is displayed as $\lambda = 10^{-3}$).
        Error bars indicate the standard deviation over five training runs.
        }
        \label{fig:supp_L96_error_vs_lambda}
    \end{center}
\end{figure}

Figure~\ref{fig:supp_L96_error_vs_lambda} shows the $\lambda$ dependence of the simulation errors for the Lorenz-96 model, the counterpart of main article Fig.~\ref{M-fig:kf040_error_vs_lambda} for the Kolmogorov flow.
The $L^2$ error depends non-monotonically on $\lambda$, with a minimum near $\lambda \approx 0.05$.
The initial decrease occurs because coarse-graining preferentially removes the small-scale components of the fast variable $Y$, which are hard to predict.
The steep rise of the spectral error for $\lambda \gtrsim 0.2$ is likely a normalization effect, as in the Kolmogorov flow (Section~\ref{M-subsec:experimental-results}).
Although the spectral error reaches approximately 0.6 at $\lambda=0.2$, it can be reduced to approximately 0.3 by increasing the number of input time steps (\Cref{subsec:dependence-on-the-number-of-input-time-steps}).
PDD and SA-PDD give nearly the same errors over the whole range, except that SA-PDD has a slightly larger $L^2$ error for $\lambda \lesssim 5 \times 10^{-3}$, consistent with Table~\ref{M-table:main_text_error_simulation}.

\FloatBarrier

\subsection{Unconditional generation}
\label{subsec:details-of-experimental-results-spatiotemporal-data-generation}

\begin{figure}[tb!]
    \begin{center}
        \includegraphics[width=\textwidth]{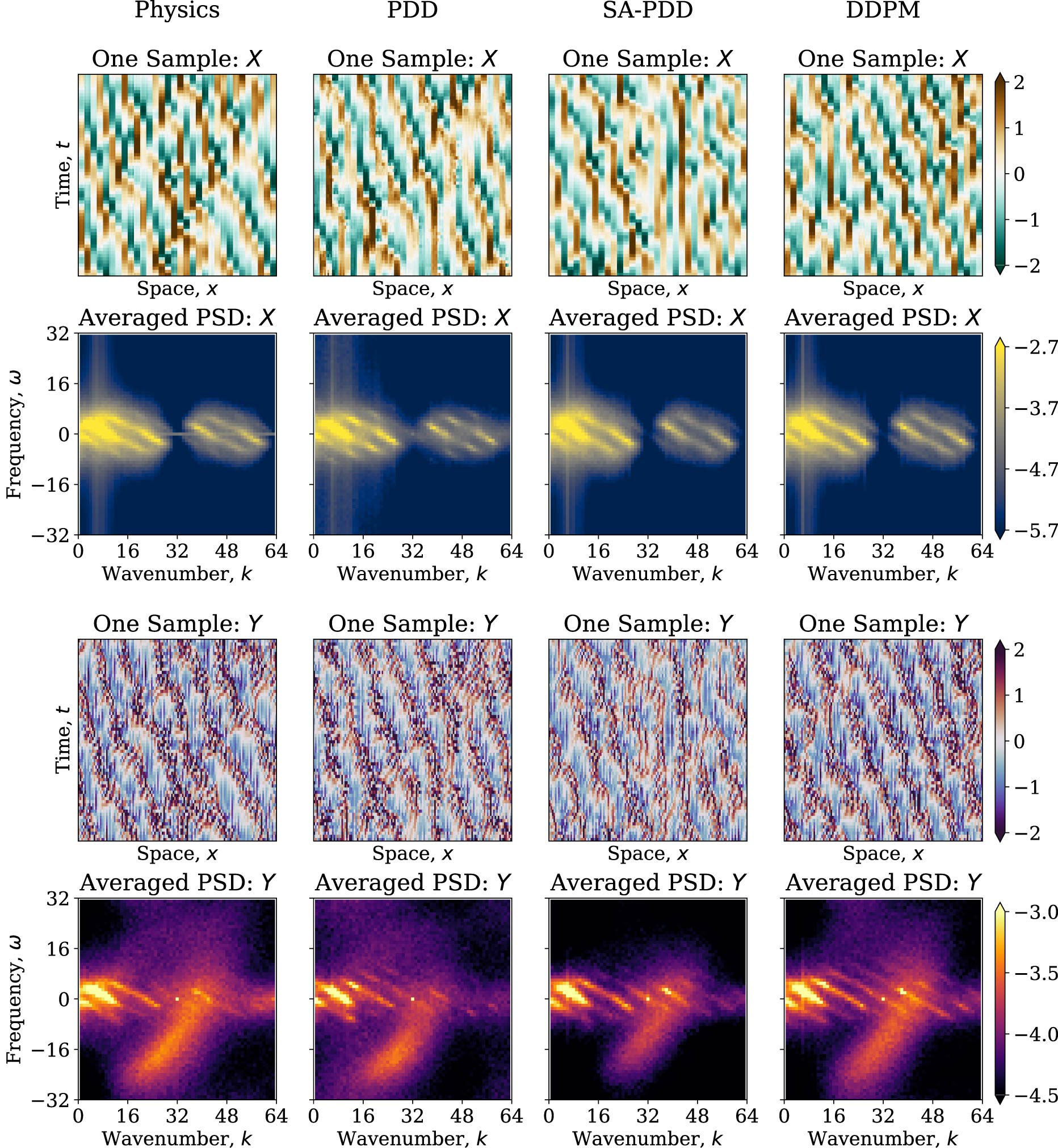} 
        \caption{Unconditional generation for the Lorenz-96 model, in the same layout as Fig.~\ref{fig:supp_L96_simulation_lambda0}.
        Columns: physics-based simulation (Physics), PDD, SA-PDD, and DDPM.
        Rows: one sample of the slow variable $X$, its spatiotemporal PSD, and the same two rows for the fast variable $Y$.
        PSDs are averaged over the 100 samples, and $\log_{10}$ is applied to the average.
        }
        \label{fig:supp_L96_generation}
    \end{center}
\end{figure}

\begin{figure}[tb!]
    \begin{center}
        \includegraphics[width=\textwidth]{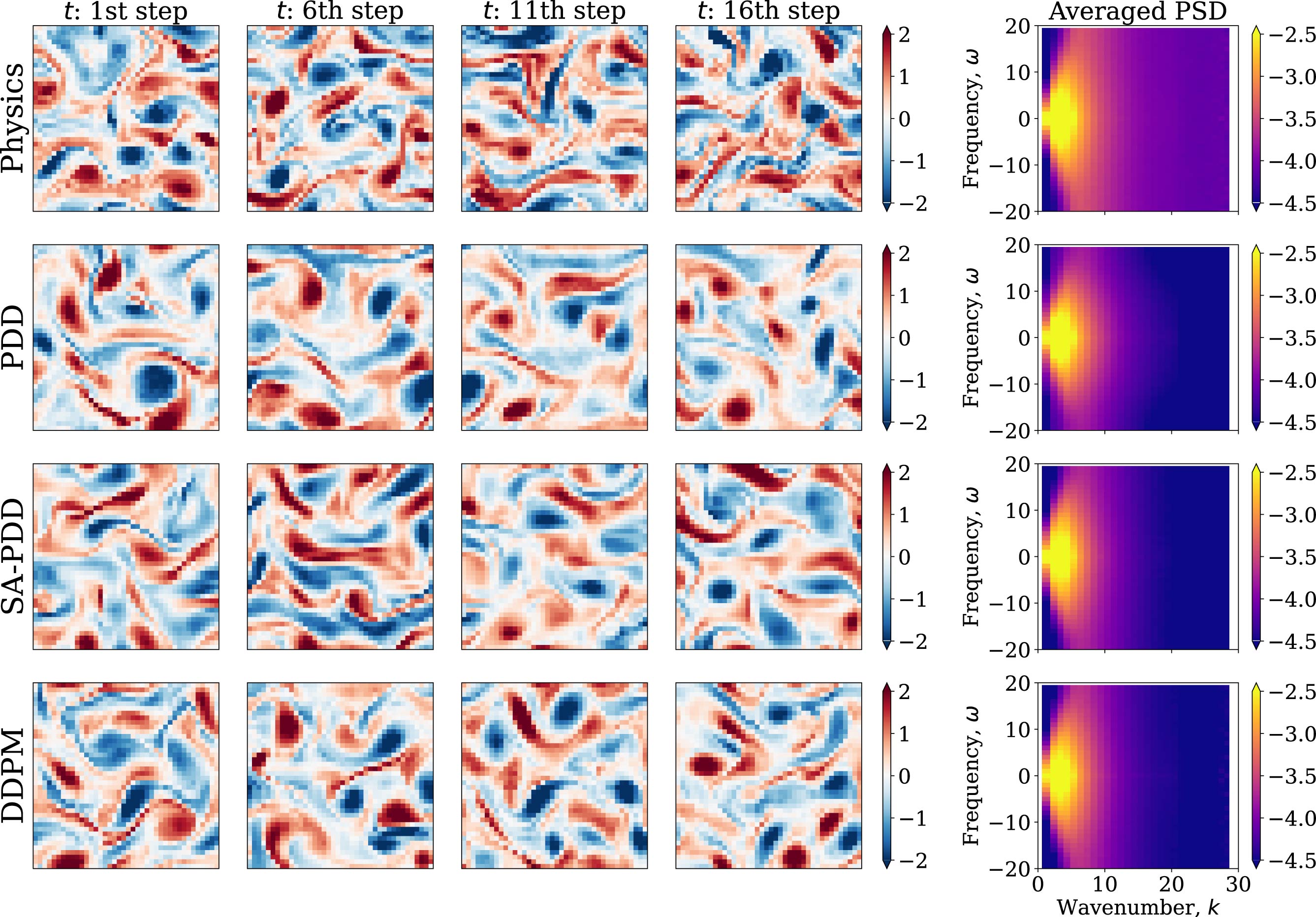} 
        \caption{Unconditional generation for the $40 \times 40$ Kolmogorov flow (vorticity).
        Rows: physics-based simulation (Physics), PDD, SA-PDD, and DDPM.
        Left: vorticity at the 1st, 6th, 11th, and 16th time steps of one sample; right: spatiotemporal PSD.
        PSDs are averaged over the 100 samples, and $\log_{10}$ is applied to the average.
        }
        \label{fig:supp_KF040_generation}
    \end{center}
\end{figure}

\begin{figure}[tb!]
    \begin{center}
        \includegraphics[width=\textwidth]{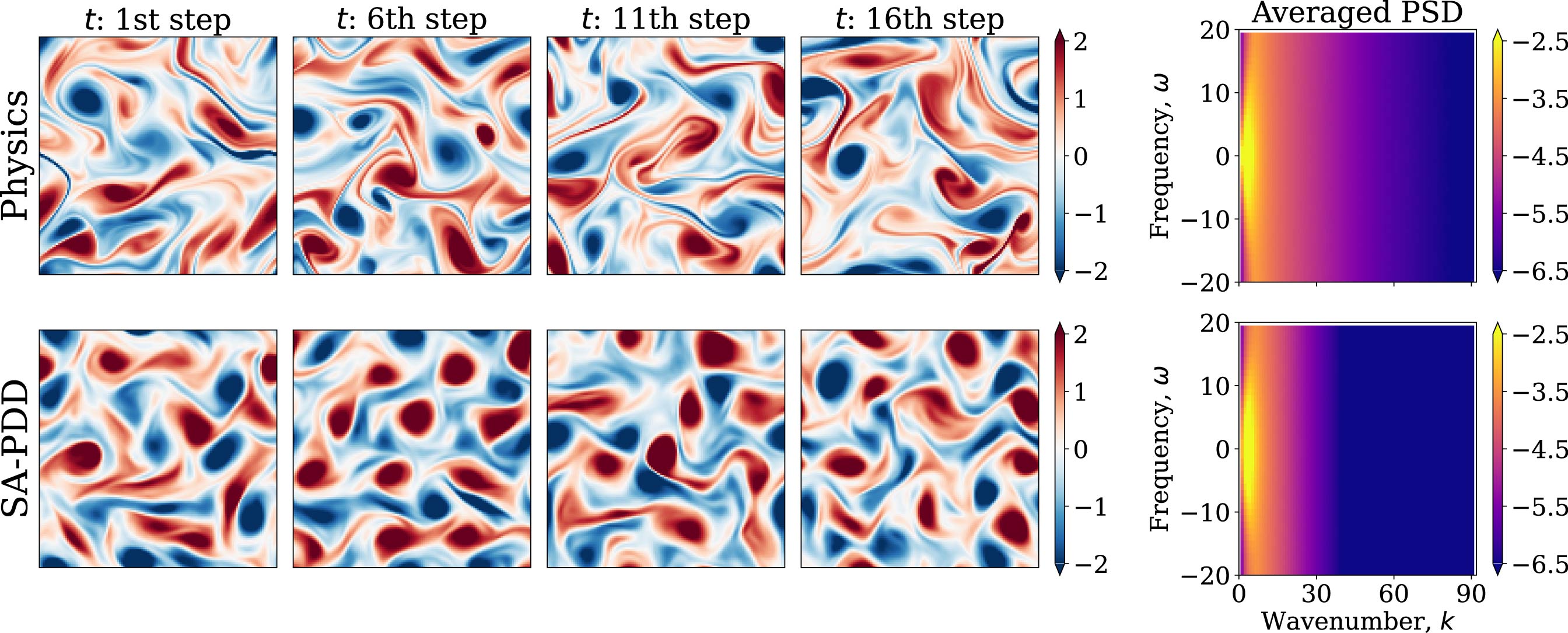} 
        \caption{Unconditional generation for the $128 \times 128$ Kolmogorov flow (vorticity) with SA-PDD, in the same layout as Fig.~\ref{fig:supp_KF040_generation}.
        Rows: physics-based simulation (Physics) and SA-PDD.
        All settings, including $\alpha$ and $\beta$, are the same as at $40 \times 40$.
        PSDs are averaged over the 100 samples, and $\log_{10}$ is applied to the average.
        }
        \label{fig:supp_KF128_generation}
    \end{center}
\end{figure}

Figures~\ref{fig:supp_L96_generation}, \ref{fig:supp_KF040_generation}, and~\ref{fig:supp_KF128_generation} show unconditional generation results for the Lorenz-96 model, the $40 \times 40$ Kolmogorov flow, and the $128 \times 128$ Kolmogorov flow, respectively.
For the first two, we compare with the DDPM trained for unconditional generation (\Cref{subsec:ddpm-models}); the $128 \times 128$ Kolmogorov flow is run with SA-PDD only.
PDD, SA-PDD, and DDPM generate samples that visually resemble those from physics-based simulations, and the spatiotemporal PSDs are well reproduced.
At $128 \times 128$, SA-PDD underestimates the spectral power at wavenumbers $k \gtrsim 30$, as in the simulation at this resolution (main article Fig.~\ref{M-fig:kf128_simulation}).
These results support the quantitative evaluation in main article Table~\ref{M-table:main_text_error_generation}.

\FloatBarrier

\subsection{Super-resolution}
\label{subsec:details-of-experimental-results-super-resolution}

\begin{figure}[tb!]
    \begin{center}
        \includegraphics[width=\textwidth]{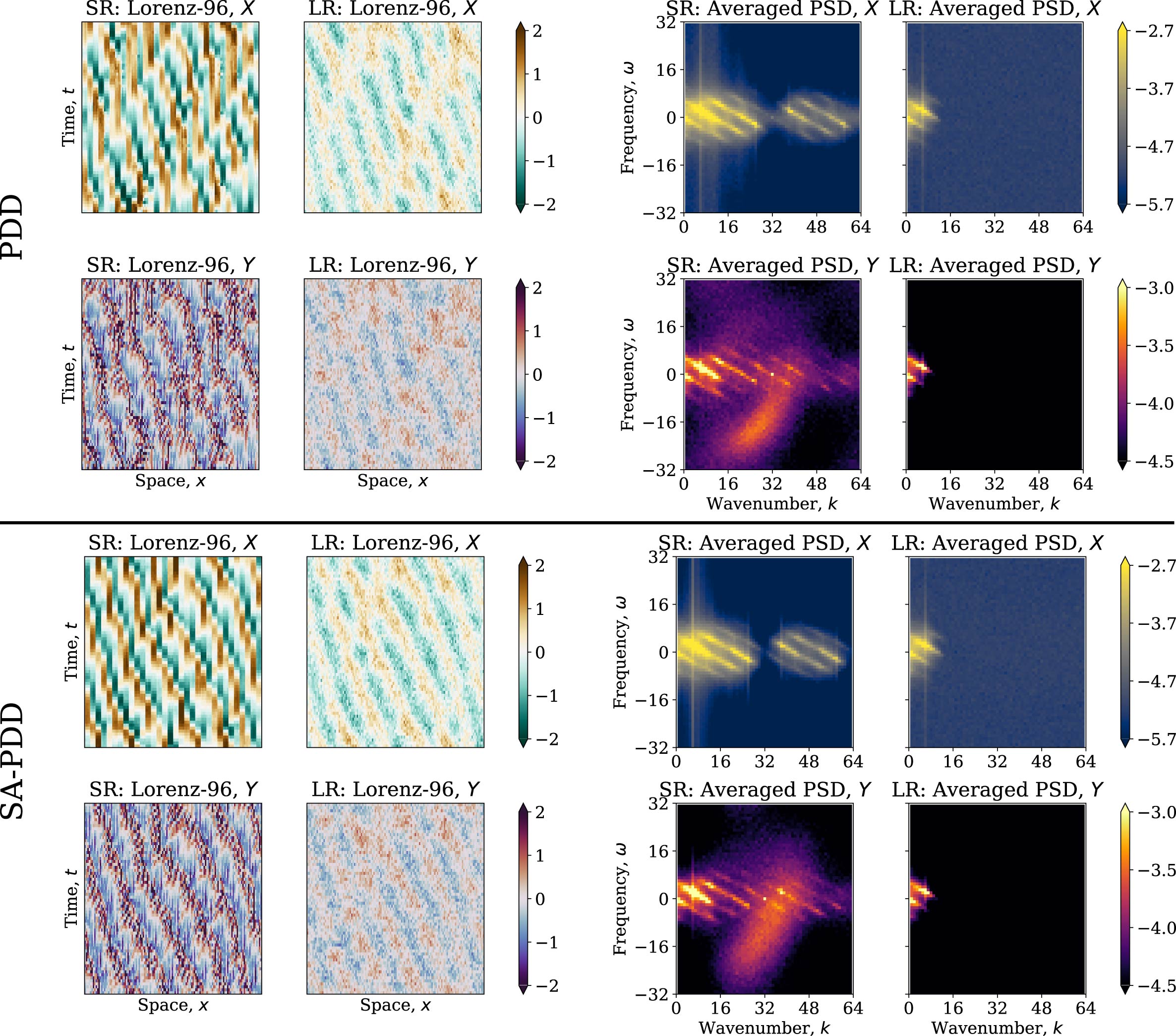} 
        \caption{
        Super-resolution of the Lorenz-96 model with PDD (top block) and SA-PDD (bottom block).
        In each block: super-resolved output at $\lambda = 0$ (SR) and low-resolution input, a spatiotemporal sample simulated at $\lambda = 0.2$ with $\sigma_\lambda > 0$ (LR), for the slow variable $X$ (top) and the fast variable $Y$ (bottom).
        Left: one sample; right: spatiotemporal PSD.
        PSDs are averaged over the 100 samples, and $\log_{10}$ is applied to the average.
        The reverse-$\lambda$ integration restores the small-scale structure absent from the input.
        }
        \label{fig:supp_L96_super_resolution}
    \end{center}
\end{figure}

Figure~\ref{fig:supp_L96_super_resolution} shows super-resolution results for the Lorenz-96 model with PDD and SA-PDD, the counterpart of main article Figs.~\ref{M-fig:kf040_super_resolution} and~\ref{M-fig:kf128_super_resolution} for the Kolmogorov flow.
As in the main article, the input is a spatiotemporal sample simulated at $\lambda = 0.2$ with $\sigma_\lambda > 0$, and the reverse-$\lambda$ integration is performed down to $\lambda = 0$ (\Cref{alg:generation}).
The super-resolved output restores the small-scale components of both $X$ and $Y$ absent from the input, and its PSD agrees with that of the physics-based simulation in Fig.~\ref{fig:supp_L96_simulation_lambda0}.
PDD and SA-PDD give visually similar outputs, consistent with their comparable super-resolution (SR) errors in main article Table~\ref{M-table:main_text_error_super_resolution}.

\FloatBarrier

The evaluation here (and in the main article) uses the spectral error and the PSD, which are insensitive to phase and do not separate a super-resolved sample from an unconditional sample with the same spectrum.
The input used in that evaluation is a spatiotemporal sample simulated at $\lambda = 0.2$, for which no physics-based reference exists.
As a control experiment, where the reference exists, we repeated super-resolution with the input built from the test data,
\begin{equation}
    u_\lambda^{\rm LR} = \mathcal{C}_\lambda u_0 + \sqrt{\Sigma_\lambda}\,\epsilon,
\end{equation}
at $\lambda = 0.2$ [restated from main article Eq.~(\ref{M-eq:ou-solution-spatiotemporal})].
There are now three fields: the truth $u_0$, the input $u_\lambda^{\rm LR}$, and the super-resolved output $u_0^{\rm SR}$.

We first confirm that the input $u_\lambda^{\rm LR}$ is identified from the output $u_0^{\rm SR}$.
For each $u_0^{\rm SR}$, we compute the distance $\| u_\lambda^{\rm LR} - \mathcal{C}_\lambda u_0^{\rm SR} \|$, the root mean square over all channels and spatiotemporal grid points, to the inputs of all 100 test samples (\Cref{subsec:construction-of-training-data}).
The distance is smallest at the correct input for every sample in every setting (i.e., 500 of 500 over the five training runs).
However, identification alone does not say how closely the output agrees with its input.

This closeness is measured by the misfit between the input $u_\lambda^{\rm LR}$ and the coarse-grained output $\mathcal{C}_\lambda u_0^{\rm SR}$,
\begin{equation}
\label{eq:input-misfit-ratio}
    R := \frac{\| u_\lambda^{\rm LR} - \mathcal{C}_\lambda u_0^{\rm SR} \|}{\| \sqrt{\Sigma_\lambda}\,\epsilon \|},
\end{equation}
where $\| \cdot \|$ denotes the root mean square over all test samples, channels, and spatiotemporal grid points, and the denominator is the coarse-graining noise in the input.
If the output $u_0^{\rm SR}$ were an exact sample of the posterior $q_\lambda(\{u_0\}_t \mid \{u_\lambda\}_t)$, the pair $(u_0^{\rm SR}, u_\lambda^{\rm LR})$ would follow the same joint distribution as $(u_0, u_\lambda^{\rm LR})$ and $R = 1$ would hold up to sampling error.
For comparison, we also compute $R_{\rm shuffle}$, the same ratio with the outputs shuffled across the samples, so that each input is paired with the output of another sample.

Table~\ref{tab:sr-consistency} reports $R$ with the shuffled control $R_{\rm shuffle}$ and the normalized distance $D$ between them.
$R$ is between 1.05 and 1.52, against 2.06 to 5.01 for $R_{\rm shuffle}$, so $D$ stays below 0.36.
SA-PDD is closer to an exact sample than PDD for both systems (i.e., L96 and KF).
The values indicate that the output uses its own input, with the coarse-grained output $\mathcal{C}_\lambda u_0^{\rm SR}$ slightly biased or over-dispersed relative to the posterior.
The results show that the super-resolved output shares the large-scale structure of its own low-resolution input, unlike an unconditional sample with the same spectrum.

\begin{table}[tb]
    \caption{
    Consistency of the super-resolved output with its low-resolution input for the Lorenz-96 model (L96) and the Kolmogorov flow (KF), in the control experiment where the input is built from the test data.
    $R$: the input misfit in Eq.~(\ref{eq:input-misfit-ratio}), which is 1 up to sampling error for an exact sample of the posterior.
    $R_{\rm shuffle}$: the same ratio with the outputs shuffled across the samples.
    $D := (R - 1)/(R_{\rm shuffle} - 1)$: 0 for an exact sample of the posterior and 1 for an output independent of its own input.
    Values of $R$, $R_{\rm shuffle}$, and $D$ are means $\pm$ standard deviations over five training runs with 100 samples each.
    }
    \label{tab:sr-consistency}
    \centering
    \small
    \begin{tabular}{llccc}
        \toprule
                            & Method  & $R$ & $R_{\rm shuffle}$ & $D$ \\
        \midrule
        L96                 & PDD     & 1.517 ($\pm$ 0.017) & 2.455 ($\pm$ 0.014) & 0.355 ($\pm$ 0.009)\\
        L96                 & SA-PDD  & 1.148 ($\pm$ 0.001) & 2.497 ($\pm$ 0.008) & 0.098 ($\pm$ 0.001)\\
        KF $40 \times 40$   & PDD     & 1.190 ($\pm$ 0.018) & 2.061 ($\pm$ 0.014) & 0.179 ($\pm$ 0.016)\\
        KF $40 \times 40$   & SA-PDD  & 1.055 ($\pm$ 0.002) & 2.072 ($\pm$ 0.007) & 0.051 ($\pm$ 0.002)\\
        KF $128 \times 128$ & SA-PDD  & 1.048 ($\pm$ 0.010) & 5.009 ($\pm$ 0.055) & 0.012 ($\pm$ 0.002)\\
        \bottomrule
    \end{tabular}
\end{table}

\FloatBarrier
\section{Additional experimental results}
\label{supp:additional-experimental-results}

This section reports additional experiments to further analyze the proposed framework.
All results in this section are for the Lorenz-96 model with PDD; qualitatively similar results were obtained with PDD for the $40 \times 40$ Kolmogorov flow.
In the figures, the columns labeled Physics and PDD show the physics-based simulation and the output of the trained predictor $\hat f_\lambda^*$, respectively, and all quantities are non-dimensionalized.

\subsection{Simulation with noise}
\label{subsec:simulation-with-noise}

\begin{figure}[tb!]
    \begin{center}
        \includegraphics[width=\textwidth]{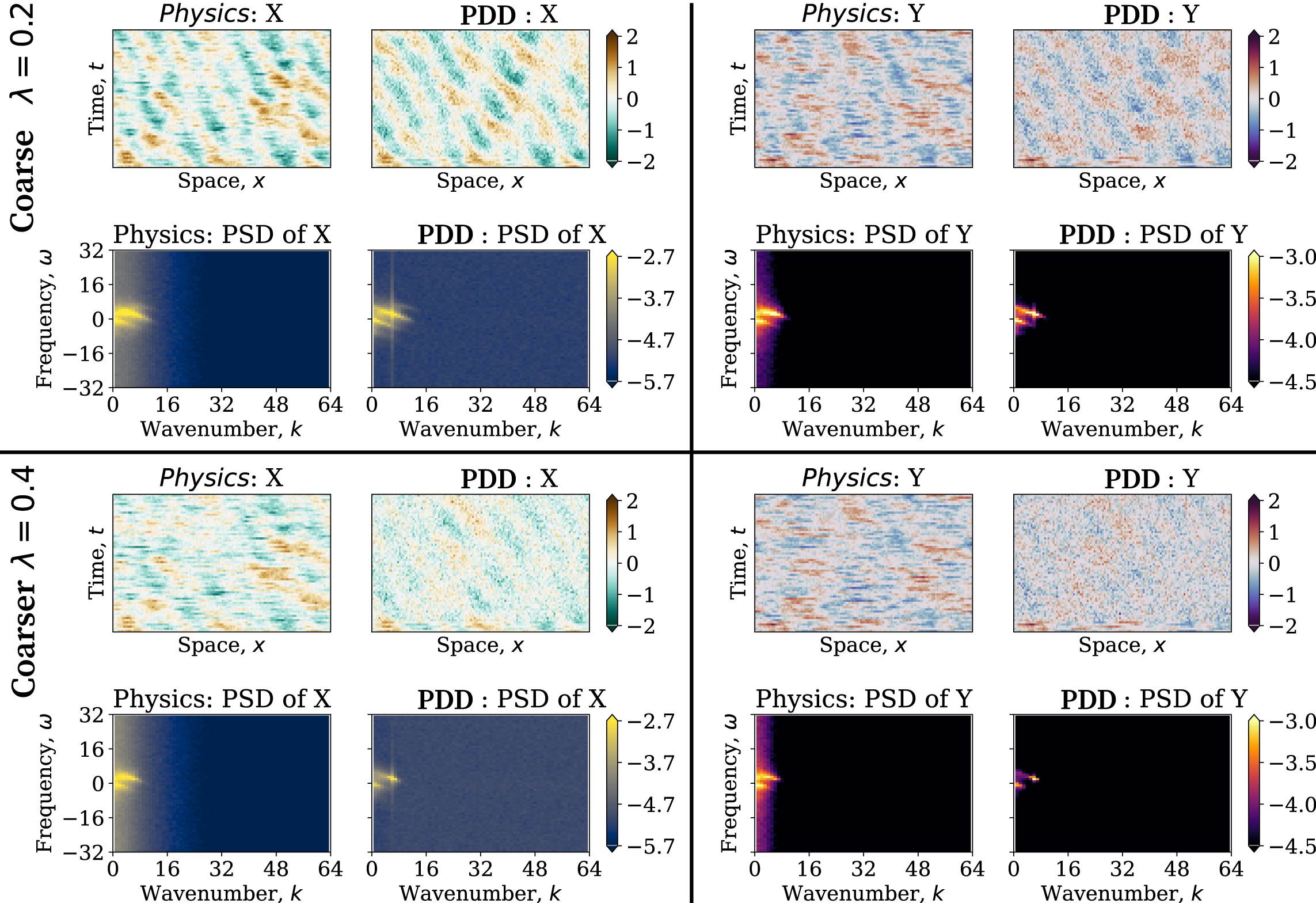} 
        \caption{
        Simulation results at $\lambda=0.2$ and 0.4 for the Lorenz-96 model, where noise is injected during simulations (i.e., $\sigma_\lambda > 0$ in main article Eq.~(\ref{M-eq:governing-eq-along-t})).
        Left: slow variable $X$; right: fast variable $Y$.
        In each half, the physics-based reference (Physics) and PDD are compared, with one sample above and the spatiotemporal PSD below.
        PSDs are averaged over the 100 samples, and $\log_{10}$ is applied to the average.
        }
        \label{fig:supp_L96_simulation_noise_injected}
    \end{center}
\end{figure}

In this subsection only, we solve the temporal evolution equation~(\ref{M-eq:governing-eq-along-t}) of the main article with noise ($\sigma_\lambda > 0$) following Algorithm~\ref{alg:prediction}.
The amplitude $\sigma_\lambda$ is not tuned here: it is given by the trace matching in Eq.~(\ref{M-eq:sigma-trace-matching}), as in all the other experiments (main article Appendix~\ref{M-subsec:specification-of-the-noise-amplitude-sigma-lambda}).
The reference data are constructed by adding noise to physics-based simulation results as in Eq.~(\ref{M-eq:ou-solution-spatiotemporal}) and Algorithm~\ref{alg:training}.
Since the noise amplitude is quite small when $\lambda \approx 0$, we illustrate the cases $\lambda=0.2$ and $\lambda=0.4$ in Fig.~\ref{fig:supp_L96_simulation_noise_injected}.

The model remains numerically stable even with noise injection.
At $\lambda=0.2$, the simulated spatiotemporal sample is similar to the reference.
At $\lambda = 0.4$, however, the signal-to-noise ratio decreases, making comparison with the reference more difficult.
Unlike scale-dependent coarse-graining (i.e., the Laplacian damping), this noise has statistically uniform amplitude across all Fourier modes, overwhelming large- and small-scale components and leaving the signal dominated by noise.

\FloatBarrier

\subsection{Dependence on the number of input time steps}
\label{subsec:dependence-on-the-number-of-input-time-steps}

It is common in time-series modeling to use multiple past time steps as inputs \citep[e.g.,][]{Tashiro+21,Yuan+24,Price+25}.
However, our choice of five input steps (history length $m = 5$) may appear large compared with those in data-driven models for more complex atmospheric flows \citep{Price+25}.
Here we discuss this choice based on the Mori--Zwanzig formalism \citep{Zwanzig61,Mori65,Kubo91}.

\begin{figure}[tb!]
    \begin{center}
        \includegraphics[width=\textwidth]{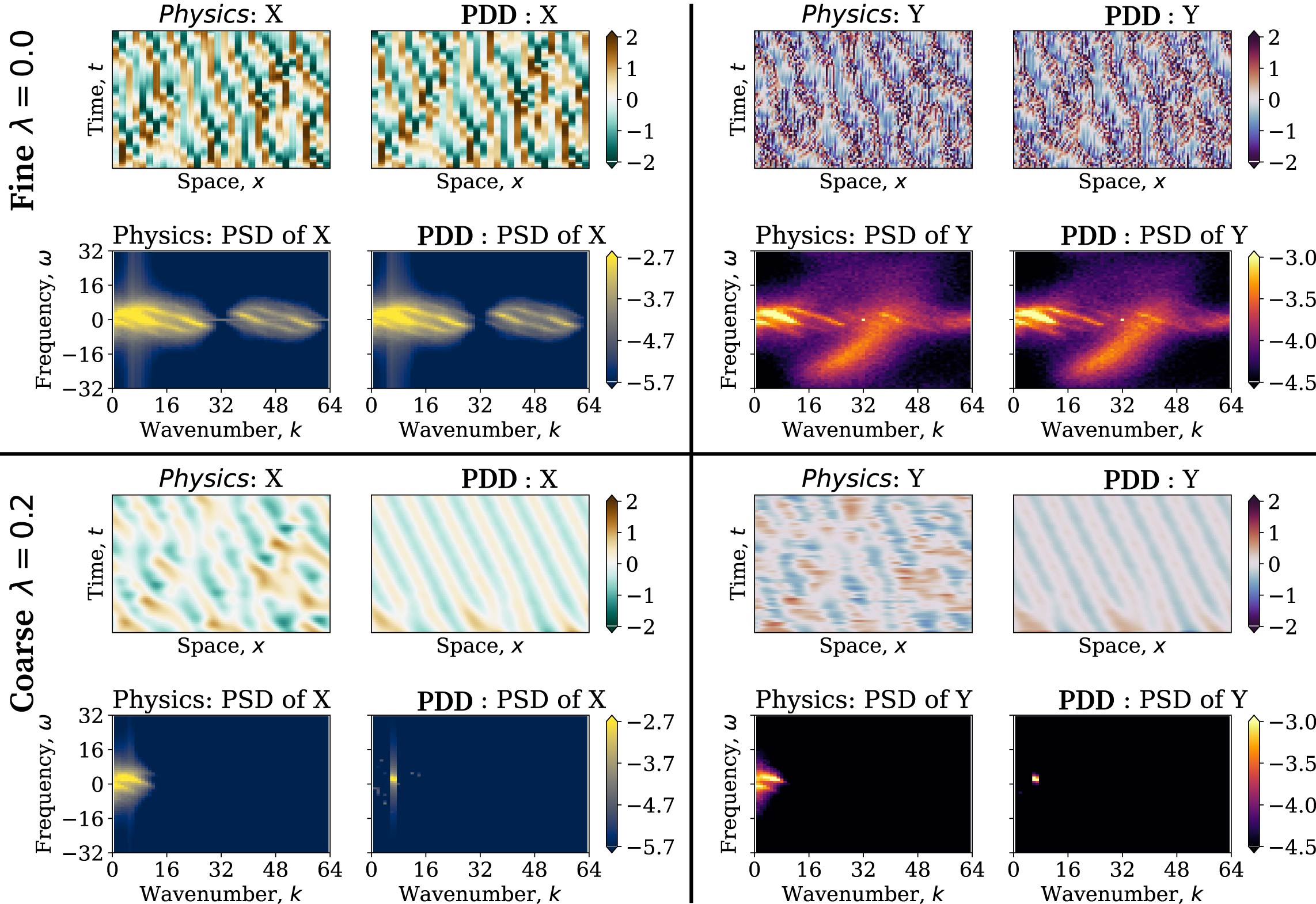} 
        \caption{
        Simulation results for the Lorenz-96 model using only the current time step as input (history length $m = 1$).
        Top two rows: $\lambda = 0$; bottom two rows: $\lambda=0.2$.
        Left: slow variable $X$; right: fast variable $Y$; in each half, physics-based simulation (Physics) and PDD, with one sample above and the spatiotemporal PSD below.
        PSDs are averaged over the 100 samples, and $\log_{10}$ is applied to the average.
        At $\lambda = 0$, simulation is accurate since the system is Markov.
        At $\lambda=0.2$, a single input step fails to capture large-scale variability, producing monotonous patterns.
        }
        \label{fig:supp_L96_simulation_window1}
    \end{center}
\end{figure}

Figure~\ref{fig:supp_L96_simulation_window1} shows results using only the current time step as input ($m = 1$).
At $\lambda = 0$, this model simulates accurately since the Lorenz-96 model is Markov at fine resolution (\Cref{subsec:construction-of-training-data}).
At $\lambda=0.2$, however, a single input step fails to reproduce large-scale variability, producing monotonous patterns instead.
In contrast, with five input steps, such variability is reproduced (Fig.~\ref{fig:supp_L96_simulation_lambda02}).

\begin{figure}[tb!]
    \begin{center}
        \includegraphics[width=0.9\textwidth]{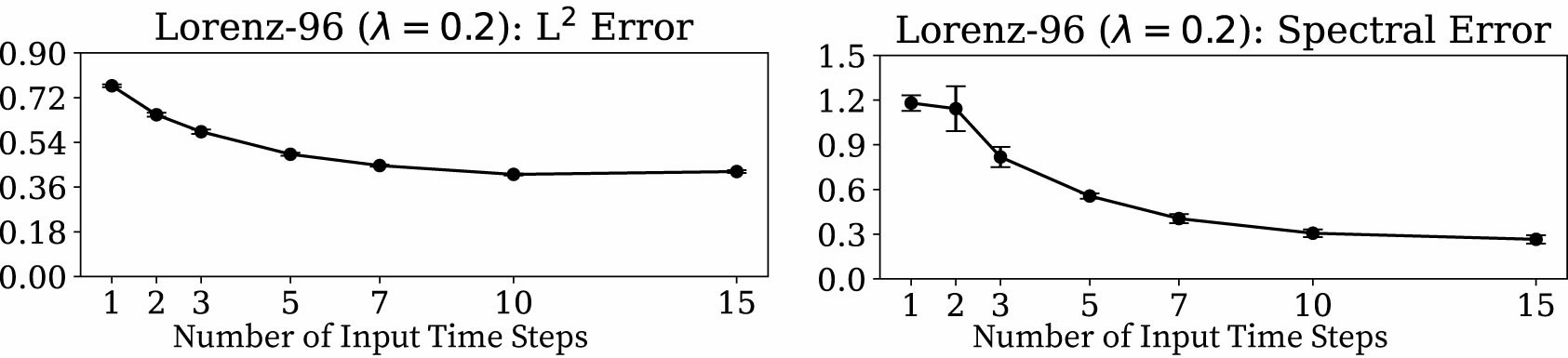} 
        \caption{Dependence of simulation accuracy on the number of input time steps (history length $m$) for the Lorenz-96 model at $\lambda=0.2$.
        Left: $L^2$ error in Eq.~(\ref{eq:supp-l2-err-definition}); right: spectral error in Eq.~(\ref{eq:supp-spec-err-definition}).
        The $L^2$ error is evaluated at the sixth time step.
        Error bars indicate the standard deviation over five training runs.
        Both metrics decrease as the history length increases, with the spectral error dropping from approximately 0.6 (5 steps) to 0.3 (15 steps).}
        \label{fig:supp_L96_window_size_dependence}
    \end{center}
\end{figure}

This result suggests that reproducing large-scale temporal evolution requires a sufficiently long history when small-scale components are eliminated.
Indeed, the simulation error decreases as the number of input steps increases (Fig.~\ref{fig:supp_L96_window_size_dependence}): when $\lambda=0.2$, increasing from 5 to 15 steps reduces the spectral error from approximately 0.6 to 0.3.

These results are consistent with the Mori--Zwanzig projection operator formalism \citep{Zwanzig61,Mori65,Kubo91}.
When small-scale components are eliminated by coarse-graining, the feedback they provide to large-scale dynamics must be represented using only the past history of large-scale variables, thereby breaking the Markov property.
In time-series modeling, multiple past states are often provided as inputs \citep[e.g.,][]{Tashiro+21,Yuan+24,Price+25}, likely because including history improves reproduction of large-scale variability.

In the main article (Section~\ref{M-sec:experiments}), we use five input time steps to adequately reproduce coarse-grained dynamics.
Using fewer steps simplifies the preparation of initial conditions and reduces computational cost, leading to a trade-off between accuracy and efficiency; we adopt five steps as a compromise.

Furthermore, unconditional generation accuracy also improves with more input steps (not shown in detail).
In unconditional generation, the reverse equation~(\ref{M-eq:reverse-rg-sde}) of the main article is integrated toward smaller $\lambda$ (from $\lambda_{\max}$ to $\lambda_{\min}$).
Since the dynamics become more strongly history-dependent at large $\lambda$, accurate score estimation requires multiple input steps.

\FloatBarrier

\subsection{Laplacian term in a variance-preserving process}
\label{subsec:effect-of-the-laplacian}

We compare two forward processes: a variance-preserving process \citep{Ho+20,Song+21} and its Laplacian-augmented extension.
This tests whether the Laplacian yields scale-selective suppression even in the presence of uniform damping.
Specifically, we consider the following forward equation,
\begin{align}
    \partial_\lambda u_\lambda(x,t) &= \left( \alpha\nabla_x^2 - \gamma \right) u_\lambda(x,t) +\beta \eta_\lambda(x,t). \label{eq:forward-sde-with-gamma}
\end{align}
We compare two cases: $\alpha = 0$ (no Laplacian) and $\alpha = 0.1$ (with Laplacian, as in the main experiments).
In both cases, we set $\gamma = 3$ and $\beta = \sqrt{2\gamma}$ following the DDPM setting (\Cref{subsec:ddpm-models}), which makes the process variance-preserving if $\alpha = 0$.
All models are trained with the path-KL objective $\mathcal{L}^{\mathrm{PDD}}$ in main article Eq.~(\ref{M-eq:loss-pdd}) (Algorithm~\ref{alg:training}), not by score-based learning.

\begin{figure}[tb!]
    \begin{center}
        \includegraphics[width=\textwidth]{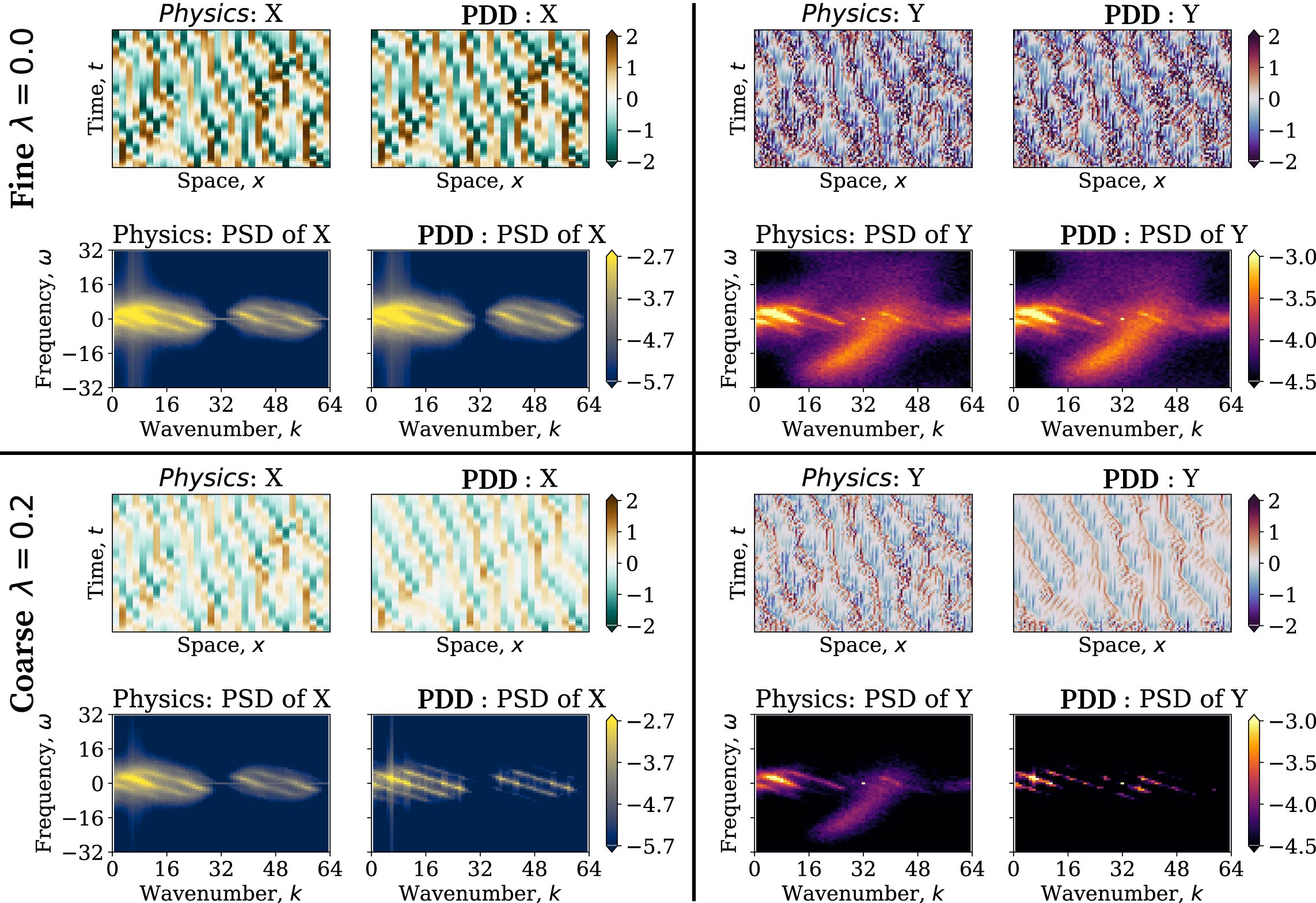} 
        \caption{
        Simulation results for the Lorenz-96 model, where the forward process is described by Eq.~(\ref{eq:forward-sde-with-gamma}) with $\alpha=0$ (no Laplacian), $\gamma=3$, and $\beta=\sqrt{2\gamma}$.
        Top two rows: $\lambda = 0$; bottom two rows: $\lambda=0.2$.
        Left: slow variable $X$; right: fast variable $Y$; in each half, physics-based simulation (Physics) and PDD, with one sample above and the spatiotemporal PSD below.
        PSDs are averaged over the 100 samples, and $\log_{10}$ is applied to the average.
        }
        \label{fig:supp_L96_simulation_ablation_without_laplacian}
    \end{center}
\end{figure}

\begin{figure}[tb!]
    \begin{center}
        \includegraphics[width=\textwidth]{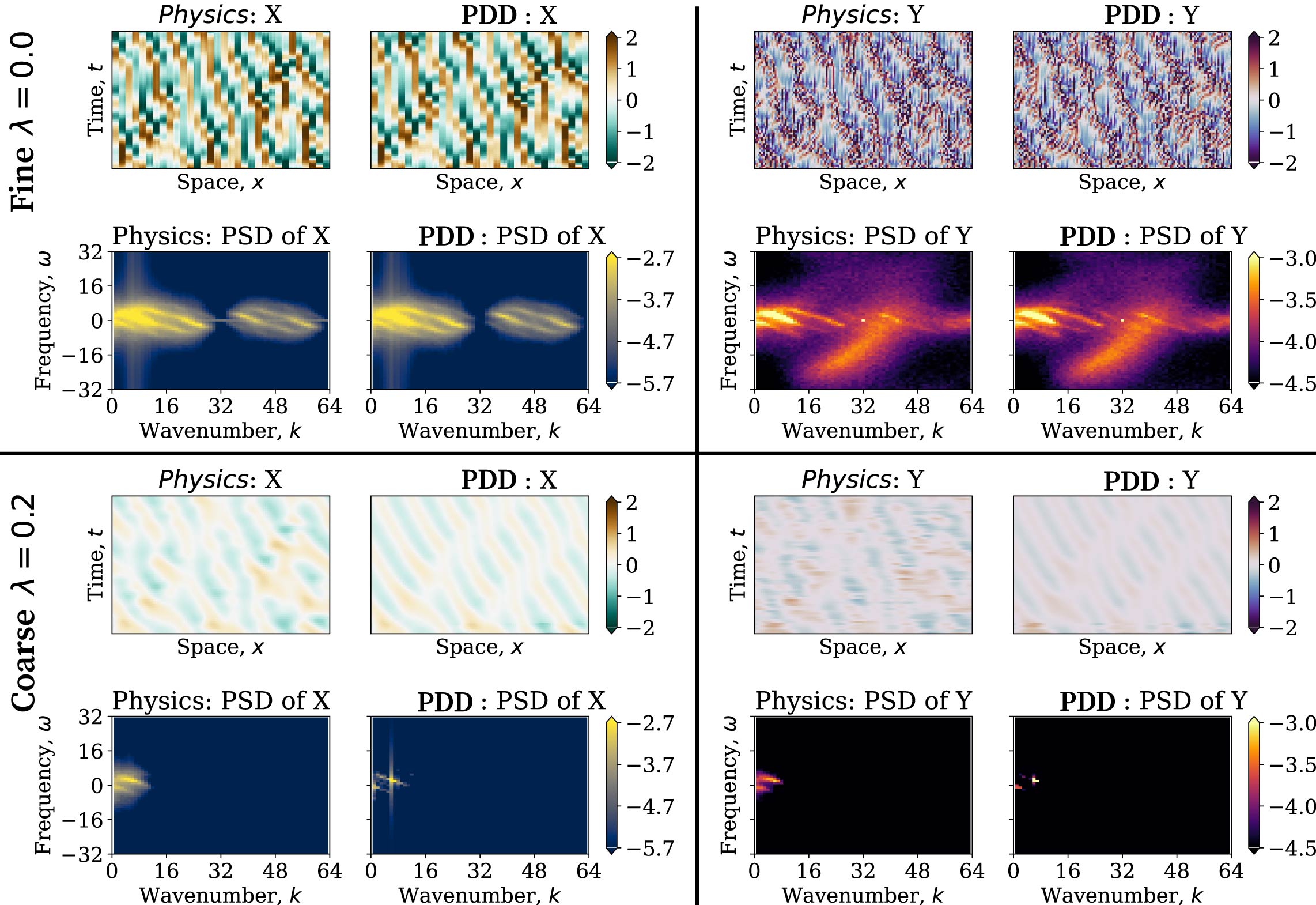} 
        \caption{
        Simulation results for the Lorenz-96 model, where the forward process is described by Eq.~(\ref{eq:forward-sde-with-gamma}) with $\alpha=0.1$, $\gamma=3$, and $\beta=\sqrt{2\gamma}$, in the same layout as Fig.~\ref{fig:supp_L96_simulation_ablation_without_laplacian}.
        PSDs are averaged over the 100 samples, and $\log_{10}$ is applied to the average.
        }
        \label{fig:supp_L96_simulation_ablation_with_laplacian}
    \end{center}
\end{figure}

Without the Laplacian ($\alpha = 0$), all Fourier modes are damped at the same rate, so the relative magnitudes of large- and small-scale components remain unchanged across $\lambda$ (Fig.~\ref{fig:supp_L96_simulation_ablation_without_laplacian}).
In contrast, with the Laplacian, small-scale components are sufficiently damped even when scale-independent damping $-\gamma u_\lambda(x,t)$ is included in Eq.~(\ref{eq:forward-sde-with-gamma}) (see Fig.~\ref{fig:supp_L96_simulation_ablation_with_laplacian}).
These results suggest that our method works even in the presence of uniform damping.

\FloatBarrier

\subsection{Effect of noise}
\label{subsec:effect-of-noise}

The forward process in main article Eq.~(\ref{M-eq:forward-rg-sde}) consists of a Laplacian term and a noise term.
In \Cref{subsec:effect-of-the-laplacian}, we studied the role of the Laplacian term; here, we ablate noise injection to examine its effect on simulation, generation, and super-resolution.
Specifically, we train a model using deterministically coarse-grained data:
\begin{align}
    u_\lambda(x,t) = \mathcal{C}_\lambda u_0(x,t),
\end{align}
which is constructed by setting $\epsilon \equiv 0$ in Eq.~(\ref{M-eq:ou-solution-spatiotemporal}), i.e., $u_\lambda = \mathcal{C}_\lambda u_0 + \sqrt{\Sigma_\lambda}\,\epsilon$ with $\epsilon=0$.
We keep $\beta>0$ as in the main experiments and use the same $\Sigma_\lambda$ (and hence $\sigma_\lambda$) for a controlled comparison.
We train this noise-free model with the objective $\mathcal{L}^{\mathrm{PDD}}$ in Eq.~(\ref{M-eq:loss-pdd}), which then reduces to a regression on deterministically smoothed fields.

From a theoretical perspective, diffusion induced by noise leads to a convolution of probability distributions, and the influence of small-scale components is reflected in the statistical properties of large-scale components through the effective theory \citep{Carosso20,Cotler+Rezchikov23a}.
In other words, noise statistically represents the eliminated small-scale degrees of freedom (main article Sections~\ref{M-subsec:rg-coarse-graining} and~\ref{M-subsec:intermediate-states}).
Here, we experimentally investigate the effect of noise.

\begin{figure}[tb!]
    \begin{center}
        \includegraphics[width=\textwidth]{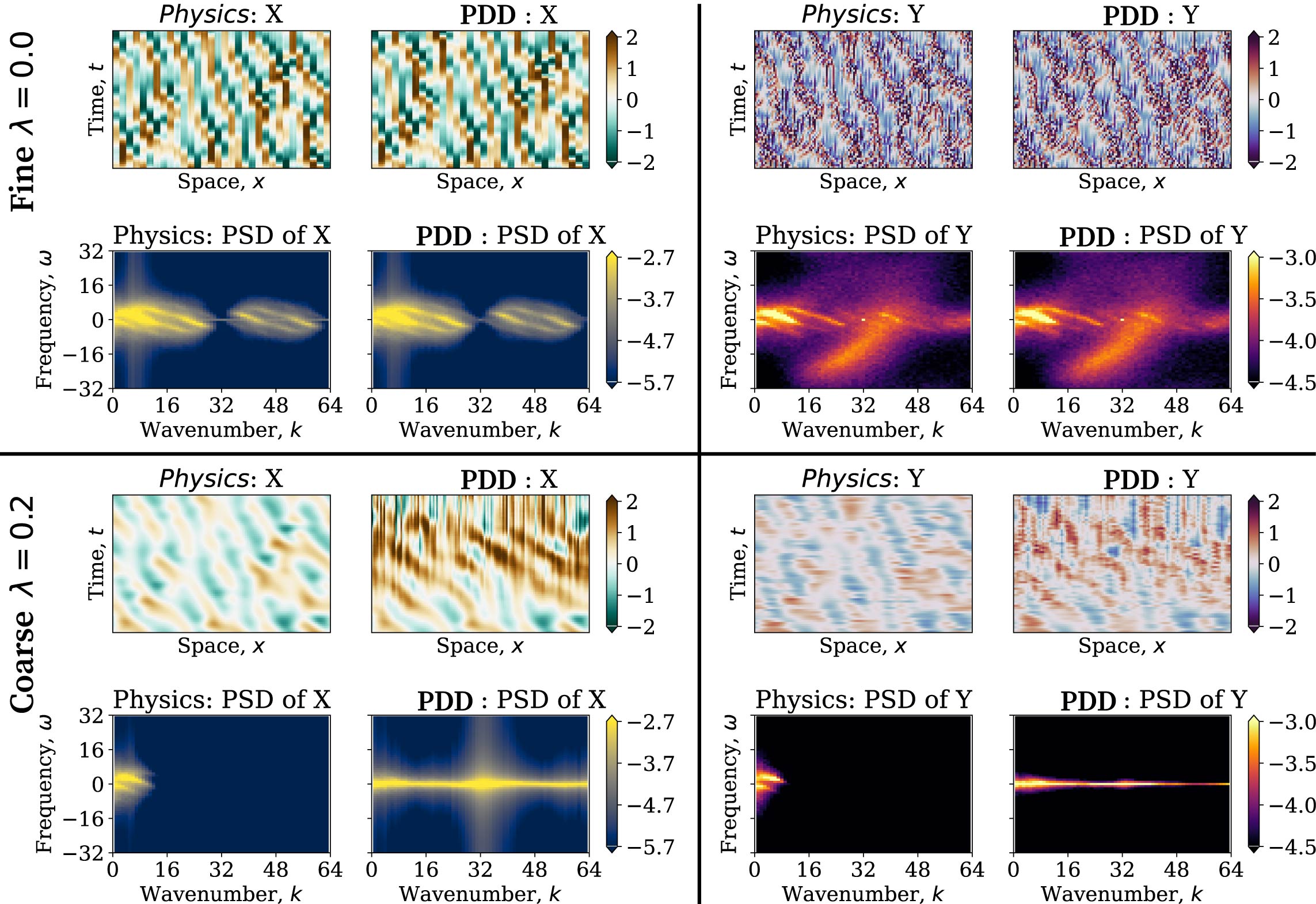} 
        \caption{Simulation results for the Lorenz-96 model trained without noise injection ($\epsilon = 0$ in main article Eq.~(\ref{M-eq:ou-solution-spatiotemporal})).
        Top two rows: $\lambda = 0$; bottom two rows: $\lambda=0.2$.
        Left: slow variable $X$; right: fast variable $Y$; in each half, physics-based simulation (Physics) and PDD.
        Each panel shows spatiotemporal evolution (top) and power spectral density (PSD; bottom).
        PSDs are averaged over the 100 samples, and $\log_{10}$ is applied to the average.
        While simulation at $\lambda = 0$ remains stable, numerical instability increases at larger $\lambda$, demonstrating that noise is essential for stable coarse-grained simulation.
        }
        \label{fig:supp_L96_simulation_ablation_without_noise}
    \end{center}
\end{figure}

Even without noise, simulations are numerically stable when $\lambda \approx 0$; however, numerical instability increases and the simulation fails as $\lambda$ becomes larger (Fig.~\ref{fig:supp_L96_simulation_ablation_without_noise}).
It is known that injecting noise during training improves the numerical stability of learned simulators \citep{Pfaff+21,Sanchez+20}, and our results are consistent with this observation.

When no noise is injected during training, generation fails (not shown in detail).
Specifically, the generated spatiotemporal samples look like noise, and the characteristic spatiotemporal patterns are not reproduced.
Theoretically, noise induces a convolution of probability distributions, which corresponds to statistically integrating out the eliminated small-scale degrees of freedom \citep{Carosso20,Cotler+Rezchikov23a}.
Data generation via the reverse process can therefore be interpreted as reconstructing small-scale components from a statistically integrated representation \citep{Cotler+Rezchikov23a}.
When noise is not injected during training, such statistical integration does not occur, and the model fails to learn how to reconstruct small-scale components.

\begin{figure}[tb!]
    \begin{center}
        \includegraphics[width=\textwidth]{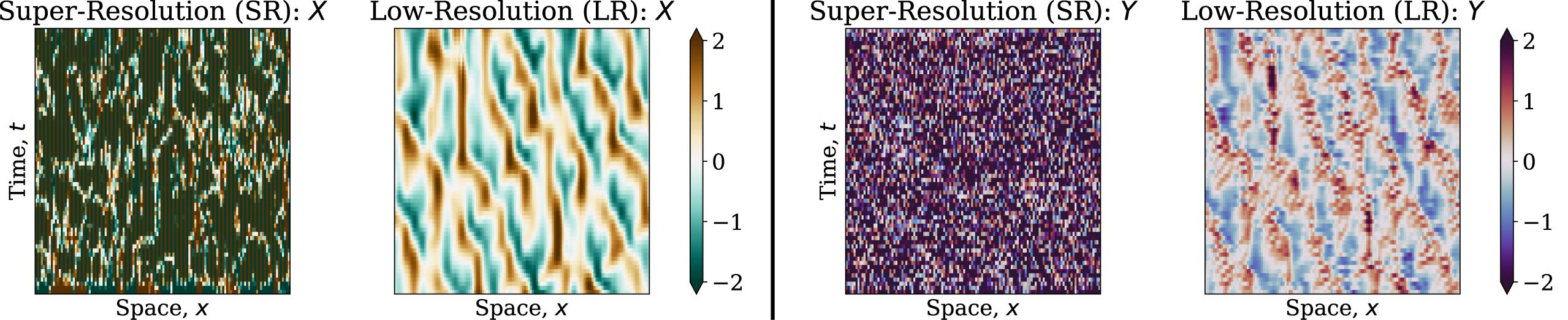} 
        \caption{
        Super-resolution results for the Lorenz-96 model, without noise during training.
        Left: slow variable $X$; right: fast variable $Y$; in each half, the super-resolved output (SR) and the coarse-grained (low-resolution, LR) input.
        The LR input is obtained from the simulation at $\lambda=0.025$.
        Using this as input, super-resolution is performed following Algorithm~\ref{alg:generation}.
        }
        \label{fig:supp_L96_super_resolution_no_noise}
    \end{center}
\end{figure}

We further examine whether super-resolution is possible in the absence of noise.
Choosing a relatively small diffusion scale for which coarse-grained (low-resolution) simulation is successful, $\lambda = 0.025$, we apply the super-resolution procedure.
As expected, super-resolution also fails in this case (Fig.~\ref{fig:supp_L96_super_resolution_no_noise}).
This result is consistent with the failure of unconditional spatiotemporal data generation and suggests that, without statistical integration induced by noise, the model does not learn to reconstruct small-scale components.

\FloatBarrier

\subsection{Experimental results with FNO}
\label{subsec:experimental-results-with-fno}

We evaluate PDD using the FNO architecture in place of the U-Net (\Cref{subsec:neural-network-architecture}).
For simulation at each $\lambda$ and unconditional generation, the FNO-based model yields results comparable to those obtained with the U-Net.
Here, we focus on super-resolution, which combines forward-$t$ simulation and reverse-$\lambda$ sampling.

\begin{figure}[tb!]
    \begin{center}
        \includegraphics[width=\textwidth]{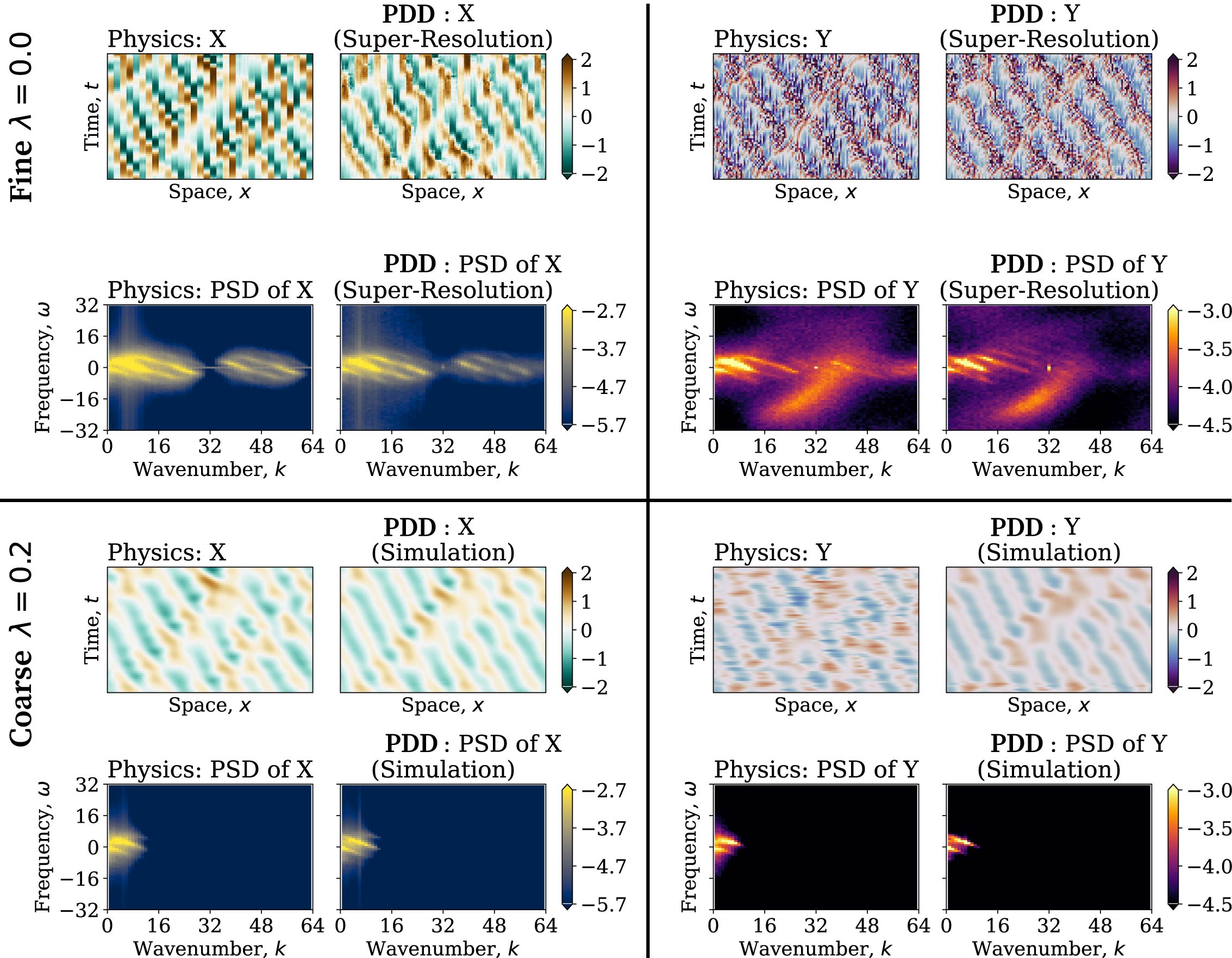} 
        \caption{
        Super-resolution and simulation results for the Lorenz-96 model using an FNO-based PDD.
        Top two rows: $\lambda = 0$ (super-resolution); bottom two rows: $\lambda=0.2$ (simulation).
        Left: slow variable $X$; right: fast variable $Y$.
        Each panel shows spatiotemporal evolution (top) and power spectral density (PSD; bottom), comparing physics-based simulation (left column) with the FNO-based PDD (right column).
        PSDs are averaged over the 100 samples, and $\log_{10}$ is applied to the average.
        The coarse-grained simulation ($\lambda=0.2$) captures large-scale dynamics, and the reverse-$\lambda$ integration super-resolves this to $\lambda = 0$ (Algorithm~\ref{alg:generation}), recovering small-scale structure consistent with the physics-based reference.
        }
        \label{fig:supp_L96_fno_simulation_and_super_resolution}
    \end{center}
\end{figure}

Following the main article (Section~\ref{M-subsec:experimental-results}), we perform simulation at $\lambda=0.2$ and apply reverse-$\lambda$ sampling to obtain super-resolved fields at $\lambda = 0$ (Fig.~\ref{fig:supp_L96_fno_simulation_and_super_resolution}).
The coarse-grained (low-resolution) simulation at $\lambda=0.2$ reproduces the coarse-grained physics-based simulation well, capturing the fluctuations of large-scale components.
The super-resolved result at $\lambda = 0$ recovers small-scale components and exhibits spatiotemporal patterns consistent with the physics-based simulation from the same initial condition.
These results suggest that our proposed method is effective across different network architectures.

\FloatBarrier 

\end{document}